\documentclass[a4paper,11pt]{article}
\pdfoutput=1 
\RequirePackage{snapshot}  

\usepackage{jheppub} 

\usepackage{bm,amsmath,amssymb,slashed,graphicx,%
            enumerate,alltt,xspace,multirow,xcolor,mathrsfs}
\usepackage{fancyvrb}
\usepackage{booktabs}
\usepackage{graphicx}
\usepackage{subcaption}
\usepackage{xspace}
\usepackage[utf8]{inputenc}
\usepackage[export]{adjustbox}

\makeatletter
\g@addto@macro\bfseries{\boldmath}
\makeatother

\definecolor{labelkey}{rgb}{0,0.5,0.0}

\usepackage{listings}
\definecolor{semiblue}{rgb}{0.3,0.3,0.8}
\newcommand{\logbook}[2]{}

\definecolor{darkgreen}{rgb}{0,0.7,0}
\definecolor{ddarkgreen}{rgb}{0,0.5,0}
\definecolor{grey}{rgb}{0.5,0.5,0.5}
\definecolor{orange}{rgb}{1.0,0.4,0.4}
\definecolor{cyan}{rgb}{0.0,1.0,1.0}
\definecolor{magenta}{rgb}{1.0,0.0,1.0}

\newcommand{\cdiff}{c_{_\delta}}

\newcommand{\GeV}{\;\mathrm{GeV}}

\newcommand{\order}[1]{{\cal O}\left(#1\right)}
\newcommand{\as}{\alpha_s}

\newcommand{\abar}{{\bar{\alpha}}}

\newcommand{\rtas}{\sqrt{\alpha_s}}

\newcommand{\qbar}{{\bar q}}
\newcommand{\ktcut}{k_{t,{\text {cut}}}}
\newcommand{\nlp}{N^{\text {(Lund)}}}
\newcommand{\avg}[1]{\left\langle #1\right\rangle}
\newcommand{\avnlp}{\langle N^{\text {(Lund)}}\rangle}

\newcommand{\avnlpi}{\langle N_i^{\text {(Lund)}}\rangle}
\newcommand{\avnlpZ}{\langle N_Z^{\text {(Lund)}}\rangle}
\newcommand{\avnlpH}{\langle N_H^{\text {(Lund)}}\rangle}

\newcommand{\coshnu}{\cosh\nu}
\newcommand{\sinhnu}{\sinh\nu}

\newcommand{\beq}{\begin{eqnarray}}
\newcommand{\eeq}{\end{eqnarray}}
\newcommand{\dis}{\displaystyle}
\newcommand{\dd}{\mathrm{d}}
\renewcommand{\l}{{\ell}}

\newcommand{\normtheta}{\bar\theta}

\newcommand{\DL}{\text{DL}\xspace}
\newcommand{\NDL}{\text{NDL}\xspace}
\newcommand{\NNDL}{\text{NNDL}\xspace}

\definecolor{colourndlsqr}{RGB}{255,128,  0} 
\definecolor{coloureloss} {RGB}{255,  0,  0} 
\definecolor{colourclust} {RGB}{255,  0,255} 
\definecolor{colourhme}   {RGB}{  0,128,  0} 
\definecolor{colourpair}  {RGB}{  0,  0,255} 
\definecolor{colourend}   {RGB}{  0,  0,  0} 

\title{Lund and Cambridge multiplicities for precision physics}

\preprint{OUTP-22-07P}

\newcommand{\OXaff}{Rudolf Peierls Centre for Theoretical Physics, Clarendon 
Laboratory, Parks Road,
  University of Oxford, Oxford OX1 3PU, UK}
\newcommand{\IPhTAff}{Universit\'e Paris-Saclay, CNRS, CEA, Institut de physique 
th\'eorique, 91191, Gif-sur-Yvette, France}

\author[a]{Rok Medves,}%
\author[b]{Alba Soto-Ontoso,}%
\author[b]{Gregory Soyez}%

\emailAdd{rok.medves@physics.ox.ac.uk}
\emailAdd{alba.soto@ipht.fr}
\emailAdd{gregory.soyez@ipht.fr}

\affiliation[a]{\OXaff}
\affiliation[b]{\IPhTAff}

\date{Received: date / Accepted: \today}

\abstract{
  We revisit the calculation of the average jet multiplicity in
  high-energy collisions. 
  First, we introduce a new definition of (sub)jet multiplicity based
  on Lund declusterings obtained using the Cambridge jet algorithm.
  We develop a new systematic resummation approach.
  This allows us to compute both the Lund and the Cambridge average
  multiplicities to next-to-next-to-double (\NNDL) logarithmic
  accuracy in electron-positron annihilation, an order
  higher in accuracy than previous works in the literature.   
  We match our resummed calculation to the exact NLO ($\order{\as^2}$)
  result, showing predictions for the Lund multiplicity at LEP
  energies with theoretical uncertainties up to $50\%$ smaller than the
  previous state-of-the-art.
  Adding hadronisation corrections obtained by Monte Carlo
  simulations, we also show a good agreement with existing Cambridge
  multiplicity data.
  Finally, to highlight the flexibility of our method, we extend the
  Lund multiplicity calculation to hadronic collisions where we reach
  next-to-double logarithmic accuracy for colour singlet production.
}

\begin{document}


\maketitle

\section{Introduction}\label{sec:intro}

Multiplicities, e.g.\ of (charged) particles or jets, are amongst the
most fundamental observables studied in collider physics.
Since they probe the full multiple branching structure of QCD, they
have been used in various forms to probe the dynamics of the
strong interaction.
This has been the case, historically, when assessing Quantum
Chromodynamics (QCD) as the fundamental theory of 
strong interactions at $e^+e^-$ colliders (see
e.g.~\cite{PLUTO:1980mrm,OPAL:1990xiz,L3:1992nwf,ALEPH:1996oqp}).
Besides this, multiplicities have been relied upon in several
contexts, including for instance the tuning of Monte Carlo event
generators (see
e.g.~\cite{Gieseke:2012ft,Skands:2014pea,Sherpa:2019gpd}) or the
tagging of quark and gluon jets
(e.g.~\cite{Gallicchio:2011xq,ATLAS:2014vax,Frye:2017yrw}).
Experimentally, charged-particle multiplicity is one of the simplest and cleanest 
observables, and its importance stretches from calibration to 
advanced tagging techniques.
In fact, the first experimental paper using LHC data reported the
measurement of charged hadron multiplicity in proton-proton
collisions~\cite{ALICE:2009wpl}.

From the theoretical viewpoint, the definition of multiplicity can be
rather subtle.
Early work considered the total hadronic multiplicity within a jet. 
This is equivalent to the first moment of the jet fragmentation function 
and was studied in perturbative QCD in several
works~\cite{Bassetto:1979nt,Furmanski:1979jx,Mueller:1981ex,Webber:1984jp,
  Malaza:1984vv,Malaza:1985jd,Dremin:1993tg,Konishi:1979cb,Dokshitzer:1991wu,Perez-Ramos:2013eba}.
These calculations played a pivotal role in understanding the
singularity structure of QCD, see Ref.~\cite{Dremin:2000ep} for a
review.
Since the total hadronic multiplicity is obviously
infrared-and-collinear unsafe, the calculations introduced an infrared
regulator $Q_0\sim\Lambda_{\text{QCD}}$ that was fitted to the data.
The probability distribution of hadronic multiplicity in QCD jets was measured 
in $e^+e^-$ annihilation~\cite{OPAL:1997dkk} and shown to be well reproduced 
by the aforementioned analytic calculations. 

A theoretically more favourable way of determining the multiplicity of
an event was established with the advent of jet clustering algorithms,
basing it on the number of reconstructed jets.
In Ref.~\cite{Catani:1991pm}, the average jet multiplicity, $\langle N\rangle$, 
was defined as the average number of reconstructed jets that pass a 
certain resolution cut, typically a transverse momentum cut 
$\ktcut$, guaranteeing infrared-and-collinear safety. As $\ktcut$ is reduced 
from the hard scale $Q$ down, more jets are resolved and 
$\langle N\rangle$ increases. 
A step further was considered in Ref.~\cite{Catani:1992tm}, where the
concept of `subjet multiplicity' was introduced. In this case, a jet
finder is run twice with two different resolution scales. The jets are
defined with $k_{t0,{\text {cut}}}$ and the subjets with a finer scale
$k_{t1,{\text {cut}}}< k_{t0,{\text {cut}}}$.  The first step is thus
to classify the event as an $n$-jet event and then, for each
individual jet, to count the number of subjets above
$k_{t1,{\text {cut}}}$.

Let us now discuss the basic structure of the average jet multiplicity
in perturbative QCD. The presence of two disparate scales in the problem, 
either $\ktcut$ and the hard scattering scale $Q$ in the case of jet multiplicity 
or $k_{t0,{\text {cut}}}$ and $k_{t1,{\text {cut}}}$ for subjet multiplicity, 
splits the phase-space into three different regimes.
If $\ktcut\sim Q$, fixed-order perturbation theory applies. Otherwise, 
when  $\Lambda_\text{QCD}\ll \ktcut\ll Q$ the logarithm $L\equiv \ln(Q/\ktcut)$ 
becomes large and must be resummed to all orders in order to guarantee the 
convergence of the perturbative expansion.
In such a case the resummation structure of jet multiplicity can be organised as follows
\begin{equation}
\label{eq:log-counting}
  \langle N(\as,L)\rangle
  = \langle N(\as,0)\rangle \bigg[
    \underbrace{h_1(\as L^2)}_{\DL}
    + \underbrace{\rtas h_2(\as L^2)}_{\NDL}
    + \underbrace{\as h_3(\as L^2)}_{\NNDL} + \dots
  \bigg]
  + \order{e^{-|L|}}\, ,
\end{equation}
where $\as$ is the strong coupling constant and the N$^k$DL function 
$\as^{k/2}h_{k+1}(\as L^2)$ resums terms of order $\as^n 
L^{2n-k}$. That is, the function $h_1$ captures the double logarithmic (\DL) 
contribution, $h_2$ the next-to-double-logarithmic (\NDL) enhancements, $h_3$ the 
next-to-next-to-double-logarithmic (\NNDL) enhancements and so on.
Typical values of $\as L^2$ at high-energy colliders are approximately 
$\as L^2 \leq 5$ (see e.g.\ Ref.~\cite{Hamilton:2020rcu}).
Finally, whenever $\ktcut \approx \Lambda_{\rm QCD}$ the jet multiplicity 
calculation enters the domain of non-perturbative QCD. 

Previous calculations of Eq.~\eqref{eq:log-counting} date back to the
early 90's. In Ref.~\cite{Catani:1991pm}, the average jet multiplicity in
$e^+e^-$ collisions was computed at \NDL accuracy
using the Durham clustering algorithm~\cite{Catani:1991hj,Catani:1992tm}
(see also Ref.~\cite{Gerwick:2012fw} for a calculation of the average jet
multiplicities using different jet algorithms).  
This resummed calculation was then used by the OPAL Collaboration to extract the 
value of the strong coupling constant at the $Z$-boson mass~\cite{OPAL:1993pnw}. 
In both theory and experimental papers it was realised through general-purpose 
Monte Carlo event generators that non-perturbative corrections affected the 
jet multiplicity for $\ktcut$ values well above $\Lambda_\text{QCD}$,
i.e.\ $\ktcut \approx 4 \GeV$~\cite{Catani:1991pm}.
This was among the motivations to develop a new jet clustering algorithm that uses 
angular distance as its metric, the so-called Cambridge
algorithm~\cite{Dokshitzer:1997in}. 
To demonstrate the benefits of the novel algorithm, the authors of
Ref.~\cite{Dokshitzer:1997in} computed the average jet multiplicity
using the Cambridge algorithm at \NDL accuracy and showed the
reduction of non-perturbative effects with respect to the $k_t$
prescription, i.e.\ the value of $\ktcut$ at which parton-level and
hadron-level Monte Carlo results start to differ significantly is reduced from
$\ktcut\sim 4 \GeV$ to $\ktcut\sim 1.5 \GeV$ (see
  Ref.~\cite{Dokshitzer:1997in} as well as Fig.~\ref{fig:v-OPAL-mc}
  below)
when using the Cambridge algorithm instead of Durham.

Concerning other collision systems, the extension of both the 
$k_t$-clustering algorithm and the multiplicity calculation to 
deep-inelastic scattering was presented shortly after the $e^+e^-$ case in 
Refs.~\cite{Catani:1992rm, Catani:1993yx}, respectively.
The first subjet multiplicity calculation in a hadron collider jet was reported 
in Refs.~\cite{Seymour:1996np, Forshaw:1999iv} and used to extract the 
multiplicity ratio between quark and gluon jets at Tevatron 
energies~\cite{D0:2001nam}. So far, average subjet multiplicity has neither been 
predicted nor measured at LHC energies.   

More recently, a rising interest in jet substructure has prompted
a family of 
analysis based on the Lund jet plane, originally introduced in
Ref.~\cite{Dreyer:2018nbf} for proton-proton ($pp$) collisions.
In such an approach, a full clustering tree, based on the
Cambridge/Aachen (C/A) clustering
algorithm~\cite{Dokshitzer:1997in,Wobisch:1998wt}, is constructed for
an event, tracking its properties at each clustering vertex.
Traversing backwards the C/A clustering sequence one
constructs a tree structure, the Lund planes, where the kinematic
properties of the C/A pairwise recombinations are tuples
called {\em Lund declusterings}. This construction provides an
angular-ordered picture similar to that of Lund
diagrams~\cite{Andersson:1988gp} used in resummation and
Monte Carlo developments.
This provides a useful tool both theoretically and experimentally.
For example, a precise calculation of the primary Lund plane
density in perturbative QCD~\cite{Lifson:2020gua} has been
successfully compared to the ATLAS experimental measurement in
Ref.~\cite{ATLAS:2020bbn}.
The Lund jet plane has also been successfully utilised to inform
machine-learning-based studies for jet tagging~\cite{Dreyer:2020brq,Fedkevych:2022mid}
and quark/gluon jet discrimination~\cite{Dreyer:2021hhr}.

The goal of this paper is to extend the programme of Lund plane physics
to the concept of (sub)jet multiplicity. To do so, we 
propose an alternative definition of subjet multiplicity based on Lund
declustering that we dub `Lund multiplicity'.
We compute Lund multiplicity at \NNDL accuracy --- i.e.\ up to and
including the $h_3$ term in Eq.~\eqref{eq:log-counting} --- in
$\ell^+\ell^- \to Z\to q\bar q$ and $\ell^+\ell^- \to H\to gg$
collisions, an order higher in logarithmic accuracy than previous
works in the literature.
We derive this result from a novel resummation approach that does not
rely on the generating functional approach.
As a byproduct, we also obtain the resummed average Cambridge multiplicity at
\NNDL accuracy.
Our full results are summarised in Sec.~\ref{sec:final-result}.
We further provide a compact analytic expression for the subjet
multiplicity at \NDL accuracy for colour singlet production in
hadronic collisions, i.e.\ Drell-Yan ($pp \to q\qbar \to Z$) and Higgs
production via gluon fusion ($pp \to gg \to H$). This would, for
example, be directly useful to test the logarithmic accuracy of new
hadron-collider showers, as was done for $e^+e^-$ collisions in
Ref.~\cite{Dasgupta:2020fwr}.

The potential impact of our result is two-fold.
First, our \NNDL result for the
Cambridge multiplicity could be matched with recent progress to
compute next-to-next-to leading order (NNLO, $\as^3$) fixed-order
distributions.
This would yield to interesting phenomenological studies either at LEP
(e.g.\ compared to measurements from the OPAL collaboration~\cite{JADE:1999zar})
or at future circular colliders.
Besides the potential phenomenological impact of this calculation, an
analytic formula for \NNDL subjet multiplicity is fundamental for
testing the logarithmic accuracy of parton showers beyond NLL
accuracy, along the lines of Ref.~\cite{Dasgupta:2020fwr}. 

This paper is organised as follows. We begin by providing the
algorithmic definition of Lund multiplicity in
Sec.~\ref{sec:lund-mult-def}.
In Sec.~\ref{sec:recap-dl-ndl}, we introduce the building blocks of
the new resummation approach for subjet multiplicity and, as a warm-up
exercise, use it to re-derive previous results in the literature at \DL
and \NDL accuracy.
Sec.~\ref{sec:nndl} is the core of this paper since it contains all
the steps of the \NNDL calculation, with ready-to-use formul\ae\
for both the Lund and Cambridge multiplicities given
in Sec.~\ref{sec:final-result}.
We cross-check the $\order{\as^2}$ expansion of our resummation against {\tt
  Event2}~\cite{Catani:1996jh,Catani:1996vz} simulations in
Sec.~\ref{sec:event2}.
In Sec.~\ref{sec:matching}, we match our resummed result to exact
next-to-leading order (NLO, $\mathcal{O}(\alpha^2_s)$) distributions
and briefly discuss Lund multiplicity at LEP energies.
In Sec.~\ref{sec:v-OPAL} we provide a short comparison to the
Cambridge multiplicity measured by the OPAL
collaboration~\cite{JADE:1999zar}, including hadronisation effects
from Monte Carlo simulations.
The extension of our calculation to initial-state radiation is
presented in Sec.~\ref{sec:ndl-pp}, where we compute the Lund
multiplicity at \NDL accuracy in colour singlet production.
Finally, we conclude and outline some potential extensions of this
work in Sec.~\ref{sec:conclusions}.
The impact of choosing a different jet clustering algorithm and/or
recombination scheme is presented in Appendix~\ref{app:mult-def}.

\section{Lund-based multiplicity}\label{sec:lund-mult-def}

We begin by introducing a novel algorithm for evaluating the
multiplicity of jets with a relative $k_t$ above a given $\ktcut$ in
$e^+e^-$ collisions.
The algorithm draws from the Lund diagram representation of the phase-space 
where the emission kinematics is visualised in the $(\eta, \ln k_t)$
plane~\cite{Andersson:1988gp,Andersson:1988ee,Dreyer:2018nbf}, where
$\eta$ is the rapidity of the emission, and $Q$ the centre of mass energy. 

The procedure to compute the Lund multiplicity, $\nlp$, is the
following. The starting point is to cluster the full event with the
Cambridge algorithm~\cite{Dokshitzer:1997in} with resolution parameter
$y_\text{cut}=1$,\footnote{or, equivalently, the generalised
  $e^+e^-$ $k_t$ algorithm with $p=0$ and
  $R>\pi$~\cite{Cacciari:2011ma}.
  Note that we use the standard convention whereby $y_\text{cut}$ is
  normalised by the squared centre-of-mass energy $Q^2$.} so as to
generate a single jet with an angular-ordered clustering sequence.
We first undo the last step of the clustering i.e.\ get the two
exclusive Cambridge jets. We work in the centre-of-mass of the
collision, where this yields a pair of back-to-back jets, splitting
the event in two hemispheres. For each jet, we proceed as follows
(still working in the centre-of-mass of the collision)
\begin{enumerate}
\item Set $\nlp=1$.
\item \label{item:lund-multiplicity-step2}
  Undo the last clustering step to generate two subjets $j_1$ and
  $j_2$, with $j_1$ the most energetic, i.e.\ $E_{1}>E_{2}$.
\item \label{item:lund-multiplicity-step3}
Calculate the relative transverse momentum of the splitting as
\begin{equation}
k_t \equiv \min(E_1,E_2)\sin\theta = E_2\sin\theta\, ,
\label{kt:def}
\end{equation}
  with $\theta$ the opening angle between the
  pair of subjets.

\item \label{item:lund-multiplicity-step4}
  If $k_t\ge\ktcut$ the splitting contributes to the Lund
  multiplicity, i.e.\ $\nlp$ is incremented by one, and we go back to
  step~\ref{item:lund-multiplicity-step2} for {\it each} of the two subjets.
\item Otherwise, if $k_t<\ktcut$, repeat from step 
\ref{item:lund-multiplicity-step2} following only the hardest subjet $j_1$.
\end{enumerate}    
The procedure terminates when there is nothing left to decluster.
The event-wide Lund multiplicity would then be the sum of the results
obtained for each of the two exclusive jets.

Naturally, the Lund multiplicity fluctuates on an event-by-event basis.
In this work, we compute the average $\avnlp$, up to \NNDL accuracy,
and leave the calculation of the full distribution (or of its moments)
for future work.

For the case of $pp$ collisions we introduce an analogous procedure
where (a) instead of using the Cambridge algorithm we cluster the
event with Cambridge/Aachen~\cite{Dokshitzer:1997in, Wobisch:1998wt}
with a finite radius $R$ of order 1 (in practice, we use $R=1$), (b)
the ordering measure in step~\ref{item:lund-multiplicity-step2} is
taken by the (sub)jet transverse momentum relative to the beam, $p_t$,
instead of its energy, and (c) the relative transverse momentum of a
declustering in step~\ref{item:lund-multiplicity-step3} is defined as
$k_t = \min{(p_{t,i}, p_{t,j})}\Delta_{ij}$ with the distance
$\Delta_{ij}=\sqrt{\Delta y_{ij}^2 +\Delta \phi_{ij}^2}$ computed in
the rapidity-azimuth plane~\cite{Dreyer:2018nbf}. Roughly speaking,
the $pp$ procedure is an iteration of the $e^+e^-$ recipe for each of
the $R=1$ jets found by the Cambridge/Aachen algorithm. These $R=1$
jets can thus be viewed as the primary event radiation, associated
with the incoming-incoming dipole, just like the primary radiation in
$e^+e^-$ stems from the leading $q\bar q$ pair, and the subjets as
subsidiary radiation.
Similarly, in deep-inelastic scattering one would follow the same
steps but clustering in the Breit frame instead of in the event frame.

\paragraph{Relation to other multiplicities.}
The Lund-based definition of multiplicity shares some similarities with
other multiplicities in the literature. Let us first discuss its
relation with previous definitions of multiplicity using the Cambridge
algorithm~\cite{Dokshitzer:1997in}. We show in
Appendix~\ref{app:cambridge-vs-lund} that, in the $e^+e^-$ case, the
Lund procedure is equivalent to running the standard Cambridge
algorithm with $y_\text{cut}=1$ and counting the total number of
clusterings for which $k_t>\ktcut$, provided that the Lund $k_t$
definition, Eq.~\eqref{kt:def}, is used in both cases. If one uses
instead the Cambridge definition for the relative transverse momentum
of a clustering
\begin{equation}
\label{kt:def:cambridge}
k_t^\text{(Cam)} = \min{(E_i, E_j)}\sqrt{2(1-\cos{\theta_{ij}})},
\end{equation}
and imposes the condition $k_t^\text{(Cam)}>\ktcut$, one obtains a
different average multiplicity.
We show in Sec.~\ref{sec:nndl} that it starts to differ from the Lund
multiplicity at \NNDL and we compute the difference.
Furthermore, the \emph{standard} definition of the Cambridge jet multiplicity
would instead count the number of jets obtained when running the
Cambridge algorithm with $y_\text{cut}=\ktcut^2/Q^2$.
We show in Appendix~\ref{app:cambridge-vs-lund} (and
Sec.~\ref{sec:nndl}) that, at our targeted \NNDL accuracy, this is also
equivalent to counting the total number of clusterings in the Cambridge
sequence with $y_\text{cut}=1$ using the
definition~(\ref{kt:def:cambridge}) for the relative transverse
momentum.

Additionally, if, in the Lund declustering procedure, we only follow
the hardest branch (in step~\ref{item:lund-multiplicity-step4} above) 
instead of iterating over both branches, we recover the 
primary Lund plane multiplicity.
Note that beyond the soft-and-collinear approximation, the primary Lund-plane
multiplicity has a more involved analytic structure than $\avnlp$
since, as we will see below, clustering logarithms only appear from
\NNDL onwards while they would start already at \NDL for the primary
case (see e.g.\ Ref.~\cite{Lifson:2020gua}).
This is also reminiscent of the iterated Soft~Drop multiplicity
$n_{\text{SD}}$~\cite{Frye:2017yrw} defined as the number of
splittings on the primary Lund plane that satisfy the Soft~Drop
condition~\cite{Larkoski:2014wba} (for the Soft~Drop parameter $\beta$
set to -1).

An approach often used in the literature is to define the jet
multiplicity using the Durham ($k_t$) algorithm (as done, for example
in Ref.~\cite{Catani:1991pm}).
At \NDL accuracy, one gets the same multiplicity whether one uses the
Cambridge or Durham algorithm.
However, the two definitions start to differ at \NNDL.
We argue in Appendix~\ref{app:kperp-vs-cambridge} that the use of the
Cambridge algorithm considerably simplifies the \NNDL calculation,
while reaching \NNDL accuracy with the $k_t$ algorithm would require
semi-numerical ingredients.

Finally, we show in Appendix~\ref{app:recomb-scheme} that using a
different recombination scheme, such as the
winner-takes-all~\cite{Larkoski:2014uqa}, would only bring differences
beyond our targeted \NNDL accuracy.

\section{Revisiting \DL and \NDL results}\label{sec:recap-dl-ndl}

In this section, we derive the average Lund and Cambridge
multiplicities at double (\DL) and next-to-double (\NDL) logarithmic
accuracy.
These agree with the \NDL result of Ref.~\cite{Catani:1991pm} based on
the Durham algorithm, although they would start to differ at \NNDL
accuracy.

For the all-order \NDL resummation, we introduce a new formalism
compared to what was initially used in Refs.~\cite{Catani:1991pm,Catani:1993yx,Dokshitzer:1997in}. 
This new formalism is, we believe, considerably simpler and can almost straightforwardly be 
extended to \NNDL accuracy as we will show in Sec.~\ref{sec:nndl}.
The \DL resummation is presented in Sec.~\ref{sec:recap-dl}, with
the \NDL resummation following in Sec.~\ref{sec:recap-ndl}.

\subsection{Double-logarithmic (\DL) accuracy}\label{sec:recap-dl}

To gain insight into the resummation structure of double logarithms, we
start by discussing the first orders in the strong coupling.
For simplicity, we calculate the multiplicity in one of the two
hemispheres, $\avnlpi \equiv \avnlp/2$, with $i=q,g$, which depends 
only on the flavour of the hard partons produced in the underlying Born-level
process (at least at the accuracy we are interested in here).
We denote by $h_{1,2,3}^{(i)}$ the corresponding \DL, \NDL and \NNDL
functions that appear in Eq.~\eqref{eq:log-counting} (for a single hemisphere).

We consider two hard processes at Born-level: (i) a back-to-back $q\qbar$ pair 
produced by the decay of a $Z$ boson and (ii) the equivalent situation but for 
gluons arising from a Higgs boson, i.e.\ $e^+e^-\to H\to gg$. In this way, we 
compute the multiplicity for both quark and gluon jets. The centre-of-mass 
energy of the $e^+e^-$ system is denoted by $Q$.

Initially, at order $\as^0$ we have a single parton in each hemisphere and thus 
$\avnlpi=1$. At $\order{\as}$, we can either have a
real emission or a 1-loop virtual correction. We thus write
\begin{equation}\label{eq:basic-order-alphas}
  \avnlpi_{\order{\as}}
  = \frac{1}{\sigma_0 + \sigma_1}
  \left\{[1]\sigma_0
    +  \int \dd \Phi \left| \mathcal{M}_R\right|^2 \,
    \left[ 1 + \Theta{(k_t > \ktcut)}\right]
  + \int \dd \Phi \left| \mathcal{M}_V\right|^2 \, \left[ 1 \right]
  \right\},
\end{equation}
where $\sigma_0$ is the Born-level cross-section, $\sigma_1$ the
$\order{\as}$ correction to the inclusive cross-section, $\dd \Phi$ is
the phase-space element and $|\mathcal{M}_{R(V)}|^2$ is the real
(virtual) matrix element. Note that this $\order{\as}$ expression is completely 
general so as to facilitate future discussions at higher logarithmic 
accuracy.

In Eq.~\eqref{eq:basic-order-alphas}, the first term is the Born-level 
contribution with a single parton in the hemisphere, the second term is the real
$\order\as$ correction which, in addition to the hard parton, can receive
an additional contribution to the multiplicity if the emitted gluon
has a $k_t$ above $\ktcut$, and the last term is the one-loop virtual
correction where there is again a single parton in the hemisphere. 
The contribution from the hard parton in the real and virtual $\order\as$ terms 
gives (by definition) the correction $\sigma_1$ to the inclusive cross-section, 
therefore cancelling the overall normalisation.
Furthermore, in the contribution from the new emission (proportional
to $\Theta(k_t>\ktcut)$) the $\sigma_1$ correction in the
normalisation can be neglected since it's beyond \NDL
accuracy,\footnote{As we will see in Sec.~\ref{sec:nndl-top}, this also
  holds at NNDL accuracy, where $\sigma_1$ can also be neglected.  }
leading to
\begin{equation}\label{eq:multiplicity-dl-alphas-one}
  \avnlpi_{\order{\as}}
  = 1 + \int \dd \Phi \frac{\left| \mathcal{M}_R\right|^2}{\sigma_0}
  \, \Theta(k_t > \ktcut).
\end{equation}
At \DL accuracy, it is sufficient to consider the emission of a
soft-and-collinear gluon, in which case the Born-level kinematics
factorises from the gluon emission and
Eq.~\eqref{eq:multiplicity-dl-alphas-one} becomes
\begin{equation}\label{eq:multiplicity-dl-alphas-two}
  \avnlpi_{\order{\as}}
  = 1 +
  \int [\dd k] \left|\mathcal{M}(k)\right|^2
  \, \Theta(k_t > \ktcut),
\end{equation}
with $[\dd k]\equiv \dd^4k\,\delta(k^2)/(2\pi)^3$.
In the soft-and-collinear limit, we have
\begin{equation}\label{eq:kt-soft-coll}
  k_t
  \approx \frac{Q}{2} z\theta
  \approx Qze^{-\eta},
\end{equation}
with $z$ the energy fraction of the emitted gluon, i.e.\ $E_k=zQ/2$,
$\theta$ its emission angle, and $\eta=-\ln\tan(\theta/2)$ its rapidity.
Furthermore, the matrix element integration becomes
\begin{equation}\label{eq:soft-coll-me}
  \int [\dd k] \left|\mathcal{M}(k)\right|^2
  = \frac{2\alpha_s C_i}{\pi}
  \int_0^\infty d\eta \int_{e^{-\eta}}^1 \frac{\dd z}{z}
  = \frac{2\alpha_s C_i}{\pi}
  \int_0^\infty d\eta \int_0^{Qe^{-\eta}} \frac{\dd k_t}{k_t}\, ,
\end{equation}
where $C_i$ is the colour factor of the emission. 
Then, the integration in Eq.~\eqref{eq:multiplicity-dl-alphas-two} yields
\begin{equation}\label{eq:dl-alphas-result}
  \langle N^{\text{(Lund)}}_i \rangle_{\order{\as}}
  = 1 + \abar \frac{C_i}{C_A}\frac{L^2}{2},
  \quad\text{ with }\quad L = \ln\frac{Q}{\ktcut},
  \quad
  \abar = \frac{2\as C_A}{\pi},
\end{equation}
with $C_i=C_F$ for quarks and $C_i=C_A$ for gluons.
As expected, we obtain the double-logarithmic enhancement
characteristic of soft-and-collinear emissions.

Moving on to $\order{\as^2}$, we restrict the discussion to the configurations
relevant in the double-logarithmic limit. That is, we need to
account for two soft-and-collinear emissions each either real or
virtual. The $\order{\as^2}$ contribution to the Lund multiplicity can
be written as
\begin{align}\label{eq:alphas2-dla}
  \avnlpi_{\order{\as^2}}
   = \left(\frac{2\as}{\pi}\right)^2
   & \int_0^\infty\dd\eta_1\int_0^1 \frac{\dd z_1}{z_1}
    \int_0^\infty\dd\eta_2\int_0^1 \frac{\dd z_2}{z_2}\Big\lbrace
    C_i^2 \big[1\big] - 
    C_i^2 \big[1+\Theta(k_{t,2}>\ktcut)\big]  \nonumber \\
  & - 
    C_i^2 \big[1+\Theta(k_{t,1}>\ktcut)\big]
    -C_i C_A \big[1+\Theta(k_{t,1}>\ktcut)\big] \Theta(\eta_2>\eta_1)
    \nonumber\\
    & + 
      C_i^2 \big[1+\Theta(k_{t,1}>\ktcut)+\Theta(k_{t,2}>\ktcut)\big]
      \nonumber \\
    &+C_i C_A \big[1+\Theta(k_{t,1}>\ktcut)+\Theta(k_{t,21}>\ktcut)\big] 
    \Theta(\eta_2>\eta_1)
    \Big\rbrace,
\end{align}
where the first line includes the virtual-virtual (VV) and
virtual-real (VR) cases, the second line corresponds to the
real-virtual (RV) case and the last two lines contain the real-real
case (RR). We have accounted for the fact that the second emission,
when being real, can be either a {\em primary} emission (i.e.\ a
radiation from the leading parton, with a colour factor $C_i$), or a
{\em secondary} emission (i.e.\ a radiation from the first gluon, with
a colour factor $C_A$).
In writing Eq.~(\ref{eq:alphas2-dla}), we have defined $\eta_2$ as the
rapidity of emission 2 with respect to its emitter, i.e.\ with respect
to the parent parton for primary emissions, $\eta_2=-\ln\tan
(\theta_{2q}/2)$, and with respect to the first emitted gluon for
secondary emissions, $\eta_2=-\ln\tan (\theta_{21}/2)$.
With this in mind, we note the presence of the angular-ordering
constrain ($\Theta(\eta_2>\eta_1)$) for the (secondary) $C_iC_A$
contribution in both the RR and RV contributions.\footnote{Both these
  terms share the same soft matrix element (modulo an overall
  sign). Angular ordering arises after azimuthal averaging.}
At \DL accuracy where we have strong angular ordering, the relative
transverse momentum is also measured with respect to the emitter.
In the soft-and-collinear limit and with our convention for $\eta_2$,
we therefore have
$k_{t,2}\approx E_2\theta_{2q}\approx Q z_2e^{-\eta_2}$ for primary
emissions and $k_{t,21}\approx E_2\theta_{21}\approx Q z_2e^{-\eta_2}$
for secondary emissions.
Finally, while in an exact $\mathcal{O}(\as^2)$ calculation the
multiplicity should be normalised by the $\mathcal{O}(\as^2)$
inclusive cross-section, at \DL accuracy it can safely be
normalised by the Born-level condition for reasons equivalent to those
discussed above at $\mathcal{O}(\as)$. 
Taking these facts into account, we find that the only terms that do
not cancel in Eq.~\eqref{eq:alphas2-dla} are
\begin{align}\label{eq:NNLO-LP-multiplicity}
\avnlpi_{\order\as^2}= 
  \abar^2 \frac{C_i}{C_A}
  \dis\int_0^1 \frac{\dd z_1}{z_1} \dis\int_0^\infty\dd\eta_1
  \dis\int_0^1 \frac{\dd z_2}{z_2} \dis\int_{\eta_1}^\infty\dd\eta_2 ~
  \Theta(k_{t,21}>\ktcut),
\end{align}
The key observation in Eq.~\eqref{eq:alphas2-dla} is that all
the real contributions cancel against virtual corrections except the
contribution from the second gluon emission in the real $C_iC_A$ term,
i.e.\ the secondary gluon radiation.
Physically, this comes from the fact that, in an angular-ordered
picture, the leading parton carries on radiating independently of
whether the first emitted gluon is real or virtual.

\begin{figure}
  \centering{\includegraphics[width=0.5\textwidth]{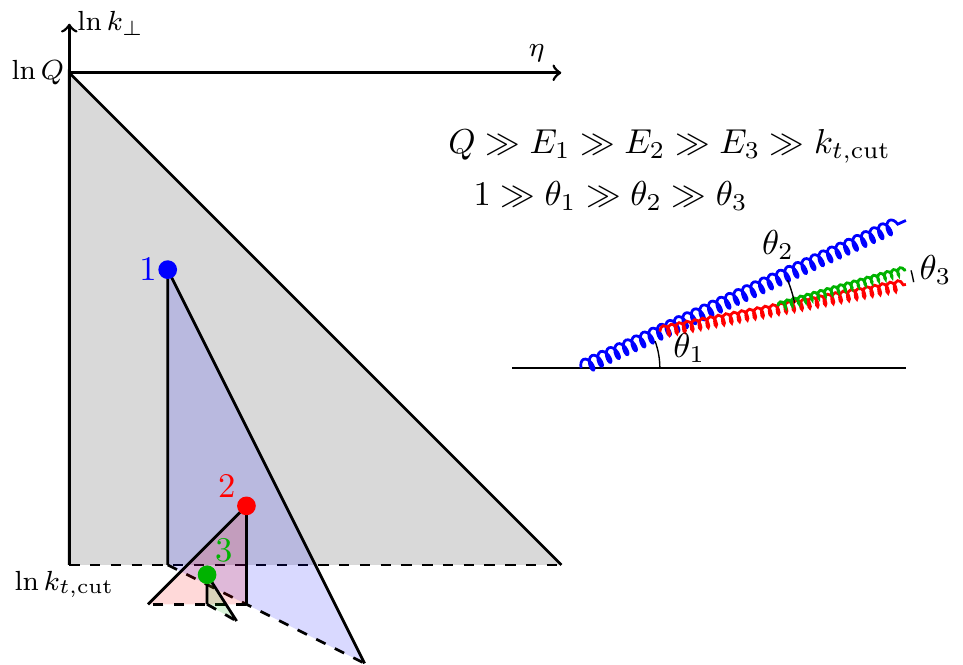}}
  \caption{Representation of the nested gluon-emission pattern,
    strongly-ordered in both energies and angles, which constitutes
    the all-order \DL contribution.}\label{fig:dl-nested}
\end{figure}

At \DL accuracy, it is relatively straightforward to show
that the above arguments at $\order{\as^2}$ generalise to all
orders.
More precisely, at order $\as^n$, the only non-zero contribution
comes from the kinematic configuration where the $n$ soft gluon
emissions are {\em nested}, i.e.\ where the second gluon is collinear
to the first, the third is collinear to the second, ... , and the
$n^\text{th}$ gluon is collinear to the $(n-1)^\text{th}$.
This is schematically represented in Fig.~\ref{fig:dl-nested}.
To write down the all-orders expression, it is useful to introduce for a given 
emission $i$ its energy fraction with respect to the initial hard
parton $x_i$, i.e.\ $x_i=2E_i/Q=\prod_{j\le i} z_j$ with $z_j$ the energy 
fraction of each emission with respect to the previous one, and the product 
running over all previous emissions. We do so since $x_i$ is the strongly 
ordered variable. We can therefore write the all-orders expression for the Lund 
multiplicity at \DL accuracy as:
\begin{align}\label{eq:final-master}
  \avnlpi_\text{DL}
  & = 1 + \frac{C_i}{C_A} \sum_{n=1}^\infty  \abar^n
    \int_{0}^\infty\dd\eta_1\int_{\eta_1}^{\infty}\dd\eta_2 \dots
    \int_{\eta_{n-1}}^\infty \dd\eta_n \nonumber\\
  & \phantom{= 1 + \frac{C_i}{C_A} \sum_{n=0}^\infty \abar^n}
    \int_0^1\frac{\dd x_1}{x_1}\int_0^{x_1}\frac{\dd x_2}{x_2} \dots
    \int_0^{x_{n-1}} \frac{\dd x_n}{x_n} \Theta(x_n 
    e^{-\eta_n}>e^{-L})\\
  & = 1 + \frac{C_i}{C_A} \sum_{n=1}^\infty \abar^n
    \int_0^L \dd\eta_n 
    \frac{\eta_n^{n-1}}{(n-1)!} \frac{(L-\eta_n)^n}{n!} 
   = 1 + \frac{C_i}{C_A} \sum_{n=1}^\infty \frac{(\abar
    L^2)^n}{(2n)!}. \nonumber
\end{align}
The series can easily be summed to get
\begin{equation}\label{eq:dl-resum}
  N^\text{(DL)}_i \equiv \avnlpi_\DL = 1 +
  \frac{C_i}{C_A}\left(\coshnu-1\right),
\end{equation}
where we introduce the following notations which will be helpful
throughout this paper:
\begin{equation}\label{eq:nu-def}
  \nu = \sqrt{\abar L^2} = \sqrt{\frac{2C_A\xi}{\pi}} =
  \sqrt{\frac{2C_A\as L^2}{\pi}}.
\end{equation}
Eq.~(\ref{eq:dl-resum}) is in agreement with earlier results in the
literature~\cite{Catani:1991pm}. 
In this soft-and-collinear limit, both hard legs in a $Z \to q\qbar$
or $H \to gg$ process are independent of one another. Thus, the
average Lund multiplicity is $\langle N^{\DL}_Z\rangle = 2N_q^{(\DL)}$ and
$\langle N^{\DL}_H\rangle = 2N_g^{(\DL)}$, respectively.

$N^\text{(\DL)}_i$ is the total number of Lund declusterings
above a fixed $\ktcut$. In future discussions we also consider the
{\em differential} distribution, i.e.\ the distribution of Lund
declusterings at a given $k_t$, which can be trivially obtained from
the cumulative result:
\begin{equation}\label{eq:DL-differential}
  n_i^\text{(\DL)} = \frac{\dd N_i^\text{(\DL)}}{\dd L} =
  \frac{C_i}{C_A}\sqrt\abar \sinhnu.
\end{equation}

\subsection{Next-to-double-logarithmic (\NDL) accuracy}\label{sec:recap-ndl}
We now move to the calculation of the \NDL $h_2$ function in
Eq.~\eqref{eq:log-counting}, where $\sqrt{\as}h_2$ resums terms of
order $\as L (\as L^2)^n$.
This result can be extracted from Ref.~\cite{Catani:1991pm},
but we here propose a more simple and explicit derivation of $h_2$,
thereby introducing a resummation procedure that will be helpful for
the \NNDL resummation as well. 

\begin{figure}
  \begin{subfigure}[t]{0.24\linewidth}
    \includegraphics[scale=1]{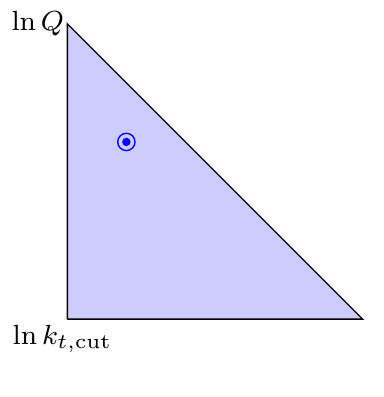}
    \caption{Running coupling}
    \label{fig:ndl-diagram-b0}
  \end{subfigure}
  \begin{subfigure}[t]{0.24\linewidth}
    \includegraphics[scale=1]{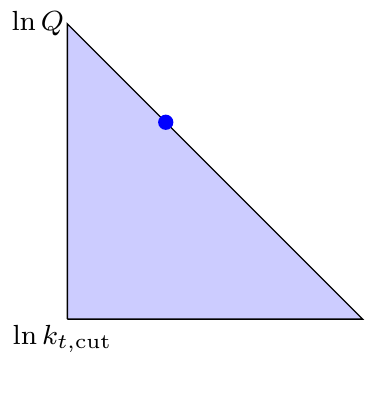}
    \caption{Hard-coll. splitting}
    \label{fig:ndl-diagram-hc}
  \end{subfigure}
  \begin{subfigure}[t]{0.24\linewidth}
    \includegraphics[scale=1]{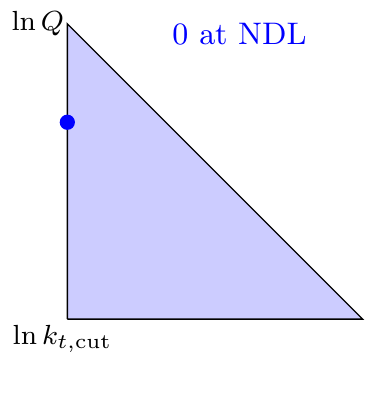}
    \caption{Large or commensurate angle splitting}
    \label{fig:ndl-diagram-LA}
  \end{subfigure}
  \begin{subfigure}[t]{0.24\linewidth}
    \includegraphics[scale=1]{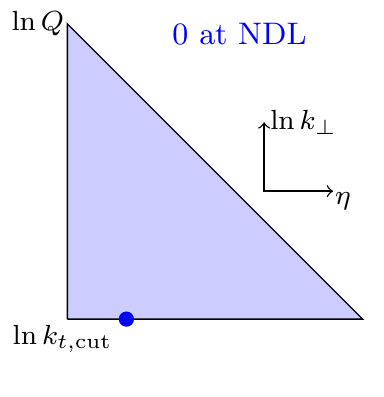}
    \caption{$k_{t, \text{emission}} \sim \ktcut$}
    \label{fig:ndl-diagram-ktcut}
  \end{subfigure}
  \caption{Lund representation of the configurations that potentially
    contribute to the multiplicity at \NDL. 
    The diagrams are drawn for the primary Lund plane but similar
    contributions exist for subsidiary planes as well.}
  \label{fig:ndl-diagrams}
\end{figure}

The central observation is that contributions of the form $\as L(\as L^2)^n$ can 
only be obtained with $n$ soft-collinear emissions, each
contributing a factor $\as L^2$, and a single correction to one of the emissions 
in the full sequence giving the extra factor $\as L$.
To find the potential configurations that can yield a factor $\as L$,
it is sufficient to go back to the $\order\as$ expression
discussed at \DL accuracy and list all possible \NDL corrections.
From Eq.~\eqref{eq:multiplicity-dl-alphas-two}, we see that we can get
corrections beyond \DL from either the matrix element, or from the 
expression of the relative transverse momentum $k_t$.
\NDL, single-logarithmic, corrections to the matrix element,
Eq.~\eqref{eq:soft-coll-me}, can be of three origins: (i) one-loop
running-coupling corrections to $\as$, (ii) hard-collinear corrections
where one would have to take into account the full Altarelli--Parisi~\cite{Altarelli:1977zs}
splitting functions at large $z$, and (iii) potential soft large-angle
corrections away from the collinear limit.
These three contributions are schematically represented in
Figs.~\ref{fig:ndl-diagram-b0}-\ref{fig:ndl-diagram-LA}.
Finally, potential corrections associated with the $k_t>\ktcut$
condition are depicted in Fig.~\ref{fig:ndl-diagram-ktcut}.

The diagrams in Fig.~\ref{fig:ndl-diagram-hc}-\ref{fig:ndl-diagram-ktcut} 
clearly exhibit a single-logarithmic behaviour, where the logarithm stems from
integrating over a line in the Lund plane, i.e.\ over a region which
extends over a distance of order $L$ in one direction and is limited
to an extent of order 1 in the other direction.

Out of the 4 possibilities sketched in Fig.~\ref{fig:ndl-diagrams},
neither the large-angle correction to the matrix element,
Fig.~\ref{fig:ndl-diagram-LA}, nor the emission near the $\ktcut$
boundary, Fig.~\ref{fig:ndl-diagram-ktcut}, contribute at \NDL.
The case of corrections with $k_t\sim\ktcut$,
Fig.~\ref{fig:ndl-diagram-ktcut}, is particularly trivial:
single-logarithmic emissions along the $\ktcut$ boundary are soft and
collinear; the \DL relations in Eq.~\eqref{eq:kt-soft-coll} as well as
the \DL simplification of the matrix-element,
Eq.~\eqref{eq:soft-coll-me}, are therefore still valid, yielding no
\NDL correction in Eq.~\eqref{eq:multiplicity-dl-alphas-two}.

Let us therefore discuss the case of emissions at large angle. The
matrix element for a soft emission, $p_k$, off a dipole with
legs $(p_i,p_j)$ is given by the well-known eikonal factor: 
\begin{align}
  \label{eq:eikonal}
  \int [\dd k] |M(k)|^2
  &=
  \frac{\as}{2\pi}(-2T_i\cdot T_j)\int \frac{\dd E_k}{E_k}
  \frac{\dd^2\theta_k}{2\pi} \,
  E_k^2 \frac{(p_i\cdot p_j)}{(p_i\cdot p_k) (p_j\cdot p_k)}
  \\
  &\equiv
  \frac{\as}{2\pi}(-2T_i\cdot T_j)\int \frac{\dd E_k}{E_k}
  \frac{\dd^2\theta_k}{2\pi}\,
  (p_k | p_i p_j),
  \nonumber
\end{align}
with $T_{i,j}$ the colour matrices associated with the dipole legs,
and $\dd^2 \theta_k/2\pi \equiv \dd \!\cos{\theta_k}\, \dd \phi_k / 2\pi$.
Since the \NDL correction can take place anywhere in the chain of \DL
emissions, we have two cases to consider: either a soft primary
emission at large angles, or any subsidiary soft emission which
happens at commensurate angle compared to another soft-collinear
emission.
If the emission is primary, the $ij$ dipole in Eq.~\eqref{eq:eikonal}
corresponds to the initial $q\bar q$ ($gg$) pair and is, therefore, in a
back-to-back configuration.
In this case, $-2T_i\cdot T_j=2C_i$ and one can easily show that
Eq.~\eqref{eq:eikonal} corresponds exactly to the soft-and-collinear
matrix element, Eq.~\eqref{eq:soft-coll-me}, and, therefore, no new \NDL
correction arises.
Conversely, if we have the emission of a soft gluon, say $k_2$, at an
angle commensurate to a previous collinear emission, $k_1$, it is
well-known that after integrating over the azimuthal angle of the soft
gluon, one recovers the property of angular
ordering~\cite{Dokshitzer:1991wu,Ellis:1991qj}, i.e.\ 
$\Theta(\theta_2<\theta_1)=\Theta(\eta_2>\eta_1)$.
This again reproduces the \DL result with no \NDL correction.
Physically, we can view the absence of large-angle contribution at
\NDL accuracy as a consequence of the fact that the Lund definition of
$k_t$ matches the physical relative transverse momentum with respect
to the emitting dipole, at least in the soft limit (see also
Eq.~(\ref{eq:kt-soft-coll})).

In summary, the \NDL $h_2$ function in Eq.~\eqref{eq:log-counting} receives only 
two non-zero contributions: 
\begin{equation}\label{eq:h2-def}
h_2(\nu) = h_{2,\beta_0}(\nu) + h_{2,\text{hc}}(\nu),
\end{equation}
that are derived in Eqs.~\eqref{eq:ndl-beta0},
\eqref{eq:ndl-hc-sum}, \eqref{eq:ndl-hc-fd} and~\eqref{eq:ndl-hc-fc}.
The resulting final expression for~\eqref{eq:h2-def} agrees with
what was originally obtained in Ref.~\cite{Catani:1991pm}.

While these two contributions could be computed in a similar fashion to 
Eq.~\eqref{eq:final-master}, we instead propose a new resummation scheme that 
relies on the above-mentioned observation that for any set of emissions, only 
one would be associated with the \NDL, yielding a contribution of order $\as L$, 
with all other emissions being treated in the \DL soft-and-collinear 
approximation.
These soft-collinear emissions can then be resummed using explicitly
the \DL results in Eqs.~\eqref{eq:dl-resum}
and~\eqref{eq:DL-differential}.

\begin{figure}
  \includegraphics[width=\textwidth]{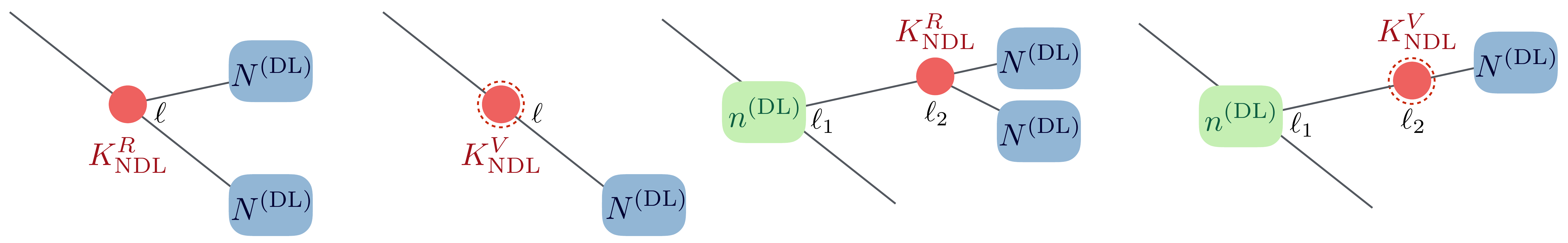}
  \caption{Sketch of the method used to compute \NDL corrections from
    two basic ingredients: a single genuine \NDL emission, the (red)
    solid vertex, and towers of \DL emissions, either before (green box, 
    Eq.~\eqref{eq:DL-differential}) or after (blue box, Eq.~\eqref{eq:dl-resum}) 
    the \NDL emission.
    The four diagrams represent the four options for the \NDL vertex: a
    real emission from the leading parton, a virtual correction from
    the leading parton, a real emission from any subsidiary \DL gluon,
    or a virtual correction from a subsidiary \DL gluon (from left to
    right).}\label{fig:ndl-sketch}
\end{figure}

To illustrate this more concretely, we define $\l\equiv\ln{Q/k_t}$, where 
$k_t$ denotes the scale at which the emission takes place.
More concretely, a generic \NDL correction can stem either from real emissions
or from virtual corrections, and can happen either for an emission
from the leading parton (i.e.\ for an emission in the primary Lund
plane) or for an emission off a previous \DL emission (at any order
in $\as$).
This leads to a total of four possible contributions that are sketched in 
Fig.~\ref{fig:ndl-sketch}. More formally, we can write the \NDL correction as 
\begin{align}
  \delta N^{(\NDL)}_i
  &= \int_0^L  \dd \l \,\Big\{
    K^R_\NDL\,
     \big[N^{(\DL)}_\text{hard}(L;\l)+N^{(\DL)}_\text{soft}(L;\l)\big]
    - K^V_\NDL N^{(\DL)}_i(L;\l)
    \Big\}  
         \label{eq:n-ndl-full}\\
   &+ \int_0^L\dd \l_1 
     n^{(\DL)}_i(\l_1) \int_{\l_1}^L  \dd \l_2\,
     \Big\{K^R_\NDL
     \big[N^{(\DL)}_\text{hard}(L;\l_2)+N^{(\DL)}_\text{soft}(L;\l_2)\big]
     - K^V_\NDL
     N^{(\DL)}_g(L;\l_2)\Big\}.\nonumber
\end{align}
This equation requires a few comments.
First, we denote by $\delta N^{(\NDL)}_i\equiv \sqrt{\as}h_{2}^{(i)}$ 
the \NDL corrections to the average Lund plane multiplicity, computed here for
a single hemisphere of flavour $i=q,g$.
Similarly, $N^{(\DL)}(L;\l)$ denotes the double-logarithmic evolution
starting from a parton of transverse momentum scale $e^{-\l}Q$ down 
to the cutoff scale $e^{-L}Q$. At this accuracy,
$N^{(\DL)}(L;\l)=N^{(\DL)}(L-\l)$.
The first line of Eq.~\eqref{eq:n-ndl-full} corresponds to the case
where the \NDL correction occurs in the primary Lund plane, i.e.\ is
associated with an emission off the leading parton as represented in
the leftmost two sketches of Fig.~\ref{fig:ndl-sketch}.
In this case, the \NDL factor $\as L$ arises after integrating over the $k_t$ of 
the emission (with $\ell=\ln Q/k_t$).
Correspondingly, the second line is associated with \NDL corrections
for an emission off a subsidiary gluon, displayed in the two rightmost
sketches of Fig.~\ref{fig:ndl-sketch}.
In this case, we have a tower of \DL emissions, 
yielding a gluon at a relative transverse momentum $k_{t,1}$, with
$\l_1=\ln(Q/k_{t,1})$. This gluon further splits at a
transverse momentum scale $k_{t,2}$, with
$\l_2=\ln(Q/k_{t,2})$ and the \NDL correction, proportional
to $\as L$, stems from the integration over $\l_2$.
Also, we have to take into account that the emission
associated with the \NDL correction can be either real or virtual. In the
former case, both offsprings can further emit \DL gluons, hence the two
contributions $N^{(\DL)}_\text{hard}(L;\l_2)$ (emissions from the
$k_{t,1}$ gluon at scales smaller than $k_{t,2}$) and
$N^{(\DL)}_\text{soft}(L;\l_2)$ (emissions off the $k_{t,2}$ gluon, the soft 
branch), while for the virtual correction, only one parton can further radiate.
This \DL radiation follows the same pattern of nested radiation as
discussed in Sec.~\ref{sec:recap-dl}, except that it covers scales
between $k_{t,2}$ and $e^{-L}Q$ as denoted explicitly by the arguments
of $N^{(\DL)}_\text{hard}(L;\l_2)$ in Eq.~\eqref{eq:n-ndl-full}.
Finally, the kernels $K^{R,V}_{\NDL}$ denote the genuine \NDL
corrections, for real and virtual emissions respectively.
Since the single-logarithmic contribution has been explicitly written
as an integration over $\l$ (or $\l_2$), the kernels are
proportional to $\as$ (potentially with extra double-logarithmic
factors proportional to $\as L^2$).
The form of Eq.~\eqref{eq:n-ndl-full} also implies that the
expressions for the \NDL kernels, $K^{R,V}_{\NDL}$, can be obtained
from the study of a single emission (treated at fixed order) and then
inserted in Eq.~\eqref{eq:n-ndl-full} to obtain the all-orders
result.\footnote{We note that there are no Sudakov factors in
  Eq.~(\ref{eq:n-ndl-full}).
  To understand this, take for example the evolution between
  $\ell_1$ and $\ell_2$ in the second line of
  Eq.~(\ref{eq:n-ndl-full}).
  By construction, all the emissions between these two scales can be
  taken in the \DL approximation.
  At this accuracy, we have already shown in Sec.~\ref{sec:recap-dl},
  that the real emissions between $\ell_1$ and $\ell_2$ exactly cancel
  the virtual corrections between these two scales and therefore no
  Sudakov factors appear in the final expression.}

In what follows, we use Eq.~\eqref{eq:n-ndl-full} to compute the two
contributions appearing in Eq.~\eqref{eq:h2-def} for one
hemisphere.

\subsubsection{Running coupling effects}\label{sec:running-coupling-NDL}

We first consider the \NDL correction associated to the running of the strong
coupling, see Fig.~\ref{fig:ndl-diagram-b0}.
At 1-loop we have
\begin{equation}
\label{eq:1loop-alphas}
\as(k_t) =
\dis\frac{\as}{1-2\as\beta_0\ln\frac{\mu_R}{k_t}}
\approx
\as + 2\as^2\beta_0 \ln{\left(\frac{\mu_R}{k_t}\right)} + 
\order{\as^3},
\end{equation}
where $\as\equiv \as(\mu_R)$ with $\mu_R$ the renormalisation scale of
order $Q$.
Unless explicitly stated otherwise, we set $\mu_R=Q$.
Furthermore $\beta_0=(11C_A - 4n_fT_R)/(12\pi)$ is the 1-loop $\beta$
function.
At \NDL accuracy, one of the emissions in the chain will get a
correction $2\as^2\beta_0 \ln(\mu_R/k_t)$, bringing a factor of
order $\as L$ compared to the dominant $\as$ contribution at \DL.

We focus on the contribution from the second line in
Eq.~\eqref{eq:n-ndl-full} --- the rightmost sketches in
Fig.~\ref{fig:ndl-sketch} --- noting that the contribution from the
first line can straightforwardly be deduced by setting $\ell_1$ to
zero.
Consider therefore that an emission with relative transverse momentum $k_{t,2}$ 
is radiated off a gluon created itself at a scale $k_{t,1}$, as in
Fig.~\ref{fig:ndl-sketch}. This new emission can be emitted over a
rapidity range $\ln(k_{t,1}/k_{t,2})\equiv \ell_2-\ell_1$ that we
can integrate out so as to find
\begin{equation}
  K_{\NDL,\beta_0} = \frac{2\as C_A}{\pi} \times (2 \as\beta_0 \ell_2 
  ) \times (\ell_2-\ell_1),
\end{equation}
valid for both real emissions and virtual corrections.

Since running coupling corrections do not alter the flavour of the
leading parton, the real contribution alongside the hard branch and
the virtual correction cancel exactly and we are only left with the
real contribution from emissions off the soft branch.
Using the fact that
$n_i^{(\DL)}=n_g^{(\DL)} C_i/C_A$, we obtain
\begin{equation}
\label{eq:n-ndl-rc}
  \delta  N^{(\NDL)}_{i,\beta_0}
  = \frac{C_i}{C_A}  \int_0^L \dd \l_1
  \left[\delta(\ell_1) + n_g^{(\DL)}(\l_1)\right] \, \abar \int_{\l_1}^L \dd 
  \l_2 (2\as\beta_0 \l_2)\, (\l_2-\l_1)\, N^{(\DL)}_g(L-\l_2),
\end{equation}
where the $\delta(\ell_1)$ term in the square 
bracket accounts for the correction on the primary Lund plane, as given by the 
first line of Eq.~\eqref{eq:n-ndl-full}.
Inserting the \DL expressions for $N^{(\DL)}_g(L-\l_2)$ and $n_g^{(\DL)}(\l_1)$, 
Eqs.~\eqref{eq:dl-resum} and \eqref{eq:DL-differential}, we can perform the 
integrations and get
\begin{equation}
\label{eq:ndl-beta0}
  h_{2,\beta_0}^{(i)}
  =\frac{C_i}{C_A} \frac{\beta_0}{2} \sqrt{\frac{\pi}{2C_A}}
  \left[(\nu^2-1)\sinhnu + \nu\coshnu \right] \, .
\end{equation}

\subsubsection{Hard-collinear correction}\label{sec:hard-collinear-NDL}

In the hard-collinear regime the emission probability given by
Eq.~\eqref{eq:soft-coll-me} has to be amended in order to incorporate
the full DGLAP splitting functions, which can be written as:
\begin{align}
\label{eq:splitting-function}
  P_{g\to gg}(z)
  &= 2C_A\left[\frac{1-z}{z}+\frac{z}{2}(1-z)\right],
    & P_{q\to qg}(z) & = 2C_F\left(\frac{1-z}{z}+\frac{z}{2}\right) \nonumber \, 
    , \\
P_{g\to q\bar q}(z) & = n_fT_R\left[z^2+(1-z)^2 \right]\, ,&&
\end{align}
with $n_f$ the number of flavours and $T_R=1/2$.\footnote{Note that we have 
symmetrised $P_{g\to gg}$ so as to only have a divergence at $z=0$. This is 
allowed since both $P_{g\to gg}$ and the observable definitions are symmetric 
under the $z\to 1-z$ transformation. This is no longer the case for 
initial-state splittings as we will discuss in Sec.~\ref{sec:ndl-pp}.}
In addition to the \DL contribution coming from the soft, $z\ll 1$, limit
of the splitting function, i.e.\ $P_{i\to ig}\approx 2C_i/z$, this
introduces hard-collinear corrections at finite $z$.
For such a hard-collinear emission happening at a rapidity $\eta$, one
can simply, at \NDL accuracy, integrate the non-divergent part of the
splitting function over $z$, yielding the familiar $B$ 
coefficients~\cite{Banfi:2004yd}:
\begin{align}
\label{eq:B-coeffs}
  B_{gg}
  & \equiv \int_0^1 \dd z \left[\frac{1}{2C_A}P_{g\to
  gg}(z)-\frac{1}{z}\right]=-\frac{11}{12}, 
  & B_{q}
  & \equiv \int_0^1 \dd z \left[\frac{1}{2C_F}P_{q\to qg}(z)-\frac{1}{z}\right]= 
  -\frac{3}{4} \nonumber \, , \\
  B_{gq}
  &\equiv \int_0^1 \dd z\, \frac{1}{2C_A}P_{g\to q\bar q}(z)= 
  \frac{n_fT_R}{3C_A},
  &&
\end{align}
and identify the $k_t$ scale of the emission, $\ell$, with $\eta_2$,
which is equivalent to neglecting any further dependence on the $z$
fraction in subsequent branchings.
The \NDL kernel associated with hard-collinear branchings, valid for both real 
and virtual contributions, is therefore
\begin{equation}\label{eq:NDL-kernel-hc}
  K_{\text{NDL,hc}} = \frac{2\as C_i}{\pi} \times B, 
\end{equation}
where $C_i$ and $B$ depend on the flavour channel under consideration.

Below, we consider separately the case of gluon emissions and the case
of a flavour-changing $g\to q\bar q$ splitting. We thus split the
flavour-diagonal (fd) and flavour-changing (fc) contributions and write
\begin{equation}\label{eq:ndl-hc-sum}
  h_{2,\text{hc}}
  = h_{2,\text{hc-fd}}
  + h_{2,\text{hc-fc}}.
\end{equation}
\paragraph{Flavour-diagonal}
Real emissions off the hard branch cancel exactly with the virtual corrections, 
as in the running coupling calculation, leaving only the contributions from \DL 
radiation off the newly emitted parton $k_{t,2}$ in Eq.~\eqref{eq:n-ndl-full}.
For primary radiation, one gets either a coefficient $B_i=B_q$ or a
coefficient $B_i=B_{gg}$ depending on whether the leading jet parton
is a quark or a gluon.
However, since subsidiary \DL emissions are always gluons, their \NDL
hard-collinear corrections come with a $B_{gg}$ factor.
We therefore obtain
\begin{equation}\label{eq:ndl-hc-fd-integral}
  \delta N^{(\NDL)}_{i,\text{hc-fd}}
  = \frac{C_i}{C_A}  \int_0^L \dd \l_1
  \left[B_i \delta(\ell_1) + B_{gg} n_g^{(\DL)}(\l_1)\right] \, \abar 
  \int_{\l_1}^L
  \dd \l_2\, N^{(\DL)}_g(L-\l_2),
\end{equation}
where we have used Eq.~\eqref{eq:B-coeffs} to integrate over the
$z$-coordinate.
Performing the remaining integrals yields
\begin{equation}
\label{eq:ndl-hc-fd}
h_{2,\text{hc-fd}}^{(i)} = \frac{C_i}{C_A} \sqrt{\frac{C_A}{2\pi}} 
\left[2B_i\sinhnu + B_{gg}(\nu\coshnu -\sinhnu)\right] \, .
\end{equation}

\paragraph{Flavour-changing}\label{sec:hard-collinear-NDL-FC}
The contribution from flavour-changing $g\to q\bar q$ splittings involves two 
modifications with respect to the flavour-diagonal part.
First, primary $g\to q\bar q$ branchings are only possible for gluon
jets.
Second, while after a real splitting both resulting subjets are
quarks, one still has a gluon after a virtual correction. As a
consequence, one no longer has an exact cancellation between the real
term $N_\text{hard}^\text{(DL)}$ and the virtual correction which
involves $N_g^\text{(DL)}$, meaning that all the contributions have to
be kept.
Taking these differences into account, one easily arrives at the
following expression:
\begin{equation}
\label{eq:ndl-hc-fc-raw}
  \delta N^{(\NDL)}_{i,\text{hc-fc}}
  = \frac{C_i}{C_A} B_{gq} \int_0^L \dd \l_1\,
  \left[\delta_{ig}\delta(\ell_1)
    +n_g^{(\DL)}(\l_1)\right]\,
   \abar \int_{\l_1}^L \dd 
   \l_2\,\left[2N^{(\DL)}_q(L\!-\!\l_2)-N^{(\DL)}_g(L\!-\!\l_2)\right],
\end{equation}
where $\delta_{ig}$ selects gluon-initiated jets.
After integration, we get
\begin{subequations}\label{eq:ndl-hc-fc}
\begin{align}
  \label{eq:ndl-hc-fc-q}
  h^{(q)}_{2,\text{hc-fc}}
  & = \frac{C_F}{C_A} \sqrt{\frac{C_A}{2\pi}} B_{gq} \left[\cdiff\nu\coshnu
    + \left(2-3\cdiff\right) \sinhnu+ 2(\cdiff-1)\nu\right]\, ,\\
  \label{eq:ndl-hc-fc-g}
  h^{(g)}_{2,\text{hc-fc}}
  & = \sqrt{\frac{C_A}{2\pi}} B_{gq} \left[\cdiff\nu\coshnu
    + (2-\cdiff) \sinhnu \right] \, ,
\end{align}
\end{subequations}
for quark and gluon-initiated jets respectively. For the sake of
conciseness, we have introduced
\begin{equation}\label{eq:cdiff}
  \cdiff\equiv \frac{2C_F-C_A}{C_A}.
\end{equation}

\section{Calculation at \NNDL accuracy}\label{sec:nndl}

\subsection{Contributions and master formula}\label{sec:nndl-list}

We start the \NNDL calculation by providing a systematic construction
of the physical contributions appearing at \NNDL accuracy. Each
contribution which is not trivially vanishing will then be discussed
in the following subsections.
Therefore, here we lay out the main physics arguments, so as to
factor them from the more technical calculations in the following
sections.
 
To reach \NNDL accuracy, we need to compute the \NNDL $h_3$ function,
resumming terms of order $\as^n L^{2n-2}$, in
Eq.~\eqref{eq:log-counting}.
A term of such form can be obtained by either having (a) a single
emission contributing a pure factor $\as$ with no logarithmic
enhancement (together with $n-1$ soft-and-collinear emissions
contributing as many factors of $\as L^2$), or (b) two
emissions contributing a factor of order $\as L$ (together with
$n-2$ soft-and-collinear emissions).
In what follows, we will consider that the logarithms in this counting
are of kinematic origin --- i.e.\ arising from integrations over the
phase-space of the emissions --- and treat as a third category, (c),
the contributions involving the running of the strong coupling.
In each category, one should take into account contributions from both
real and virtual emissions.

\begin{figure}[ht!]
  \begin{subfigure}[t]{0.23\linewidth}
    \includegraphics[width=\textwidth]{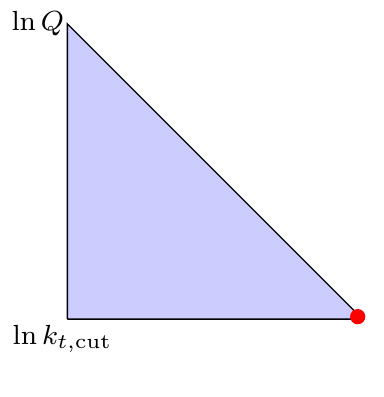}
    \caption{collinear endpoint}\label{fig:nndl-diagram-tip}
  \end{subfigure}
  \hfill
  \begin{subfigure}[t]{0.23\linewidth}
    \includegraphics[width=\textwidth]{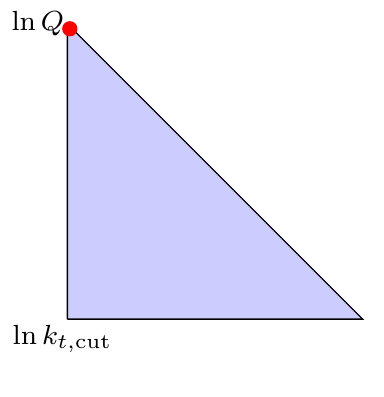}
    \caption{hard matrix-element}\label{fig:nndl-diagram-hardme}
  \end{subfigure}
  \hfill
  \begin{subfigure}[t]{0.23\linewidth}
    \includegraphics[width=\textwidth]{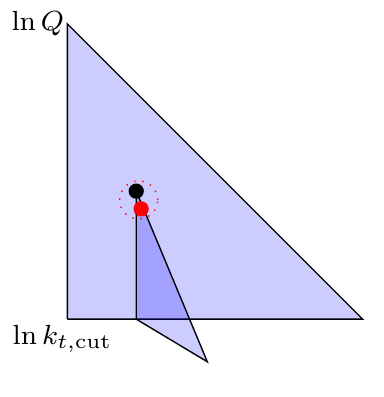}
    \caption{close-by pair}\label{fig:nndl-diagram-double-soft}
  \end{subfigure} 
  \hfill
  \begin{subfigure}[t]{0.23\linewidth}
    \includegraphics[width=\textwidth]{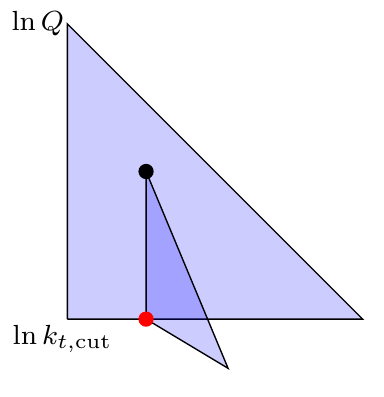}
    \caption{clustering}\label{fig:nndl-diagram-clust}
  \end{subfigure}\\
  \begin{subfigure}[t]{0.23\linewidth}
    \includegraphics[width=\textwidth]{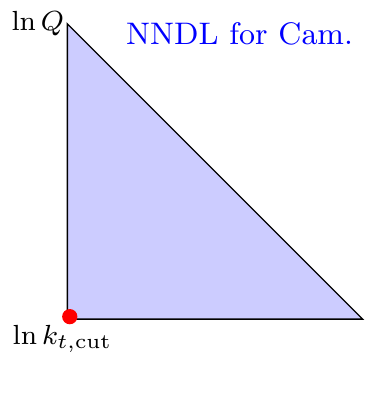}
    \caption{large-angle,$\ktcut$}\label{fig:nndl-diagram-lakt}
  \end{subfigure}
  \hfill
  \begin{subfigure}[t]{0.23\linewidth}
    \includegraphics[width=\textwidth]{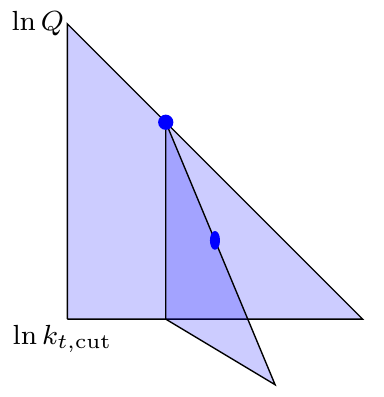}
    \caption{two hard-collinear}\label{fig:nndl-diagram-hcxhc}
  \end{subfigure}
  \hfill
  \begin{subfigure}[t]{0.23\linewidth}
    \includegraphics[width=\textwidth]{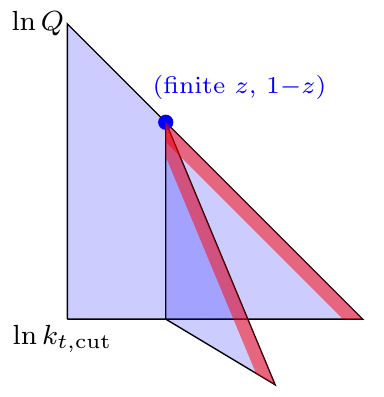}
    \caption{energy loss}\label{fig:nndl-diagram-eloss}
  \end{subfigure}
  \hfill
  \begin{subfigure}[t]{0.23\linewidth}
    \includegraphics[width=\textwidth]{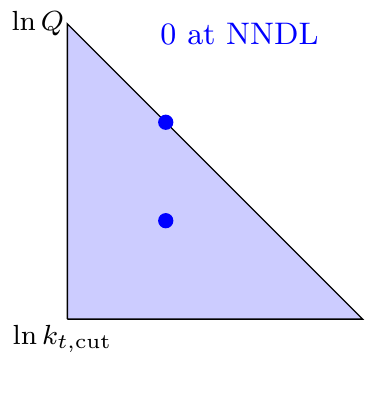}
    \caption{commensurate angle to hard-collinear}\label{fig:nndl-diagram-hcxcomm}
  \end{subfigure}\\
  \begin{subfigure}[t]{0.23\linewidth}
    \includegraphics[width=\textwidth]{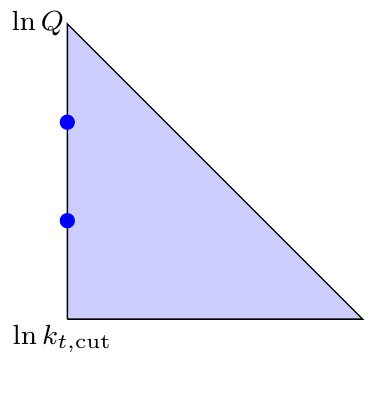}
    \caption{two large-angle}\label{fig:nndl-diagram-la}
  \end{subfigure}  
  \hfill
  \begin{subfigure}[t]{0.23\linewidth}
    \includegraphics[width=\textwidth]{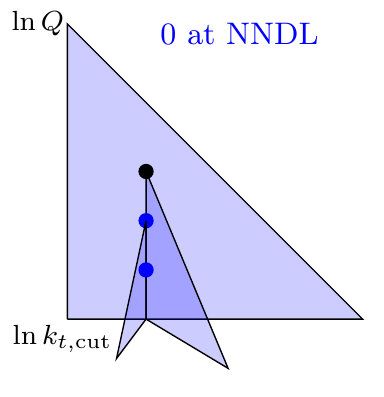}
    \caption{3 commensurate angles}\label{fig:nndl-diagram-3comm}
  \end{subfigure}
  \hfill
  \begin{subfigure}[t]{0.23\linewidth}
    \includegraphics[width=\textwidth]{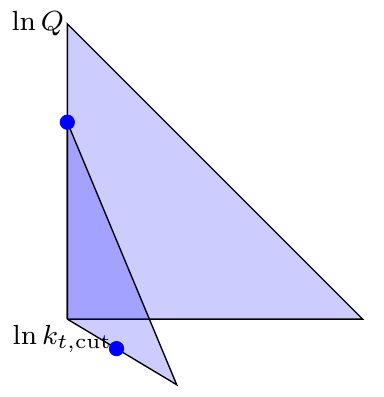}
    \caption{large-angle $\times$ 
    $k_{t,\text{cut}}$}\label{fig:nndl-diagram-laxL}
  \end{subfigure}
  \hfill
  \begin{subfigure}[t]{0.23\linewidth}
    \includegraphics[width=\textwidth]{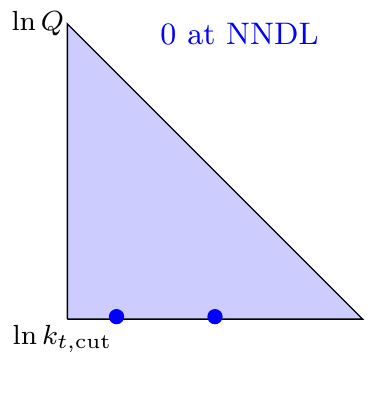}
    \caption{two $k_{t,\text{cut}}$ boundary}\label{fig:nndl-diagram-Lsqr}
  \end{subfigure}\\
  \begin{subfigure}[t]{0.485\linewidth}
    \includegraphics[width=0.485\textwidth]{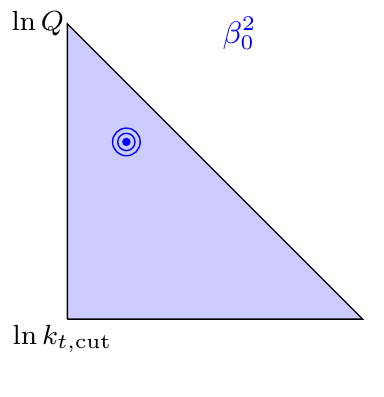}
    \hfill
    \includegraphics[width=0.485\textwidth]{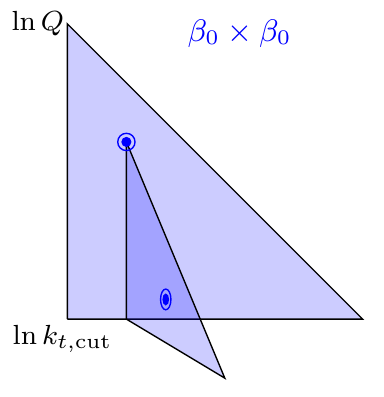}
    \caption{squared running coupling}\label{fig:nndl-diagram-b0sqr}
  \end{subfigure}
  \begin{subfigure}[t]{0.485\linewidth}
    \includegraphics[width=0.485\textwidth]{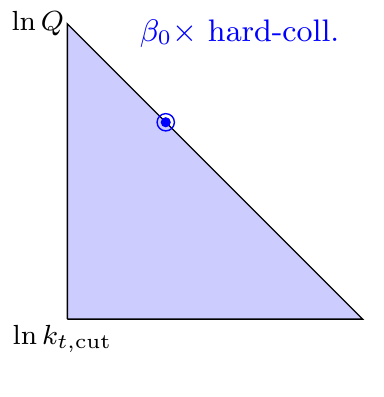}
    \hfill
    \includegraphics[width=0.485\textwidth]{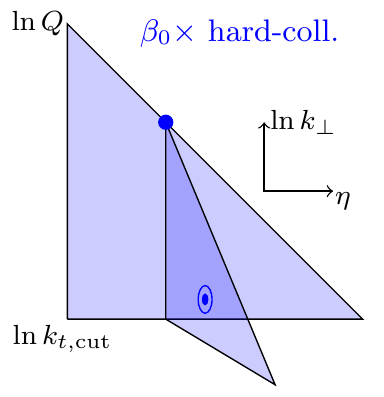}
    \caption{running coupling $\times$ hard-collinear}\label{fig:nndl-diagram-hcb0}
  \end{subfigure}
  \caption{Lund representation of the configurations that
    contribute to the multiplicity at \NNDL.
    Black dots indicate soft-collinear emissions ($\propto \as L^2$), blue dots
    emissions which contribute a factor $\as L$, and red
    dots emissions which only contribute a factor $\as$ with no
    logarithmic enhancement.
    Running-coupling corrections ($\propto \as L$) are represented by
    open circles.  }\label{fig:nndl-diagrams-all}
\end{figure}

We first discuss the category (a), whose contributions are represented
in Fig.~\ref{fig:nndl-diagram-tip}-\subref{fig:nndl-diagram-lakt}.
Following the discussion at \NDL in
Sec.~\ref{sec:recap-ndl}, for a contribution to be of
$\order \as$ with no logarithmic enhancements, it should be
kinematically confined to a region of $\order 1$ in the Lund plane.
This means either in the vicinity of another emission or in one of the three 
corners of a (primary or subsidiary) Lund plane: (i) at the top of the plane, 
(ii) at the collinear end of the $\ktcut$ boundary, or (iii) close to the
$\ktcut$ boundary at an angle commensurate to a previous
emission.\footnote{As for our \NDL discussion, we include in `emissions at 
commensurate angles' the case of primary emissions at large angles, technically
commensurate with the angle $\pi$ of the initial $q\bar q$ (or $gg$)
system.}
This directly yields five \NNDL-like kinematic configurations compared to what 
was needed at \NDL accuracy. These contributions involve just one emission 
that is located at:
\begin{itemize}
\item The collinear endpoint of the Lund plane (primary or
  subsidiary), as represented in Fig.~\ref{fig:nndl-diagram-tip}. This is related
  to the exact $k_t$ definition and discussed in
  Sec.~\ref{sec:nndl-end}.
\item The top of the primary Lund plane, see
  Fig.~\ref{fig:nndl-diagram-hardme}. This is associated with
  fixed-order matrix-element corrections, see Sec.~\ref{sec:nndl-top}.
\item The top of a subsidiary Lund plane or in the
  vicinity of another emission, i.e.\ associated with a (close-by) pair
  of emissions at commensurate $k_t$ and angles. This is shown in 
  ~Fig.~\ref{fig:nndl-diagram-double-soft} and related to the soft limit of the 
  $1\to 3$ splitting function (or to the collinear limit of the
  double-soft matrix element) as discussed in 
  Sec.~\ref{sec:nndl-double-soft} and App.~\ref{app:triple-collinear}.
\item Close to the $\ktcut$ boundary at angles commensurate to
  a previous emission, see Fig.~\ref{fig:nndl-diagram-clust}. This is
  associated with the details of the Cambridge clustering and
  discussed in Sec.~\ref{sec:nndl-clustering}.
\item At large angles with $k_t\sim\ktcut$ in the primary Lund plane,
  see Fig.~\ref{fig:nndl-diagram-lakt}.  This correction vanishes for
  the Lund multiplicity but is non-zero for Cambridge, as
  discussed in Sec.~\ref{sec:nndl-cam}.
\end{itemize}
We now move on to the kinematic configurations, (b), where two
emissions contribute a factor $\as L$. Following the discussion
about \NDL multiplicity, for an emission to contribute a factor
$\as L$ it has to be either hard-and-collinear, or at an angle
commensurate to a previous emission (including large-angle emissions in
the primary Lund plane), or close to the $\ktcut$ boundary.
This results in the following set of configurations, cf.\
Fig.~\ref{fig:nndl-diagram-hcxhc}-\subref{fig:nndl-diagram-Lsqr},
required at \NNDL accuracy:
\begin{itemize}
\item Two hard-collinear emissions: this includes of course the
  contribution where two emissions receive \NDL-like corrections from
  the finite splitting function in the hard-collinear limit, as shown
  in Fig.~\ref{fig:nndl-diagram-hcxhc} and discussed in
  Sec.~\ref{sec:nndl-squared-hc}.
  There is however an additional effect to take into account: after a first
  hard-collinear splitting both daughter partons carry a finite
  fraction of the energy of the parent parton. This affects the
  phase-space available for subsequent emissions, shifting the
  hard-collinear boundary of their Lund planes.
  This contribution, dubbed {\em energy loss}, is represented in
  Fig.~\ref{fig:nndl-diagram-eloss}, where the (red) bands indicate
  that after the hard-collinear (blue) emission, this phase-space
  is no longer accessible. The calculation is presented in
  Sec.~\ref{sec:nndl-eloss}.
\item One hard-collinear emission and one commensurate-angle emission.
  For this, assume we have a set of emissions where one of them,
  $k_1$, is a hard-collinear splitting and one of them, $k_2$, occurs
  at an angle commensurate with that of a previous emission.
  We need to consider two cases depending on whether the
  commensurate-angle emission happens at an angle commensurate to the
  hard-collinear one, i.e.\ $\theta_1\approx\theta_2$, or to any other
  soft-collinear emission, i.e.\ $\theta_1\ll\theta_2$ or
  $\theta_1\gg\theta_2$.
  We show in Sec.~\ref{sec:nndl-hardcoll-comm-angle} that the first
  case yields no \NNDL corrections. 
  For the second case, it is sufficient to realise that emissions well
  separated in angle are independent of one another (in particular
  due to our use of the Cambridge clustering).
  This would thus result in factorised \NDL contributions from $k_1$
  and $k_2$. This vanishes as well since commensurate-angle emissions
  have no \NDL corrections (see Sec.~\ref{sec:recap-ndl}).
\item One hard-collinear emission and one emission close to the
  $\ktcut$ boundary (not shown in Fig.~\ref{fig:nndl-diagrams-all}).
  This vanishes since a hard-collinear emission does not change the
  $k_t$ of following emissions in the Lund planes of either daughter
  partons.\footnote{To understand this, consider a hard-and-collinear
    emission followed by an emission close to the $\ktcut$
    boundary. This soft emission would set the minimum energy in the
    definition of $k_t$, Eq.~\eqref{kt:def}, regardless of whether the
    previous emission is hard or not.}
\item Two emissions at commensurate angles.
  Here one should distinguish two cases: firstly, the case where the
  first emission is a primary large-angle emission,
  Fig.~\ref{fig:nndl-diagram-la} and
  Sec.~\ref{sec:nndl-large-angle-first}; secondly the case of three
  emissions at commensurate angles (i.e.\ a first soft-collinear
  emission followed by two commensurate-angle soft emissions),
  Fig.~\ref{fig:nndl-diagram-3comm}, which actually vanishes as
  explained in Sec.~\ref{sec:nndl-three-comm-angles}.\footnote{ This
    does not cover the case where the two commensurate-angle emissions
    are emitted off different partons, at widely separated angles
    (i.e.\ in independent Lund planes).
    This contribution however vanishes 
    as the contributions from each commensurate-angle emission would
    factorise into a product of two commensurate-angle \NDL
    contributions which are themselves zero.}
\item One emission at an angle commensurate to that of a previous
  emission and one emission close to the $\ktcut$ boundary.  For this
  to have an non-zero effect, one should be in a configuration where
  the presence of the commensurate-angle emission leads to a
  difference between the physical $k_\perp$ of the second emission and
  its reconstructed Lund $k_t$, i.e.\ configurations where
  Eq.~\eqref{eq:kt-soft-coll} has to be revisited.
  This is only possible when the commensurate-angle emission is a
  primary emission at large angle,
  cf.~Fig.~\ref{fig:nndl-diagram-laxL}.
  In practice, this contribution exactly cancels the one from
  diagram~\ref{fig:nndl-diagram-la} and we show this explicitly in
  Sec.~\ref{sec:nndl-large-angle-first}. 
\item Two emissions close to the $\ktcut$ boundary,
  Fig.~\ref{fig:nndl-diagram-Lsqr}.
  Due to our choice of using the Cambridge algorithm to define
  the Lund multiplicity, two emissions close to the $\ktcut$
  boundary and well-separated in angles would never cluster with one
  another: if they belong to the same Lund plane, the most collinear
  would cluster with the plane's leading parton before the
  larger-angle does, and two collinear emissions in different Lund
  planes would cluster with the leading parton of their respective
  plane.
  This choice of a Cambridge-based multiplicity definition, and
  the consequence that the contributions from two commensurate-$k_t$
  emissions vanish is crucial for the feasibility of the \NNDL
  calculation.
  For the more standard definition based on the Durham
  algorithm~\cite{Catani:1991pm}, this diagram would not only be non-zero 
  at \NNDL, but it would also depend in an intricate way on the full
  angular structure of the event, making it extremely difficult to
  calculate analytically. We elaborate more on this point in 
  App.~\ref{app:kperp-vs-cambridge}.
\end{itemize}

\noindent Finally, we discuss the contributions associated with the running of
the strong coupling constant.
Since the 1-loop running contributes at the \NDL accuracy, one would
naively expect to have two-loop corrections at \NNDL.
However, if we write the two-loop running coupling at a scale $k_t$ as
a function of $\as\equiv\as(Q)$ (taking $\mu_R=Q$ without loss of
generality), i.e.
\begin{equation}\label{eq:running-coupling}
  \as(k_t)
  = \frac{\as}{1-2\as\beta_0\ln(Q/k_t)}
  - \frac{\as^2\beta_1}{\beta_0}
    \frac{\ln[1-2\as\beta_0\ln(Q/k_t)]}{[1-2\as\beta_0\ln(Q/k_t)]^2},
\end{equation}
one sees that the two-loop $\beta_1$ correction starts as
$\beta_1\as^3\ln(Q/k_t)$ which, after integration over the phase-space
for emissions with $k_t>\ktcut$, would bring contributions starting at
order $\beta_1\as^3\ln^3(Q/\ktcut)$, i.e.\ only arising from
N$^3$DL.
This means that the running coupling corrections all come from
squared \NDL corrections and can be split into:
\begin{itemize}
\item Terms proportional to $[\as\beta_0\ln(Q/k_t)]^2$ where the
  two $\beta_0$ factors can either come from a single emission by
  expanding the first term of Eq.~\eqref{eq:running-coupling} to
  second order in $\beta_0$, or by having two different emissions,
  each receiving a $\beta_0$ correction. These contributions are
  depicted in Fig.~\ref{fig:nndl-diagram-b0sqr} and computed in
  Sec.~\ref{sec:nndl-squared-rc}.
\item Terms involving one \NDL running coupling correction and one \NDL
  hard-collinear correction. In this case, the running-coupling
  correction can coincide with the hard-collinear emission or can
  appear for any other soft-collinear emission,
  Fig.~\ref{fig:nndl-diagram-hcb0}. These are calculated in
  Sec.~\ref{sec:nndl-rc-times-hc}.
\end{itemize}

\begin{table}
  \begin{center}
    \begin{tabular}{l c c c c}
      \toprule
      Contribution
      & $h_3$
      & Diagrams
      & Section
      & Result \\
      \midrule
      Collinear endpoint
      & $h_{3,\text{end}}$
      & \ref{fig:nndl-diagram-tip}
      & \ref{sec:nndl-end}
      & \eqref{eq:NNDL-endpoint-resummation} \\
      Hard matrix-element
      & $h_{3,\text{hme}}$
      & \ref{fig:nndl-diagram-hardme}
      & \ref{sec:nndl-top}
      & \eqref{eq:nndl-hme-resummation}\\
      Commensurate $k_t$ and angle
      & $h_{3,\text{pair}}$
      & \ref{fig:nndl-diagram-double-soft}
      & \ref{sec:nndl-double-soft}
      & \eqref{eq:NNDL-double-soft-K-resummation}\\
      Clustering
      & $h_{3,\text{clust}}$
      & \ref{fig:nndl-diagram-clust}
      & \ref{sec:nndl-clustering}
      & \eqref{eq:NNDL-clustering-resummed-result}\\
      Large angle and $k_t\sim\ktcut$
      & $h_{3,\text{la}\&k_t}$
      & \ref{fig:nndl-diagram-lakt}
      & \ref{sec:nndl-cam}
      & 0 (Lund), \eqref{eq:NNDL-cam-result} (Cam)\\
      \midrule
      Energy loss
      & $h_{3,\text{eloss}}$
      & \ref{fig:nndl-diagram-eloss}
      & \ref{sec:nndl-eloss}
      & \eqref{eq:NNDL-E-loss-resummation-result}\\
      (Hard-collinear)$^2$
      & $h_{3,\text{hc}^2}$
      & \ref{fig:nndl-diagram-hcxhc}
      & \ref{sec:nndl-squared-hc}
      & (\ref{eq:NNDL-HC-squared-resummed-result})\\
      Hard-coll. $\times$ comm. angles
      & ---
      & \ref{fig:nndl-diagram-hcxcomm}
      & \ref{sec:nndl-hardcoll-comm-angle}
      & 0 \\
      First emission at large angles
      & ---
      & \ref{fig:nndl-diagram-la},~\ref{fig:nndl-diagram-laxL}
      & \ref{sec:nndl-large-angle-first} 
      & 0 \\
      3 commensurate angles
      & ---
      & \ref{fig:nndl-diagram-3comm}
      & \ref{sec:nndl-three-comm-angles}
      & 0 \\
      \midrule
      $\beta_0^2$
      & $h_{3,\beta_0^2}$
      & \ref{fig:nndl-diagram-b0sqr}
      & \ref{sec:nndl-squared-rc}
      & (\ref{eq:NNDL-beta0-squared-resummed})\\
      $\beta_0\times$hard-collinear      
      & $h_{3,\beta_0\times\text{hc}}$
      & \ref{fig:nndl-diagram-hcb0}
      & \ref{sec:nndl-rc-times-hc}
      & (\ref{eq:nndl-beta0-HC-FC-result})\\
      \midrule
      Total
      & $h_3$
      & Fig.~\ref{fig:nndl-diagrams-all}
      & \ref{sec:final-result}
      & (\ref{eq:final-result-h3q})-(\ref{eq:final-nndl-cam})\\
      \bottomrule
    \end{tabular}
  \end{center}
  \caption{Relevant coefficients for the \NNDL function
    $h_3$.}\label{table:nndl-coefficients}
\end{table}

\noindent Putting all the non-vanishing contributions together, we get
the following expression for the \NNDL function $h_3$:
\begin{equation}\label{eq:nndl-master}
  h_3
  = h_{3,\text{end}} + h_{3,\text{hme}} + h_{3,\text{pair}} + h_{3,\text{clust}}
  + h_{3,\text{eloss}} + h_{3,\text{hc}^2}
  + h_{3,\beta_0^2} + h_{3,\beta_0\times\text{hc}},
\end{equation}
where all the functions depend only on $\xi=\as L^2$. For Cambridge multiplicity, 
we instead find
\begin{equation}\label{eq:nndl-master-cambridge}
  h^\text{(Cam)}_3 = h^\text{(Lund)}_3 + h_{3,\text{la}\& k_t} 
\end{equation}
with $h^\text{(Lund)}_3$ given by Eq.~\eqref{eq:nndl-master}.

Most of the calculations below are more easily addressed if we
consider a single hemisphere, therefore computing the $h_3^{(i)}$
function with $i=q,g$.
Although the presence of large-angle contributions,
$h_{3,\text{hme}}$ in particular, makes it less obvious that the two
hemispheres can be treated independently, one is always entitled to
write the \NNDL correction to the average Lund multiplicity in
$Z \to q\qbar$ and $H \to gg$ events as
$\delta N^{(\NNDL)}_Z=\as h_{3,Z}=2\as h_3^{(q)}$ and
$\delta N^{(\NNDL)}_H=\as h_{3,H}=2\as h_3^{(g)}$.

Table~\ref{table:nndl-coefficients} summarises the physical origin of
the various \NNDL corrections together with the section discussing them
and, when non-zero, the equation where to find the final result.
Eq.~\eqref{eq:nndl-master}, together with the individual
contributions listed in Table~\ref{table:nndl-coefficients} are the
main results of this paper.
For completeness, the full result for $h_3(\nu)$ is summarised in
Sec.~\ref{sec:final-result}.

\subsection{Corrections involving a pure $\as$ 
contribution}\label{sec:nndl-alphas}

All the pure $\as$ contributions involve physical configurations
which are new compared to the NDL result. For each of these new kinematic
configurations, the $\as$-factor arises through a phase-space
integration which is only non-exponentially vanishing in a constant
region of the Lund plane. 
This integration typically produces a coefficient
$\tfrac{\as}{2\pi}D$ with a different constant $D$ for each
configuration.
This constant can be extracted from a fixed-order calculation before
we use it to derive the all-order behaviour by adding an arbitrary
number of soft-and-collinear emissions, i.e.\ following a similar
procedure as for the \NDL calculation in Sec.~\ref{sec:recap-ndl},
where the NDL kernel in Eq.~\eqref{eq:n-ndl-full} and in
Fig.~\ref{fig:ndl-sketch} is now an \NNDL kernel.
%

\subsubsection{Hard-collinear endpoint}\label{sec:nndl-end}
If we consider a parton of flavour $i$ splitting into two collinear
daughter partons of flavours $b$ and $c$, the contribution of the
additional parton to the average multiplicity can be written as
\begin{align}\label{eq:NNDL-endpoint-FO}
  \langle N^{\text{(Lund)}}_{i \to bc}\rangle_{\order{\as}} = \abar
  \int_0^\infty \dd \eta
  \int_0^{1} \dd z \frac{1}{2C_A} P_{i\to bc}(z) \,
  \Theta{\left(\min{\left(z, 1-z\right)} e^{-\eta}> e^{-L}\right)},
\end{align}
with $P_{i \to bc}(z)$ the DGLAP splitting kernel (see
Sec.~\ref{sec:hard-collinear-NDL}) for the flavour channel under
consideration.
Say we write
$P_{i\to bc}(z) = \frac{2C_i}{z}\,\delta_{cg}+ p^\text{(finite)}_{i\to bc}(z)$,
separating explicitly the soft-enhanced and finite contributions.
For the contribution proportional to $1/z$, an \NNDL correction arises
from keeping explicitly the $\min(z,1-z)e^{-\eta}$ in the $k_t$ constraint
instead of approximating it by $ze^{-\eta}$ (which would produce the \DL
result, with no \NDL corrections).
Similarly, for the finite piece of the splitting function,
$p^\text{(finite)}_{i\to bc}(z)$, keeping explicitly the
$\min(z,1-z)e^{-\eta}$ in the $k_t$ constraint instead of
approximating it by $e^{-\eta}$ (which would produce the
hard-collinear \NDL contribution) induces an additional \NNDL
correction.
Both contributions, which come from taking into account the exact
expression for the reconstructed $k_t$ of the emission, are relevant
in the $\order 1$ phase-space region where $z$ is finite and the
reconstructed $k_t$ is close to the cut $\ktcut$, cf.\
Fig.~\ref{fig:nndl-diagram-tip}.

Considering the two contributions and performing the $\eta$
integration, one gets an \NNDL correction of the form
\begin{equation}
  \delta N_{i\to bc}^{(\NNDL)}
  = \abar \int_{1/2}^1 \frac{\dd z}{z} \frac{C_i}{C_A} \delta_{cg}\,\ln\!\Big(\frac{1-z}{z}\Big)
   + \abar \int_0^1 \dd z\, \frac{p^\text{(finite)}_{i\to bc}(z)}{2C_A}\,\ln(\min(z,1-z))
  = \frac{\as}{2\pi}D_{\text{end}}^{i \to bc}.
\end{equation}
Inserting the explicit expressions from
Eq.~\eqref{eq:splitting-function} for the splitting kernels, we get
\begin{subequations}
  \label{eq:NNDL-endpoint-FO-D-coeff}
\begin{align}
  \label{eq:NNDL-endpoint-FO-qq}
  D_{\text{end}}^{q\to qg} &= 
  \, C_F\left( 3 + 3\ln{2} - \frac{\pi^2}{3} \right), \\
  \label{eq:NNDL-endpoint-FO-gg}
  D_{\text{end}}^{g\to gg} &= 
  \, C_A \left(\frac{137}{36}+\frac{11}{3}\ln 2-\frac{\pi^2}{3}\right),\\
  \label{eq:NNDL-endpoint-FO-gq}
  D_{\text{end}}^{g\to q\qbar} &= 
  \, n_f T_R \left(-\frac{29}{18}-\frac{4}{3}\ln 2\right).
\end{align}
\end{subequations}
We now proceed with the all-order resummation of the collinear
endpoint correction.
As discussed in Sec.~\ref{sec:nndl-list}, this is achieved by
considering a collinear endpoint emission together with a series of
\DL soft-and-collinear emissions.
Since an emission with $k_t \sim \ktcut$ has no phase-space to further
emit a soft-and-collinear gluon --- as is evident from
Fig.~\ref{fig:nndl-diagram-tip} --- the collinear endpoint emission
should occur as the last step in the series of (nested) \DL emissions.
We therefore write
\begin{equation}\label{eq:NNDL-endpoint-resummation-integral}
  \as h_{3, \text{end}}^{(i)} = \frac{\as}{2\pi}
  \int_0^L \dd \l \,
  \left[\delta(\l) (D_{\text{end}}^{ i \to ig} +
  \delta_{ig}D_{\text{end}}^{g \to q\qbar}) \,
  + \frac{C_i}{C_A}n_g^{(\DL)}(\l) 
  \left( D_{\text{end}}^{g\to gg} + D_{\text{end}}^{g\to q\qbar} \right)\right].
\end{equation}
The first term in this expression, proportional to $\delta(\l)$,
includes corrections from the leading parton, i.e.\ from the
hard-collinear endpoint of the primary Lund plane.
The second term includes the contributions from the hard-collinear
endpoints of all the secondary gluon emissions.
Performing the integrations explicitly, we find
\begin{subequations}
  \label{eq:NNDL-endpoint-resummation}
\begin{align}
  \label{eq:NNDL-endpoint-q}
  h_{3, \text{end}}^{(q)}
  &= \frac{1}{2\pi}
    D_{\text{end}}^{q\to qg}
  + \frac{1}{2\pi} \,
    \left(D_{\text{end}}^{g\to gg}+D_{\text{end}}^{g\to q\bar q}\right)
    \frac{C_F}{C_A}(\coshnu-1), \\
  \label{eq:NNDL-endpoint-g}
  h_{3, \text{end}}^{(g)}
  &= \frac{1}{2\pi}
  \left( D_{\text{end}}^{g\to gg} + D_{\text{end}}^{g\to q\qbar} \right)
  \coshnu,
\end{align}
\end{subequations}
for quark and gluon jets, respectively.
%

\subsubsection{Hard matrix-element corrections (top of the Lund
  plane)}\label{sec:nndl-top}
We next consider the \NNDL contribution, $h_{3,\text{hme}}$, stemming
from finite corrections to the Born-level matrix element. As shown in
Fig.~\ref{fig:nndl-diagram-hardme}, this accounts for the $\order{1}$
region at the top of the primary Lund plane where a first primary
splitting is both hard and wide-angle.

Let us first consider the calculation at $\order{\as}$, for which we
go back to the full expression for the Lund plane multiplicity,
Eq.~\eqref{eq:basic-order-alphas}.
The contribution from the leading parton (the `1' in all three
terms) cancels exactly the $\sigma_0+\sigma_1$, and, for the remaining
contribution off the extra emission (the $\Theta(k_t>\ktcut)$ term),
$\sigma_1$ only contributes at higher orders. As a consequence,
Eq.~\eqref{eq:multiplicity-dl-alphas-one} is still valid.
At order $\as$ we should therefore consider the exact matrix element
for $Z\to q\qbar g$, and for $H\to ggg$ and $H\to gq\qbar$. 

An important technical remark is in order. By definition, the exact
matrix element represents the emission probability for an emission
anywhere in the Lund plane. Therefore, integrating the matrix element
with the $\ktcut$ condition includes three NNDL corrections: (i) the
top of the Lund plane, (ii) the collinear endpoint (see
Sec.~\ref{sec:nndl-end}) and (iii) the large
angle, $k_t\sim\ktcut$ corner. The latter contribution has been shown
to vanish in the previous section.
To avoid double counting we thus have to subtract the contribution
from the collinear endpoint from the full integration.

The exact matrix elements are best written in terms of the energy
fractions $x_i=2 E_i/Q$, with $x_1+x_2+x_3=2$. For a generic decay of
a boson $X$ into three particles of momenta $k_i$, we
therefore write the full-event result as
\begin{equation}\label{eq:hme-integration}
  \langle N^\text{(Lund)}_{X\to k_1k_2k_3}\rangle_{\order{\as}} = 
  \int\! \dd x_1 \dd x_2 \dd x_3\,
  \frac{|\mathcal{M}^{X\to 
  k_1k_2k_3}|^2}{\sigma_0}\Theta(k_{t,\text{exact}}>\ktcut) 
  \delta(x_1+x_2+x_3-2).
\end{equation}
For the different processes we consider, the matrix elements (for the
full event, i.e.\ for both hemispheres) are given by
~\cite{Ellis:1991qj,Badger:2004ty}:
\begin{align}
  \frac{|\mathcal{M}^{Z\to q\bar q g}|^2}{\sigma_0} 
  &= \frac{\as C_F}{2\pi}
    \frac{x_1^2+x_2^2}{(1-x_1)(1-x_2)},
    \label{eq:ME-Zqqg}
\end{align}
and
\begin{subequations}
  \begin{align}
    \frac{|\mathcal{M}^{H\to ggg}|^2}{\sigma_0} 
    &= \frac{\as C_A}{6\pi}
      \frac{1+(1-x_1)^4+(1-x_2)^4+(1-x_3)^4}{(1-x_1)(1-x_2)(1-x_3)},
      \label{eq:ME-Hggg}\\
      \frac{|\mathcal{M}^{H\to g q\bar q}|^2}{\sigma_0} 
    &= \frac{\as n_fT_R}{\pi}
      \frac{(1-x_2)^2+(1-x_3)^2}{1-x_1},
      \label{eq:ME-Hqqg}
  \end{align}
\end{subequations}
where the last line refers to the case of a $H\to gg$ splitting
followed by a $g\to q\bar q$ branching, neglecting the situation where
the Higgs boson would directly decay into a $q\bar q$ pair.
The exact $k_t$ in Eq.~\eqref{eq:hme-integration} depends on which pair of 
particles gets clustered first by the Cambridge algorithm:
\begin{align}
  \Theta(k_{t,\text{exact}} > \ktcut)
  &= \Theta(\theta_{12} < \theta_{13}, \theta_{23})\,
    \Theta\!\left(\min(x_1, x_2) \sin(\theta_{12}) > e^{-L}\right)
    \nonumber \\
  &+ \Theta(\theta_{13} < \theta_{12}, \theta_{23})\,
    \Theta\!\left(\min(x_1, x_3) \sin(\theta_{13}) > e^{-L}\right)
    \nonumber \\
  &+ \Theta(\theta_{23} < \theta_{12}, \theta_{13})\,
    \Theta\!\left(\min(x_2, x_3) \sin(\theta_{23}) > e^{-L}\right).
  \label{eq:kt-exact-definition}
\end{align}
Explicit integration of~(\ref{eq:hme-integration}) gives the known \DL
and \NDL contributions, respectively proportional to $L^2$ and $L$, negligible 
terms suppressed by powers of $e^{-L}$ and a constant.
This constant is the sum of the collinear endpoints and hard
matrix-element \NNDL contributions.
Subtracting the collinear endpoint computed in
Sec.~\ref{sec:nndl-end}, we find\footnote{Note the pre-factor
  $1/\pi$ instead of $1/(2\pi)$ such that the
  coefficient $D$ corresponds to the contribution from a single
  hemisphere, as done for the other \NNDL coefficients.}
\begin{equation}
  \delta N^{(\NNDL)}_{X\to k_1k_2k_3}= \frac{\as}{\pi}
  D_\text{hme}^{k_1k_2k_3}
\end{equation}
with
\begin{equation}\label{eq:nndl-D-hme}
  D_\text{hme}^{qqg} = C_F \left(\frac{\pi^2}{6}-\frac{7}{4}\right),
  \qquad
  D_\text{hme}^{ggg} = C_A \left(\frac{\pi^2}{6}-\frac{49}{36}\right),
  \qquad
  D_\text{hme}^{gq\bar q} = n_f T_R \frac{2}{9}.
\end{equation}
Note that these \NNDL hard matrix-element coefficients are independent
of the details of the observable.
This simply follows from the observation that the hard matrix-element
contribution comes from an $\order{1}$ region at the top of the Lund
plane while the cut on the observable only affects the soft and/or
collinear dynamics.
In particular, we have cross-checked Eq.~\eqref{eq:nndl-D-hme} using
other event shapes like the JADE~\cite{JADE:1986kta},
Durham~\cite{Catani:1991hj}, or Cambridge~\cite{Dokshitzer:1997in} jet
multiplicities.

Let us now move to the all-order treatment of the hard matrix-element
correction.
As we have argued above, the virtual corrections, $|\mathcal{M}_V|^2$,
and the NLO correction to the inclusive cross-section, $\sigma_1$, do
not contribute at $\order{\as}$.
This holds at all orders, as we prove in what follows.
Since we want to compute the correction arising at the top of the Lund
plane, it is sufficient to consider explicitly the emission (real of
virtual) with the largest $k_t$.
For this emission, the real and virtual contributions can be of two
types: (i) associated with an infrared divergence and, (ii) finite
hard matrix-element corrections.
Up to a sign, the former are the same for the real and virtual terms
and, dressed with additional soft-and-collinear emissions, give rise
to the \DL and \NDL multiplicity.
The latter are different for real and virtual terms.
For simplicity, let us consider a single hemisphere of flavour
$a$. For a given flavour channel, the real contribution is given by
$\sigma_0\tfrac{\as}{2\pi}D^{abc}$ with the $D^{abc}$ coefficient
computed above. Let us denote the corresponding virtual contribution
by $\sigma_0\tfrac{\as}{2\pi}D_V^{abc}$ where the actual value of
$D^{abc}_V$ is irrelevant for the following discussion.
Given that these contributions arise from an $\order{1}$ region at the
top of the primary Lund plane, they should be dressed by additional
\DL emissions down to the boundary scale $\ktcut$.
We can therefore write
\begin{equation}\label{eq:hme-intermediate}
  N^\text{(NNDL)}_a(L)
  = \frac{\sigma_0}{\sigma_0 + \sigma_1}\Big\{
    N_a^\text{(NDL)}(L)
    + \frac{\as}{2\pi} \sum_{b,c}D^{abc}
    \big[N_b^\text{(DL)}(L) + N_c^\text{(DL)}(L)\big]
    -D_V^{abc}N_a^\text{(DL)}(L)
  \Big\},
\end{equation}
where, for real emissions, both products of the splitting are dressed
by \DL emissions, while virtual corrections only have a single series
of \DL emissions.
Since (again considering a single hemisphere for simplicity)
\begin{equation}
  \sigma_1 = \frac{\as}{2\pi} \sigma_0 \sum_{b,c}D^{abc}-D_V^{abc},
\end{equation}
Eq.~\eqref{eq:hme-intermediate} can be simplified to
\begin{equation}
  N^\text{(NNDL)}_a(L)
  = N_a^\text{(NDL)}(L)
    + \frac{\as}{2\pi} \sum_{b,c}D^{abc}
    \Big[N_b^\text{(DL)}(L) + N_c^\text{(DL)}(L) - N_a^\text{(DL)}(L)\Big],
\end{equation}
where the virtual contribution $D_V^{abc}$ has disappeared as
anticipated. 
Applying this result to the quark and gluon cases, we get
\begin{subequations}\label{eq:nndl-hme-resummation}
  \begin{align}
    h^{(q)}_{3, \text{hme}}
    & = \frac{1}{2\pi} D^{qqg}_{\text{hme}} \coshnu,\\
    h^{(g)}_{3, \text{hme}}
    & = \frac{1}{2\pi}  \left[
      D^{ggg}_{\text{hme}} \coshnu
      + D^{gq\qbar}_{\text{hme}} \left(\cdiff \coshnu + 1 - \cdiff  \right)
      \right].
  \end{align}
\end{subequations}

\subsubsection{Soft-and-collinear emissions at commensurate angles and 
$k_t$}\label{sec:nndl-double-soft}

Similarly to the \NNDL contribution which appears at the top of the
primary Lund plane (Sec.~\ref{sec:nndl-top}), we must also account
for similar corrections on subsidiary planes.
Concretely, we must account for contributions where a first
soft-and-collinear emission is followed by another at commensurate
$k_t$ and angle, as is displayed in Fig.~\ref{fig:nndl-diagram-double-soft}, 
alongside its virtual corrections.

To help understanding the structure of this correction we first analyse the
$\order{\as^2}$ contribution to the Lund plane multiplicity.
We therefore consider the radiation of two partons, soft and collinear
compared to the hard leg.
It is helpful to decompose the matrix element for emitting two soft
particles with momenta $k_1$ and $k_2 $ into an uncorrelated,
factorised, piece and a correlated piece~\cite{Banfi:2004yd}, i.e.
\begin{equation}
\label{eq:matrix-element-decomp}
\big|\mathcal{M}(k_1,k_2)\big|^2 = \big|\mathcal{M}(k_1)\big|^2 
\big|\mathcal{M}(k_2)\big|^2 + 
\big|\widetilde{\mathcal{M}}(k_1,k_2)\big|^2.
\end{equation}
The uncorrelated piece does not bring any correction at \NNDL since
$|\mathcal{M}(k_1)|$ reduces to the soft-and-collinear \DL expression
in Eq.~\eqref{eq:soft-coll-me} and cancels against the
  virtual-real contribution.
The full \NNDL correction is therefore in $|\widetilde{\mathcal{M}}(k_1,k_2)|$.
In this case, we can assume that $k_1$ is emitted first --- and is
therefore real --- and $k_2$ later (i.e.\ $k_2$ is a secondary
emission from $k_1$).
Accounting for both cases where $k_2$ is either real or virtual, the
$\order{\as^2}$ correction to the Lund multiplicity is given by
\begin{align} 
  \langle N^\text{(Lund)}_{\text{corr}} \rangle_{\order{\as^2}} 
  = \frac{1}{\sigma_0}\int[\dd k_1]\int [\dd k_2]
    &\big|\widetilde{\mathcal{M}}_{RR}(k_1,k_2)\big|^2 \big[\Theta(k_{t, 1} > 
    \ktcut) 
    + \Theta(k_{t, 2} > \ktcut)  \big]\nonumber \\
     -&  
    \big|\widetilde{\mathcal{M}}_{RV}(k_1,k_2)\big|^2\;\Theta(k_{t, 1} > \ktcut)
    \label{eq:NNDL-double-soft-K-FO}
\end{align}
where $[\dd k_i]\equiv \dd^4 k_i\, \delta(k_i^2)/(2\pi)^3$. Combining the terms
involving $\Theta(k_{t, 1} > \ktcut)$, we recover a well-known
relationship in the literature~\cite{Catani:1990rr,Dasgupta:2021hbh}:
\begin{equation}
  \label{eq:K-term-definition}
  \,\int [\dd k_2] 
  \big|\widetilde{\mathcal{M}}_{RR}(k_1,k_2)\big|^2 -
  \big|\widetilde{\mathcal{M}}_{RV}(k_1,k_2)\big|^2
  =
  \big| \mathcal{M}(k_1)\big|^2 \as
  \left(
  -2\beta_0 \ln{\left( {k_t}/{Q}\right)} + \frac{K}{2\pi}
  \right),
\end{equation}
with $K$ the CMW factor~\cite{Catani:1990rr}, or equivalently the
1-loop cusp anomalous dimension, given by
\begin{align}\label{eq:K-definition}
  K = \left(\frac{67}{18}-\frac{\pi^2}{6} \right)C_A - \frac{10}{9} n_f T_R.
\end{align}
The $\order{\beta_0}$ term was already accounted for 
as the running-coupling \NDL contribution
(Sec.~\ref{sec:running-coupling-NDL}), while the $K$ term presents a
new contribution at \NNDL accuracy.
After integrating over the soft-and-collinear emission $k_1$ with the
phase-space given by Eq.~\eqref{eq:soft-coll-me} we find
\begin{align}
  \label{eq:NNDL-K-term-FO}
  \delta N^{(\NNDL)}_{i,K}
  = \left(\frac{C_i}{C_A} \frac{\bar\alpha L^2}{2}\right)
  \times
  \left(\frac{\as}{2\pi}K\right),
\end{align}
for a parent parton of flavour $i$.

\begin{figure}[t]
  \centering
  \includegraphics[scale=0.4]{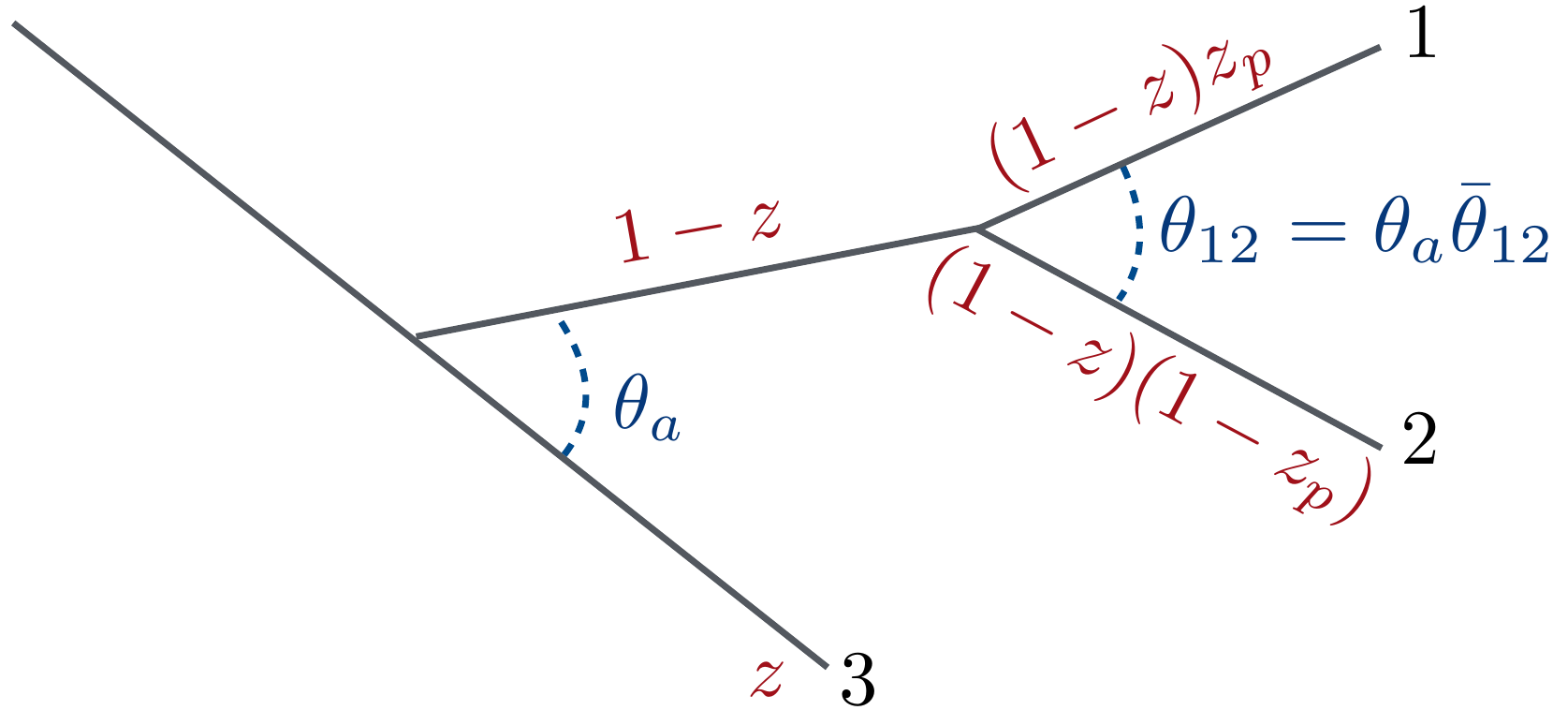}
    \caption{
    Parametrisation of an $i \to 123$ splitting.
    After a first $i \to 3+a$ splitting $a$ participates in a further $a \to 12$
    splitting with characteristic angle $\theta_{12}$ and energy
    fraction $z_p$. The angle $\theta_a$ is that between the parent parton $a$ 
    and the final-state parton $3$. Note that the soft limit corresponds to $z\to 1$ 
    while in all previous sections it was parameterised by $z\to 0$.}
  \label{fig:double-soft-parametrisation}
\end{figure}

We are now left with the computation of the term proportional to
$\Theta(k_{t,2}>\ktcut)$ in Eq.~\eqref{eq:NNDL-double-soft-K-FO}.
In practice, $|\widetilde{\mathcal{M}}_{RR}|^2$ can be obtained from
the soft limit of the $1\to 3$ splitting function, $\hat P_{1\to 3}$, or,
equivalently, from the collinear limit of the double-soft matrix
element~\cite{Dokshitzer:1992ip,Catani:1999ss}.
In both cases, one should make sure to remove potential uncorrelated
contributions.\footnote{E.g.\ for $g\to ggg$ splittings where the
triple-collinear splitting function includes an uncorrelated
contribution in the $C_A^2$ colour channel.}
The resulting expressions are explicitly given in
Appendix~\ref{app:triple-collinear} for all flavour channels.
In practice, it is helpful to use the integration variables depicted
in Fig.~\ref{fig:double-soft-parametrisation}, where the splitting is
viewed as a first soft emission, at an angle $\theta_a$ and with
a momentum fraction $1-z\ll 1$, followed by a second splitting
at an angle $\theta_{12}$ with momentum fraction $z_p$.
In the limit where $1-z\ll 1$, one has four flavour channels to
consider corresponding to the leading parton, 3, being a quark or a
gluon, and to the radiated partons, 1 and 2, being either two gluons
or a $q\bar q$ pair.
In all cases, the remaining part of
Eq.~(\ref{eq:NNDL-double-soft-K-FO}), proportional to
$\Theta(k_{t,2}>\ktcut)$, then takes the form
\begin{align} 
  \int \! \frac{\dd z}{1-z} \frac{\dd \theta_a^2}{\theta_a^2} \,
  \int \! \dd z_p \frac{\dd \normtheta_{12}^2}{\normtheta_{12}^2} 
  \frac{\dd\phi_{12}}{2\pi}
  \frac{z_p (1-z_p)(1-z)^2\normtheta_{12}^2}{\left[1+z_p(1-z_p)\normtheta_{12}^2\right]^2}
  \left(\frac{\as}{2\pi}\right)^2 \hat P_{1\to 3}^{f_1f_2f_3}
  \Theta(k_{t,21} > \ktcut).
  \label{eq:NNDL-double-soft-RR-FO}
\end{align}
for final flavours $f_{1,2,3,}$ and with
$k_{t,21}=\tfrac{Q}{2}(1-z)\min(z_p,1-z_p)\theta_a\normtheta_{12}$ the
relative transverse momentum of $k_2$ with respect to
$k_1$.\footnote{This assumes that the pair ($k_1$,$k_2$) of partons
  clusters first. Strictly speaking, this is only true in the
  strongly-angular-ordered limit. For commensurate angles, one could
  have situations where $k_1$ or $k_2$ first clusters with the leading
  parton, in which case $k_{t,21}$ should be replaced by $k_{t,2}$,
  the relative transverse momentum of $k_2$ with respect to the
  leading parton.
  This difference only matters in the region where $k_{t,2}$ is close
  to $\ktcut$ (cf.\ Fig.~\ref{fig:nndl-diagram-clust}). This
  configuration is kinematically different from the one discussed here
  and is computed in the next section.
}
From this expression, one can extract the pure \NNDL contribution from
emissions at commensurate angles and $k_t$ by simply subtracting what
is obtained in the strongly-angular-ordered limit:
\begin{align} \label{eq:NNDL-double-soft-FO}
& \delta N^{(\NNDL)}_{i,RR}\\
  & = \left(\frac{\as}{2\pi}\right)^2
    \int \! \frac{\dd z}{1-z} \frac{\dd \theta_a^2}{\theta_a^2} \,
    \int \! \dd z_p \frac{\dd \normtheta_{12}^2}{\normtheta_{12}^2} 
    \frac{\dd\phi_{12}}{2\pi}
    \left\{\frac{z_p(1-z_p)(1-z)^2\normtheta_{12}^2}{\left[1+z_p(1-z_p)\normtheta_{12}^2\right]^2}
    \hat P_{1\to 3}^{f_1f_2f_3} - \mathcal{P}\right\}\,
    \Theta (k_{t,21} > \ktcut),\nonumber
\end{align}
where $\mathcal{P}$ is the appropriate combination of DGLAP splitting
function contributing at \NDL (see Appendix~\ref{app:triple-collinear}
for practical details).
Note that since we use the exact expression for $k_{t,21}$, the
subtraction also gets rid of the contribution computed in
Sec.~\ref{sec:nndl-end} where $k_2$ is at the hard-collinear
endpoint.
Eq.~\eqref{eq:NNDL-double-soft-FO} follows the expected kinematic
behaviour: the integrations over $z_p$ and $\normtheta_{12}$ are only
non-zero in an $\order{1}$ region of finite $z_p$ and
$\normtheta_{12}\sim 1$ ($\theta_{12}\sim
\theta_a$), and the integrations over $z$ and $\theta_a$
give a double-logarithmic factor.
After integration, we get
\begin{equation}
  \delta N^{(\NNDL)}_{i,RR}
  = \left(\frac{C_i}{C_A} \frac{\bar\alpha L^2}{2}\right)
  \times
  \left(\frac{\as}{2\pi}D_\text{pair}^{123}\right),
\end{equation}
with
\begin{subequations}\label{eq:NNDL-double-soft-D-coeffs}
  \begin{align}
    \label{eq:NNDL-double-soft-D-qq}
    D^{q\qbar}_\text{pair}\equiv D_\text{pair}^{g \to q\qbar g} = 
    D_\text{pair}^{q \to q \qbar' q} 
    & = \frac{13}{9} n_f T_R, \\
    \label{eq:NNDL-double-soft-D-gg}
    D^{gg}_\text{pair}\equiv D_\text{pair}^{g \to ggg} = D_\text{pair}^{q\to 
    ggq} 
    & = \left(\frac{\pi^2}{6}-\frac{67}{18}\right)C_A.
  \end{align}
\end{subequations}
Surprisingly, the contribution coming from $i \to ggi$
splittings given by Eq.~\eqref{eq:NNDL-double-soft-D-gg} exactly compensates
the $C_A$ contribution to the CMW $K$ factor in
Eq.~\eqref{eq:NNDL-K-term-FO}. This means that the correction
corresponding to Fig.~\ref{fig:nndl-diagram-double-soft} will only
have an $n_f T_R$ colour factor.

To resum these effects to all orders, we follow the usual strategy of 
considering that the correction can appear at any point in the chain of soft-and 
-collinear particles. The main subtlety concerns the $i \to q\qbar i$ splitting 
that, as in previous cases, generates a miscancellation between real and virtual 
contributions. The resummation formula then reads
\begin{align}
  \as h^{(i)}_{3, \text{pair}}
  & = \frac{C_i}{C_A}\int_0^L \dd\l_1
    \left[\delta(\l_1) + n_g^{(\DL)}(\l_1)\right]
    \abar \int_{\l_1}^L  \dd\l_2 \int_0^{\l_2-\l_1}\dd\eta_2
    ~\frac{\as}{2\pi}
    \label{eq:NNDL-double-soft-K-resummation-formula} \\
    &\phantom{=\frac{C_i}{C_A}}
    \left[
    \left(D_\text{pair}^{gg}+K\right)N_g^{(\DL)}(L-\l_2)
    + D_\text{pair}^{q\bar q}\left(2N_q^{(\DL)}(L-\l_2)-N_g^{(\DL)}(L-\l_2) 
    \right)
    \right].
    \nonumber
\end{align}
Evaluating the integrals in
Eq.~\eqref{eq:NNDL-double-soft-K-resummation-formula}, we get
\begin{align}
  h^{(i)}_{3, \text{pair}}
  =
  \frac{C_i}{C_A}\frac{1}{2\pi}
  \left[
  (1-\cdiff) D_{\text{pair}}^{q\qbar} (\coshnu-1)
  +\left(K + D_{\text{pair}}^{gg} + \cdiff D_{\text{pair}}^{q\qbar}\right)
  \frac{\nu}{2}\sinhnu
  \right].
  \label{eq:NNDL-double-soft-K-resummation}
\end{align}
with $\cdiff$ given by Eq.~\eqref{eq:cdiff}.

\subsubsection{Clustering corrections}\label{sec:nndl-clustering}
We now discuss the kinematic configuration where we have two emissions
(in any Lund plane) at commensurate angles,
Fig.~\ref{fig:nndl-diagram-clust}.
More specifically, say we have a `parent parton', $k_0$ (either the
leading parton or any emission spawning its own Lund plane) in colour 
representation $C_i$, accompanied by two extra gluon emissions: a first, $k_1$, 
soft-and-collinear in the bulk of the Lund plane, and a second one, $k_2$, much 
softer and with a transverse momentum of order $\ktcut$.

In the angular-ordered limit we have two well-separated contributions:
(i) `primary' radiation, where $k_2$ is radiated off $k_0$. This
comes with a $C_i^2$ colour factor, and the transverse momentum of $k_2$ is 
measured with respect to $k_0$, i.e.\ $k_{t,2}\equiv k_{t,20}$;
(ii) `secondary radiation', where $k_2$ is radiated off $k_1$. This
comes with a $C_iC_A$ colour factor, and the transverse momentum of
$k_2$ is measured with respect to $k_1$, i.e.\
$k_{t,2}\equiv k_{t,21}$.
When the emissions have commensurate angles, i.e.\ when
$\theta_{01}\sim\theta_{02}\sim\theta_{12}$, one must take exactly
into account the Cambridge clustering used to build the Lund
multiplicity.
One can therefore be in a situation where a primary emission $k_2$ is
first clustered with $k_1$, meaning that its relative $k_t$ would
actually be $k_{t,21}$ instead of $k_{t,20}$, or in a situation where
a secondary emission $k_2$ is first clustered with $k_0$, meaning that
its relative $k_t$ would be $k_{t,20}$ instead of $k_{t,21}$.
When $k_{t,2}$ is close to $\ktcut$, finite differences between
$k_{t,20}$ and $k_{t,21}$ can put the emission below or above the cut,
introducing an \NNDL correction compared to what strict angular
ordering would give.
We call this a {\em clustering} correction since it is associated with
details of the Cambridge clustering.

As usual, we first compute the \NNDL correction at fixed order, here
$\order{\as^2}$.
The emission of $k_1$ off $k_0$, at any angle, can be described using the
soft-collinear matrix element in Eq.~\eqref{eq:soft-coll-me}.
Since the second emission is soft and at angles commensurate
with $\theta_{01}$, we instead describe it using the eikonal factor given 
by Eq.~\eqref{eq:eikonal}. We further assume that all three angles 
$\theta_{01}$, $\theta_{02}$ and $\theta_{12}$ are small and strongly separated
from all the other angles in the event.
In this case, $k_2$ can be emitted from any of the three dipoles
formed by $k_0$, $k_1$ and the recoiling system, $k_r$,\footnote{At our
  accuracy, the recoiling system can be viewed as a unique object as
  long as it is at much larger angles.}
and the sum over dipoles
\begin{equation}
\label{eq:k2-emission}
  C_A(k_2|k_0k_1) + C_A(k_2|k_1k_r)
  +(2C_i-C_A) (k_2|k_0k_r),
\end{equation}
simplifies considerably.
One gets to the following expression for the \NNDL clustering
correction:
\begin{align}
  \delta N^{(\NNDL)}_{i,{\text{clust}},\order{\as^2}}
  = \left(\frac{2\as}{\pi}\right)^2 C_i
 & \int\frac{\dd k_{t,1}}{k_{t,1}}
  \dd \eta_1
    \int \frac{\dd E_2}{E_2} \frac{\dd^2\theta_2}{2\pi}
    \bigg\{
    \bigg[
    \frac{C_A}{2}\frac{1}{\theta_{12}^2}
    +\frac{C_A}{2}\frac{\theta_{01}^2}{\theta_{02}^2\theta_{12}^2}
    +\Big(C_i-\frac{C_A}{2}\Big)\frac{1}{\theta_{02}^2}
    \bigg]\nonumber\\
  & \times\left[
     \Theta_{\text{clust},0}\Theta(k_{t,20}>\ktcut)
    +\Theta_{\text{clust},1}\Theta(k_{t,21}>\ktcut)
    \right]\label{eq:nndl-clust-as2-base}\nonumber\\
  & - \frac{C_i}{\theta_{02}^2} \Theta(k_{t,20}>\ktcut)
    - \frac{C_A}{\theta_{12}^2} \Theta(\theta_{21}<\theta_{01})
    \Theta(k_{t,21}>\ktcut)\bigg\},
\end{align}
where $\Theta_{\text{clust},i}$ encodes the condition for $k_2$ to be
reconstructed as an emission from $k_i$:
\begin{equation}
  \Theta_{\text{clust},1} \equiv
  \Theta(\theta_{12}<\theta_{01},\theta_{02})
  \quad\text{ and }
  \quad
  \Theta_{\text{clust},0} = 1-\Theta_{\text{clust},1}.
\end{equation}
When writing Eq.~\eqref{eq:nndl-clust-as2-base}, the \NNDL clustering
correction is explicitly written as the difference between the full
result including the exact clustering (the first two lines) and the
already-computed \NDL expression with angular ordering (the last
line).

Treating separately the $C_i^2$ and $C_iC_A$ contributions one sees
that the integration over $k_2$ in Eq.~\eqref{eq:nndl-clust-as2-base}
is finite.
The constraints on $k_{t,20}$ or $k_{t,21}$ can be used to perform the
integration over $E_2$ and are non-zero only in a region where
$\theta_{02}\sim\theta_{12}\sim\theta_{01}$. We can then do the
$\theta_{2}$ integration, yielding
\begin{equation}\label{eq:NNDL-clustering-as2}
  \delta N^{(\NNDL)}_{i,{\text{clust}},\order{\as^2}}
  = \left(\frac{2\as C_i}{\pi} \frac{L^2}{2}\right)
  \times \left[\frac{\as}{2\pi}\left(
      C_i D_\text{clust}^\text{(prim)} + C_AD_\text{clust}^\text{(sec)}
    \right)\right],
\end{equation}
with
\begin{equation}\label{eq:NNDL-clustering-D-coeff}
  D_{\text{clust}}^{\text{(prim)}} = -\frac{5\pi^2}{54},
  \quad\text{ and } \quad
  D_{\text{clust}}^{\text{(sec)}} = \frac{\pi^2}{27}.
\end{equation}
The generalisation to all orders is relatively straightforward: for
each \DL gluon emission $k_2$, emitted from a parent $k_1$, one should
include an \NNDL correction of the form of
Eq.~\eqref{eq:NNDL-clustering-as2}.
One should just properly separate the case where $k_2$ is a primary
emission, i.e.\ $k_1$ is the leading parton, for which $C_i$ is given
by the flavour of the leading parton, from all subsidiary emissions
where $C_i=C_A$.
We therefore write
\begin{align}
  \as h_{3,\,\text{clust}}^{(i)}
  =
  \frac{\as C_i}{2\pi}
  \left[
  \frac{\abar L^2}{2}
  \left(\frac{C_i}{C_A} D_{\text{clust}}^{\text{(prim)}} +
  D_{\text{clust}}^{\text{(sec)}}\right)
  +
  \int_0^L \dd \l
  \left( n_g^{(\DL)}(\l) - \abar \l \right)
  \left(D_{\text{clust}}^{\text{(prim)}} +D_{\text{clust}}^{\text{(sec)}}\right)
  \right],
  \label{eq:NNDL-clustering-resummed}
\end{align}
where the first (second) term corresponds to primary (subsidiary)
emissions.
In particular, one recognises $\int \dd\ell (\abar\ell)=\abar L^2/2$ as 
the multiplicity of primary emissions.
Eq.~(\ref{eq:NNDL-clustering-resummed}) evaluates to:
\begin{align}\label{eq:NNDL-clustering-resummed-result}
  h_{3,\,\text{clust}}^{(i)}
  & = \frac{C_i}{2\pi}
    \left\{\left[
    \coshnu-1-\left(1-\frac{C_i}{C_A}\right)\frac{\nu^2}{2}\right]
    D_\text{clust}^\text{(prim)}
    + (\coshnu-1) D_\text{clust}^\text{(sec)}\right\}.
\end{align}

\subsubsection{A primary large-angle emission with
  $k_t\sim\ktcut$: the Cambridge multiplicity}\label{sec:nndl-cam}
The last pure $\as$ contribution that we consider is that of a first,
primary, emission at large angles and with transverse momentum of
order $\ktcut$. The precise definition of $k_t$ plays a central role
in this correction and, therefore, differences between the
Lund~\eqref{kt:def} and Cambridge~\eqref{kt:def:cambridge} $k_t$
definitions are expected.

As already discussed in our \NDL calculation, the definition of the
Lund-plane transverse momentum $k_t$, Eq.~\eqref{kt:def}, coincides exactly with the
relative transverse momentum of the emission with respect to the
initial $q\bar q$ (or $gg$) dipole, i.e.\ the relation in
Eq.~\eqref{eq:soft-coll-me} is also valid for soft emissions at large
angles. Hence, this contribution vanishes for the Lund
multiplicity.
However, this is not the case for the Cambridge definition of $k_t$,
which is related to the Lund one by
\begin{equation}
\label{eq:kt-cam-vs-lund}
k_t^{(\text{Cam})} = k_t^{({\text{Lund}})} \sqrt{1+e^{-2\eta}}. 
\end{equation}
The last factor in Eq.~\eqref{eq:kt-cam-vs-lund} results in an \NNDL
correction. At $\order\as$, the contribution of this large-angle,
$k_t\sim\ktcut$ primary emission is, after subtracting the \DL
contribution:
\begin{equation}\label{eq:Cambridge-Large-angle-soft-NNDL}
  \langle N^{(\text{Cam})}_{i, \text{la} \& k_t}\rangle_{\order{\as}}
  = \bar\alpha \frac{C_i}{C_A} 
  \int_0^Q \frac{\dd k_\perp}{k_\perp}
  \int_0^\infty\dd\eta\, \Theta(k_\perp\sqrt{1+e^{-2\eta}}>Qe^{-L}>k_\perp)
  = \frac{\alpha_s}{2\pi} C_i \frac{\pi^2}{12}.
\end{equation}
Noting that, first, a large-angle emission has to be the last
clustering in the Cambridge sequence and, second, an emission with
$k_t\sim\ktcut$ leaves no \DL phase-space for subsequent emission,
this result is valid to all orders.
We therefore find that the difference between the Cambridge and Lund
multiplicities is (see Appendix~\ref{app:cambridge-vs-lund} for
further details)
\begin{equation}\label{eq:NNDL-cam-result}
  h^{(i,\text{Cam})}_{3}
  - h^{(i,\text{Lund})}_{3}
  = h^{(i,\text{Cam})}_{3,\text{la}\& k_t}
  = \frac{C_i}{2\pi} \frac{\pi^2}{12}.
\end{equation}
%

\subsection{Corrections involving two $\as L$
factors}\label{sec:nndl-asL-squared}

\subsubsection{Energy loss}\label{sec:nndl-eloss}
In Sec.~\ref{sec:hard-collinear-NDL} we accounted for the \NDL
correction corresponding to a single hard-and-collinear splitting.
In the \NDL case, once a hard-collinear splitting has happened, 
further emissions from both branches can be treated 
as if the branch was carrying the same energy as the parent parton.
In other words, if $z\sim \order{1}$ is the momentum fraction of the
hard-collinear splitting, the factors $z$ and $1-z$ in the daughter
branches can be neglected and set to $1$.
These factors, which we discuss in this section, can however not be
neglected at \NNDL accuracy.  
Conceptually, finite $z$ and $1-z$ effects limit the available
phase-space for further emissions, as indicated by the red region in
Fig.~\ref{fig:nndl-diagram-eloss}. Consequently we name this
correction {\em energy loss}.

The energy loss \NNDL contribution can be computed following our
generic approach of describing \NNDL contributions as subleading
corrections dressed by an arbitrary number of double-logarithmic, nested,
soft-and-collinear emissions.
One therefore starts either with the leading parton or with any (\DL)
gluon emitted from it via nested soft-collinear radiation at a
transverse momentum scale $\ell_1$.
This parton then undergoes a (real or virtual) hard-collinear
splitting at a transverse scale $\ell_2$, with a momentum fraction
$z$.
For a real splitting, the resulting ``hard'' and ``soft'' branches
will then radiate \DL soft-collinear emissions from (relative)
transverse momentum $k_{t,\text{hard}}=(1-z)e^{-\ell_2}$ and
$k_{t,\text{soft}}=z e^{-\ell_2}$, respectively, taking into account
the exact energy sharing in the hard-collinear branching.
One therefore writes, summing over all possible flavour configurations,
\begin{align}
 \langle N^\text{(Lund)}_{i, \text{e-loss}}\rangle = &\sum_{abc} \int_0^L \dd\ell_1
  \Big[
  \delta_{ai} \delta(\ell_1) + \delta_{ag} n_i^{(\DL)}(\ell_1)
  \Big]
  \int_{\ell_1}^L \dd\ell_2 \int \dd z \frac{\as}{\pi} P_{a\to bc}(z) \\
  & \left[
  N_b^{(\DL)}(L - \ell_2+\ln(1-z))
  +N_c^{(\DL)}(L - \ell_2+\ln(z))
  -N_a^{(\DL)}(L - \ell_2)  
  \right].\nonumber
\end{align}
Again, it is convenient to separate the splitting function into a
soft-divergent component (for gluon emissions) and a finite piece,
i.e.\
$P_{a\to bc}(z)=\tfrac{2C_a}{z}\delta_{cg} + p^\text{(finite)}_{a\to
  bc}(z)$.
In the small-$z$ limit, only the singular part contributes. Combined
with the $N_c^{(\DL)}(L - \ell_2+\ln(z))$ term in the second line,
it gives rise to the \DL behaviour.
For the remaining terms, the $z$ integral has no logarithmic
enhancements.
Next, one recognises the \NDL contribution coming from the finite
part of the splitting function, neglecting the $\ln(1-z)$ and
$\ln(z)$ offsets in $N_b^{(\DL)}$ and $N_c^{(\DL)}$.
These offsets however start contributing at \NNDL accuracy, either from the
soft part of the splitting function together with
$N_b^{(\DL)}(L-\ell_2+\ln(1-z))$, or from the finite part of the
splitting function with both $N_b$ and $N_c$ in the second line. Up to
subleading corrections, we can use the expansion:
\begin{subequations}
  \begin{align}
    N_b^{(\DL)}(L - \ell_2+\ln(1-z))
    & \simeq N_b^{(\DL)}(L-\ell_2)
    + n_b^{(\DL)}(L-\ell_2)\,\ln(1-z),\\
    N_c^{(\DL)}(L - \ell_2+\ln(z))                                    
    & \simeq N_c^{(\DL)}(L-\ell_2)
    + n_c^{(\DL)}(L-\ell_2)\,\ln(z),
  \end{align}
\end{subequations}
with \NNDL corrections associated with the right-most terms.
The $z$ integration can then be carried out for each flavours
 yielding ($n_q^{(\DL)}= C_F/C_A n_g^{(\DL)}$)
\begin{equation}
  \langle N^\text{(Lund)}_{i, \text{e-loss}}\rangle
  = \frac{C_i}{C_A}\frac{\as}{2\pi} \int_0^L \dd\ell_1
  \Big[
  D_\text{e-loss}^{i}  \delta(\ell_1) + D_\text{e-loss}^{g}  n_g^{(\DL)}(\ell_1)
  \Big]
  \int_{\ell_1}^L \dd\ell_2 \, n_g^{(\DL)}(L-\ell_2).
\end{equation}
with, the following \NNDL coefficients
\begin{subequations}
  \begin{align}
    D_\text{e-loss}^{q}
    & = 2 \int \dd z \Big[P_{q\to qg}(z)\ln(1-z) + \frac{C_A}{C_F}p^\text{(finite)}_{q\to qg}(z)\ln(z)\Big]
      = \frac{7}{2}C_A + \left(\frac{5}{2} - \frac{2\pi^2}{3} \right) C_F,\\
    D_\text{e-loss}^{g}
    & = 2 \int \dd z \Big\{P_{g\to gg}(z)\ln(1-z) + p^\text{(finite)}_{g\to gg}(z)\ln(z)
      + \frac{C_F}{C_A} P_{g\to q\bar q}(z)[\ln(1-z) + \ln(z)]\Big\}\nonumber\\
    & = \left(\frac{67}{9} - \frac{2\pi^2}{3}\right)C_A -\frac{26}{9}\frac{C_F}{C_A}n_f T_R.
  \end{align}
\end{subequations}
The remaining integrations are easily performed using the \DL results to give
\begin{subequations}\label{eq:NNDL-E-loss-resummation-result}
  \begin{align}
    \label{eq:NNDL-E-loss-resummation-result-q}
     h_{3, \text{e-loss}}^{(q)}
    &=
    \frac{C_F}{C_A}\frac{1}{2\pi}
    \left[
    D_{\text{e-loss}}^g \, \frac{1}{2}\nu \sinhnu
    +
    \left(D_{\text{e-loss}}^q  - D_{\text{e-loss}}^g\right)(\coshnu - 1)
    \right], \\
    h_{3, \text{e-loss}}^{(g)}
    &=
    \frac{1}{2\pi}D_{\text{e-loss}}^g \, \frac{1}{2}\nu \sinhnu.
    \label{eq:NNDL-E-loss-resummation-result-g}
  \end{align}
\end{subequations}

\subsubsection{Squared hard-collinear correction}\label{sec:nndl-squared-hc}
We continue by considering the \NNDL correction when two emissions in
the nested chain receive a contribution from the finite part of the
splitting function in the hard-collinear limit. In this case we directly perform 
the resummation. Since the two \NDL corrections are strongly-ordered in angle 
(and hence in $k_t$ as they happen at the hard-collinear edge of the Lund 
plane), we can even simplify Eq.~\eqref{eq:final-master} a bit further by 
inserting explicitly only the \NDL correction associated with the larger angle
(larger $k_t$) emission and rely on the (resummed) \NDL expressions,
Eqs.~\eqref{eq:ndl-hc-fd} and~\eqref{eq:ndl-hc-fc} to handle the
second \NDL correction.

We split the calculation in three parts, according to whether (i) the two
\NDL emissions are both flavour-diagonal, (ii) one is flavour-diagonal and
the other flavour-changing, or (iii) both emissions are flavour-changing:
\begin{align}
  \label{eq:NNDL-HC-squared-resummed}
  h_{3, \text{hc}^2} =
  h_{3, \text{fd}^2} + h_{3, \text{fd+fc}} + h_{3, 
  \text{fc}^2}.
\end{align}
For the sake of brevity we give the integral form of each contribution
below and write the full, integrated, result for the sum at the end of
the section.

\paragraph{Two flavour diagonal.}
Here, one has a first hard-collinear splitting occurring either on the
primary branch or on a subsidiary one. The next \NDL correction has to
happen in the secondary (gluon) branch of the first splitting for which we can 
recycle the \NDL multiplicity results from Sec.~\ref{sec:recap-ndl}.\footnote{As 
in the \NDL case, the primary branch has no corrections as the real and virtual 
contributions from the first hard-collinear splitting cancel.}
We thus write 
\begin{equation} 
  \as h_{3, \text{fd}^2}^{(i)}
  = 
  \frac{C_i}{C_A}
  \int_0^L \dd \l_1  \left[ \delta(\ell_1)B_i+n_g^{(\DL)}(\l_1)B_{gg}\right]\, 
  \abar
  \int_{\l_1}^L \dd \l_2  \, N^{(\NDL)}_{g,\text{hc-fd}}(L; \ell_2),
  \label{eq:nndl-HC-FD-FD} 
\end{equation}
with $B_i=B_q$ ($B_{gg}$) for a quark (gluon) given by Eq.\eqref{eq:B-coeffs}, and
$N^{(\NDL)}_{g,\text{hc-fd}}(L;
\l_2)=N^{(\NDL)}_{g,\text{hc-fd}}(L-\l_2)$ obtained from
Eq.~\eqref{eq:ndl-hc-fd}.

\paragraph{One flavour changing, one flavour diagonal.}
We move on to the case when one of the two hard-collinear emissions is
a gluon splitting to a $q\qbar$ pair. This proceeds as for the
flavour-diagonal case except for two modifications: (i) the first
hard-collinear splitting can either be the hard gluon emission or the
$g\to q\bar q$ splitting, and, (ii) after a $g\to q\bar q$ splitting a 
miscancellation between real and virtual splittings occur as we have discussed 
in several previous cases (cf. Sec.~\ref{sec:nndl-double-soft}).
This yields
\begin{align}\label{eq:nndl-HC-FD-FC} 
   \as h_{3, \text{fd-fc}}^{(i)}
  = 
  & \frac{C_i}{C_A} { \bigg \{ } 
    \int_0^L \dd \l_1   \left[\delta(\l_1)B_i
    +n_g^{(\DL)}(\l_1)B_{gg}\right] \,
    \abar\int_{l_1}^L \dd \l_2 \, N^{(\NDL)}_{g,\text{hc-fc}}(L; \ell_2) \\
  & + \left.
    \int_0^L \dd \l_1  \left[\delta(\l_1)\delta_{ig}
    + n_g^{(\DL)}(\l_1)\right]B_{gq} \, \abar\int_{l_1}^L \dd \l_2 
    \left[2N^{(\NDL)}_{q,\text{hc-fd}}(L;\ell_2) -
    N^{(\NDL)}_{g,\text{hc-fd}}(L; \ell_2)\right] 
  \right\}.\nonumber
\end{align}
The first (resp.\ second) line corresponds to the flavour-diagonal
(resp.\ flavour-changing) splitting happening first.
The two terms in the square brackets under the $\ell_1$ integration
describe, as above, emissions from the primary branch or from any
subsidiary one.
Finally, the $\delta_{ig}$ factor on the last line ensures that
$g\to q\bar q$ splitting on the primary branch are only included for
gluon jets.

\paragraph{Two flavour changing.}
Finally we consider the case of two flavour-changing $g \to q\qbar$
splittings. Following the same logic and accounting again for
the first splitting being either on the primary branch or on any
subsidiary one, we get
\begin{equation} 
  \as h_{3, \text{fc}^2}^{(i)}
  = 
  \frac{C_i}{C_A}
  \int_0^L \dd \l_1 \left[\delta(\l_1)\delta_{ig}+n_g^{(\DL)}(\l_1)\right] \abar
   \int_{\l_1}^L \dd \l_2 \, B_{gq}\,
   \left[2N^{(\NDL)}_{q, \,\text{hc-fc}}(L; \l_2) - N^{(\NDL)}_{g, \, 
   \text{hc-fc}}(L; \l_2)\right].
  \label{eq:nndl-HC-FC-FC} 
\end{equation}
where
$N^{(\NDL)}_{g,\text{hc-fc}}(L;\l_2)=N^{(\NDL)}_{g,\text{hc-fc}}(L-\l_2)$,
obtained from Eq.~(\ref{eq:ndl-hc-fc}).

\paragraph{Final result for $h_{3, \text{hc}^2}^{(i)}$.}
The overall contribution to $h_3$ from two hard-collinear branchings
is the sum of the above three terms,
Eqs.~(\ref{eq:nndl-HC-FD-FD})-(\ref{eq:nndl-HC-FC-FC}). Using the
explicit \NDL results, the integrations are relatively straightforward
and give
\begin{subequations}  \label{eq:NNDL-HC-squared-resummed-result}
  \begin{align}
    h_{3, \text{hc}^2}^{(q)}
    = \frac{C_F}{4\pi}
    & \left\{
      (B_{gg}+\cdiff B_{gq})^2\nu^2 \coshnu
      +8\left[2\cdiff B_{gg}- 2\cdiff B_q
      -(1-3\cdiff^2)B_{gq}\right]B_{gq}\coshnu\right.\nonumber\\
    & \left.+\left[4 B_q(B_{gg}+(2 \cdiff+1)B_{gq})-(B_{gg}+\cdiff
      B_{gq})(B_{gg}+9\cdiff B_{gq})\right]\nu\sinhnu\right.\nonumber\\
    & \left. +4(1-\cdiff^2)B_{gq}^2\nu^2+8\left[2\cdiff B_q-2\cdiff 
    B_{gg}+(1-3\cdiff^2)B_{gq}\right]B_{gq}\right\}
    \\
    h_{3, \text{hc}^2}^{(g)}
    = \frac{C_A}{4\pi}
    & \big\{
      (B_{gg}+\cdiff B_{gq})^2\nu^2 \coshnu
      -8(1-\cdiff^2)B_{gq}^2(\coshnu-1)\nonumber\\
    & +\left[(B_{gg}+\cdiff B_{gq})(3B_{gg}-5\cdiff
      B_{gq})+4(1+\cdiff)B_{gq}B_q\right]\nu\sinhnu\big\}.
  \end{align}
\end{subequations}
%

\subsubsection{A hard-collinear emission with a soft emission at
  commensurate angle}\label{sec:nndl-hardcoll-comm-angle}
We investigate the case of a first hard-and-collinear
emission which is followed by a soft-and-commensurate-angle emission, 
as depicted in Fig.~\ref{fig:nndl-diagram-hcxcomm}.
To see that this contribution does not contribute at the \NNDL
accuracy, it is sufficient to show it at $\order{\as^2}$.
We therefore consider a parton $k_0$ with colour factor $C_i$ which
radiates a hard-and-collinear gluon $k_1$ followed by a subsequent
emission of a soft gluon $k_2$ at a commensurate angle.
The arguments below can also be carried on with an initial
flavour-changing splitting.
The contribution of this system to the Lund multiplicity at
$\order{\as^2}$ is\footnote{This expression already accounts for the
  cancellation of the term proportional to $C_i(k_2|k_0k_r)$ in the
  square bracket including both real and virtual contributions for the
  $k_1$ gluon.}
\begin{align}
  \langle N^\text{(Lund)}_{i,\,\text{hc}\times\text{comm}}\rangle_{\order{\as^2}}
  &=
  \left(\frac{\as}{\pi}\right)^2
  \int_0^\infty \dd \eta_1\!\!
  \int_0^{1} \dd z_1 \, P_{i \to ig}(z_1) \!\!
  \int \frac{\dd E_2}{E_2}\frac{\dd^2\theta_2}{2\pi} \nonumber \\
  &\times
   \left[ \frac{C_A}{2}(k_2|k_1k_0) 
   +\frac{C_A}{2}(k_2|k_1k_r) - \frac{C_A}{2}(k_2|k_0k_r)
  \right]
  \Theta{(k_{t,21} > \ktcut)},
  \label{eq:NNDL-HC-comm-angle-FO}
\end{align}
where $P_{i \to ig}(z)$ is the appropriate DGLAP splitting function,
defined in Eq.~\eqref{eq:splitting-function}, $k_r$ is the recoiling
system (at angles much larger than the ones between $k_0$, $k_1$ and
$k_2$) and the eikonal functions $(k|ij)$ were defined in
Eq.~\eqref{eq:eikonal}.
Integrating over the azimuthal angle of $k_2$ one easily recovers, in
the limit $\theta_{10},\theta_{20},\theta_{21}\ll 1$, the
standard angular-ordered result, namely,
\begin{align}
  \langle N^\text{(Lund)}_{i,\,\text{hc}\times\text{comm}}\rangle_{\order{\as^2}}
  &=
  \abar
  \frac{\as}{\pi}
  \int_0^\infty \dd \eta_1\!\!
  \int_0^{1} \dd z_1 \, P_{i \to ig}(z_1) \!\! 
  \int_0^{z_1} \frac{\dd x_2}{x_2} \,  \!\!
  \int_0^{\eta_1} \dd \eta_{21} ~
  \Theta{(x_{2}e^{-\eta_{21}} > e^{-L})},
  \label{eq:NNDL-HC-comm-angle-FO-simplified}
\end{align}
where $\eta_{21} = -\ln{(\theta_{21}/2)}$, and $x_2 = 2E_{2}/Q$.
Due to exact angular ordering being recovered,
Eq.~\eqref{eq:NNDL-HC-comm-angle-FO-simplified} represents precisely
the \NDL accurate contribution to the multiplicity described in
Sec.~\ref{sec:hard-collinear-NDL}, with no \NNDL correction.

A key point in the above argument is that the $\theta_{21}$ angle in
Eq.~\eqref{eq:NNDL-HC-comm-angle-FO-simplified} is bounded by the
physical angle $\theta_{10}$, i.e.\ the angle between the daughter
partons after the $k_0k_1$ branching.
Technically, working in the collinear limit, one could choose to
impose angular-ordering with a different definition of the angle
(e.g.\ the angle of the emission with respect to the parent parton).
In this case, the contribution discussed in this section would no
longer vanish.
However, such a change would also affect the definition of $\eta$ in
the hard-collinear endpoint, Eq.~(\ref{eq:NNDL-endpoint-FO}),
reshuffling a contribution between the hard-collinear and
hard-matrix-element coefficients.
We have checked explicitly that this does not change the final
resummed result. 

\subsubsection{A primary large-angle emission}\label{sec:nndl-large-angle-first}
In Sec.~\ref{sec:recap-ndl} we showed that a first soft-and-large-angle 
primary emission does not yield a \NDL correction.
We now study the effect of such a soft-and-large-angle
emission on subsequent ones. Concretely, following the discussion in 
Sec.~\ref{sec:nndl-list} we investigate two possible kinematic configurations 
for the second emission: (i) soft and at commensurate angles to the first,
Fig.~\ref{fig:nndl-diagram-la}, and (ii) soft and collinear with
$k_t \sim \ktcut$, Fig.~\ref{fig:nndl-diagram-laxL}.
We therefore write
\begin{equation}
  D_{{\text {la}}}
  = D_{{\text {la}}^2}
  + D_{{\text {la}}\times k_t}.
\end{equation}
We show below that the contributions cancel each other exactly.

We start with the configuration in
Fig.~\ref{fig:nndl-diagram-la}, where an initial system $(k_0k_r)$ emits a 
soft-and-large-angle parton, $k_1$, which then radiates a soft
emission $k_2$ at a commensurate angle.
The contribution to the Lund multiplicity is (after cancelling the real and virtual
contributions for the first emission)
\begin{align}
  \label{eq:NNDL-two-LA-FO}
  \langle N^\text{(Lund)}_{i, \text{la}^2}\rangle_{\order{\as^2}}
  &= \left(\frac{\as}{\pi}\right)^2\int\frac{\dd 
  E_1}{E_1}\frac{\dd^2\theta_1}{2\pi}\int\frac{\dd 
  E_2}{E_2}\frac{\dd^2\theta_2}{2\pi} \\ 
   &\times C_i(k_1|k_0k_r)\left[\frac{C_A}{2}(k_2|k_1k_0) + 
   \frac{C_A}{2}(k_2|k_1k_r) -\frac{C_A}{2} (k_2|k_0k_r)    
   \right]
  \Theta{(k_{t,21} > \ktcut)}\nonumber .
\end{align}
After integrating over the solid angle of the second emission and subtracting 
the \DL term in which the two emissions are angular-ordered, we obtain the 
following \NNDL correction
\begin{align}
  \label{eq:NNDL-two-LA-FO-correction}
  \delta N^{(\NNDL)}_{i, \text{la}^2} =
  \frac{C_i}{C_A}\frac{\abar L^2}{2} \, \frac{\as}{2\pi}D_{\text{la}^2} 
  \quad {\text{with}} \quad 
  D_{{\text {la}}^2} = -\frac{\pi^2}{6}C_A.
\end{align}
Let us now move to the case where the first soft-and-large-angle emission is
followed by a  soft-and-collinear emission with $k_t \sim \ktcut$, as in
Fig.~\ref{fig:nndl-diagram-laxL}.
The Lund multiplicity at $\order{\as^2}$ is given by
\begin{align}
  \label{eq:NNDL-LA-kt-FO}
  \langle N^\text{(Lund)}_{i, \text{la} \times k_t}\rangle_{\order{\as^2}}
  &= \frac{4\as^2 C_i C_A}{\pi^2}
    \int\dd\eta_1
    \int\frac{\dd  k_{t,1}}{k_{t,1}}
    \int_{\eta_1}^\infty \dd\eta_2
    \int \frac{\dd z_2}{z_2}\Theta{(k_{t,21} > \ktcut)},
\end{align}
where we have replaced the eikonal factor for the second emission
by its soft-and-collinear limit.
The \NNDL correction stems from taking into account the exact
expression for $k_{t,21}$, namely
\begin{equation} 
  k_{t,21}
  = E_2 \sin(\theta_{21})
  \approx E_2 \theta_{21}
  \approx z_2 E_1 \theta_{21}
  \approx z_2 k_{t,1}\, 2 \cosh\eta_1\,e^{-\eta_{21}},
  \label{eq:kt-large-angle}
\end{equation}
where, for the last equality, we have used the generalisation of
Eq.~(\ref{eq:kt-soft-coll}) beyond the collinear limit,
$E_1=k_{t,1}\cosh\eta_1$, and $\theta_{21}\approx 2e^{-\eta_{21}}$.
This expression differs from the soft-collinear limit by a factor
$2e^{-\eta_1}\cosh\eta_1=1+e^{-2\eta_1}$, so that, after subtracting
the \DL contribution from Eq.~\eqref{eq:NNDL-LA-kt-FO}, one is left
with the \NNDL correction (with $\ell_1=\ln(Q/k_{t,1})$)
\begin{align}
  \label{eq:NNDL-LA-kt-FO-correction}
  \langle N^\text{(Lund)}_{i, \text{la} \times k_t}\rangle_{\order{\as^2}}
  & = \frac{4\as^2 C_i C_A}{\pi^2}
    \int_0^Ld\ell_1 \int_0^{L-\ell_1} d\eta_{21}
    \int_0^\infty \dd\eta_1 \ln(1+e^{-2\eta_1})\\
  & =
  \frac{C_i}{C_A}\frac{\abar L^2}{2} \,
    \frac{\as}{2\pi}D_{\text{la}\times k_t} 
  \qquad {\text{with}} \qquad 
  D_{{\text {la}}\times k_t} = \frac{\pi^2}{6}C_A = -D_{{\text {la}}^2}. \nonumber
\end{align}
The sum of~\eqref{eq:NNDL-two-LA-FO-correction} and
~\eqref{eq:NNDL-LA-kt-FO-correction} therefore vanishes and it is
relatively straightforward to see that this relation holds to all
orders.\footnote{One can for example realise that both contributions
  shift the boundary of the secondary Lund plane spawned by the first
  soft-large-angle emission by an equal amount proportional to
  $\int_0^\infty \dd\eta \ln(1+e^{-2\eta})$. The ``la$^2$'' shifts
  the large-angle boundary of the secondary plane, reducing the
  phase-space, while the ``la$\times k_t$'' contributions shifts the
  lower edge, enhancing the phase-space by the same amount. This
  results in an overall shift which has no net effect on the \NNDL
  multiplicity.}
%
\subsubsection{Three emissions at commensurate
angles}\label{sec:nndl-three-comm-angles}
We consider the case where a first soft-and-collinear emission $k_1$
is followed by two subsequent emissions, $k_2$ and $k_3$, strongly
ordered in energy, $E_1\gg E_2\gg E_3$, and at commensurate angles,
$\theta_1\sim \theta_2\sim\theta_3$, as depicted in
Fig.~\ref{fig:nndl-diagram-3comm}.
In the \NDL limit, once both real and virtual
emissions are taken into account, the only left-over contribution
comes from the angular-ordered \DL result,
$\theta_{23}<\theta_{12}<\theta_1$, proportional to $C_iC_A^2$.
We show here that this is still the case at \NNDL accuracy, i.e.\ that
the contribution of Fig.~\ref{fig:nndl-diagram-3comm} vanishes.

It is sufficient to consider this configuration at $\order{\as^3}$.
The first gluon, $k_1$, is emitted off a ($k_0, k_r$) dipole according to
the eikonal factor $(k_1|k_0,k_r)$ in Eq.~\eqref{eq:eikonal}, where
for a collinear $k_1$ emission the specific choice of the large-angle
recoiling momentum $k_r$ is irrelevant.
The second gluon $k_2$ can then be emitted from any of the three dipoles
($k_0, k_r$), ($k_0, k_1$) or ($k_1, k_r$) resulting in the
combination in Eq.~\eqref{eq:k2-emission}.
Finally, the third emission $k_3$ can be emitted from any dipole in
the ($k_0$, $k_1$, $k_2$, $k_r$) system.
The full matrix element for all three emissions can be deduced from the
expression in Refs.~\cite{Catani:1999ss,Hamilton:2021dyz}, valid for
$E_3\sim E_2$, which, after taking the limit $E_3\ll E_2$ can be cast
into the following form:
\begin{align}
  & C_i\frac{C^2_A}{4}(1|0r)
    \Big\lbrace (2|10)\big[
    (3|20)+(3|21)+(3|1r)-(3|0r)
    \big] + (2|1r)\big[
    (3|2r)+(3|21)+(3|10)-(3|0r)
    \big]\nonumber\\
  & \phantom{C_iC^2_A(1|0r)}
    - (2|0r) \big[
    (3|20)+(3|2r)+(3|10)+(3|1r)
    \big]\Big\rbrace\nonumber \\
  & + C_i^2\frac{C_A}{2} (1|0r) \Big\lbrace (3|0 r)
    \big[(2|1 0)+(2|1 r)\big] + (2|0r) \big[
    (3|20)+(3|2 r)+(3|10)+(3|1r)-2(3|0r)
    \big]\Big\rbrace \nonumber \\
  &+C^3_i(1|0r)(2|0r)(3|0r),
\label{eq:3-comm-angle-alphas3-rrr}
\end{align}
where for the sake of readability we used the shorthand notation
$(c|ab)$ to denote the factor $(k_c|k_ak_b)$.

Besides the matrix element proportional to
Eq.~\eqref{eq:3-comm-angle-alphas3-rrr} we need to account for the
fact that each of the three emissions can be either real or virtual.
Summing over all possible emission configurations we find that the
Lund multiplicity at $\order{\as^3}$ is\footnote{The contributions
  proportional to $\Theta(k_{t,2}>\ktcut)$ and
  $\Theta(k_{t,1}>\ktcut)$ vanish when summing the cases where the
  third soft emission is either real or virtual.
}
\begin{align} \label{eq:NNDL-3-comm-FO-simplified}
  \langle N^\text{(Lund)}_{i, 3\text{comm}}\rangle_{\order{\as^3}}
  & = \left(\frac{\as C_A}{2\pi}\right)^3 
  \frac{2C_i}{C_A}\int \frac{\dd 
  E_1}{E_1}\frac{\dd^2\theta_1}{2\pi} \int \frac{\dd 
  E_2}{E_2}\frac{\dd^2\theta_2}{2\pi}
  \int \frac{\dd E_3}{E_3}\frac{\dd^2\theta_3}{2\pi} (k_1|k_0k_r) \nonumber \\ 
  &\times \Big\lbrace (k_2|k_1 k_0)\left[(k_3|k_2k_0)+ 
  (k_3|k_2k_1)-(k_3|k_1k_0)\right] \nonumber \\ 
  &+ (k_2|k_1 k_r)\left[(k_3|k_2k_r)+ 
  (k_3|k_2k_1)-(k_3|k_1k_r)\right] \nonumber \\
  &- (k_2|k_0 k_r)\left[(k_3|k_2k_0)+ 
  (k_3|k_2k_r)-(k_3|k_0k_r)\right] \Big \rbrace \times \Theta(k_{t,3}>\ktcut).
\end{align}
A few comments about Eq.~\eqref{eq:NNDL-3-comm-FO-simplified} are in
order.
Since all emissions are above the $\ktcut$, we can ignore differences
in the $k_{t,3}$ definition, i.e.\ we can treat
$k_{t,30},k_{t,31},k_{t,32}$ indistinctly.
More interestingly, only the term proportional to $C_iC_A^2$ survives,
as was already the case at \DL accuracy.

At small angles, all the eikonal factors can be simplified according
to $(c|ab)\approx 2\theta_{ab}^2/(\theta_{ac}^2\theta_{bc}^2)$
for $b,c\neq r$ and $(c|ar)\approx 2/\theta_{ac}^2$.
After performing the (2-dimensional) integrations over $\theta_2$ and
$\theta_3$, one finds that Eq.~\eqref{eq:NNDL-3-comm-FO-simplified}
reproduces angular-ordered result with no \NNDL correction.
Thus, the contribution corresponding to
Fig.~\ref{fig:nndl-diagram-3comm} vanishes to all orders.
Although the appearance of angular ordering seems to appear somewhat
fortuitously, one wonders if it extends beyond three
commensurate-angle emissions.

\subsection{Corrections involving the running of the strong 
coupling}\label{sec:nndl-rc}

The last family of \NNDL corrections we have to compute is the one
involving the running of the QCD coupling.
Since one-loop running is sufficient, we can follow a similar strategy as for
two hard-collinear emissions (Sec.~\ref{sec:nndl-squared-hc}) and insert an 
extra subleading correction in the \NDL calculation.

Naturally, such a calculation would make use of the \NDL multiplicity 
$N_{i,\text{rc}}^{(\NDL)}(L;\ell)$ of a parton of flavour $i$
initially created at a transverse momentum scale $e^{-\ell}Q$ down to a
transverse momentum cut $e^{-L}Q$.
However, since the QCD running coupling is not scale invariant, 
$N_{i,\text{rc}}^{(\NDL)}(L;\ell) \neq N_{i,\text{rc}}^{(\NDL)}(L-\ell)$ 
one cannot simply reuse Eq.~\eqref{eq:ndl-beta0} as we did for 
the hard-collinear \NDL corrections.
Instead, $N_{i,\text{rc}}^{(\NDL)}(L;\ell)$ has to be explicitly
computed. Following the same strategy as in Sec.~\ref{sec:running-coupling-NDL}, 
we get
\begin{align}
  \label{eq:NDL-RC-cascade}
  N_{i, \text{rc}}^{(\NDL)}(L;\ell)
  &= \frac{C_i}{C_A} \int_{\ell}^L \dd \l_1 
  \left[\delta(\ell_1)+n_g^{(\DL)}(\l_1;\ell)\right]\,
    \bar\alpha \int_{\l_1}^L \dd \l_2 (2\as\beta_0 \l_2)\,
    (\l_2-\l_1)\, N^{(\DL)}_g(L;\l_2), \nonumber \\
  & = \frac{C_i}{C_A}\frac{\beta_0}{2}\sqrt{\frac{\as\pi}{2C_A}}
    \left[(\nu^2-\nu_0^2-1)\sinh(\nu-\nu_0)+(\nu-\nu_0)\cosh(\nu-\nu_0)\right],
\end{align}
with $\nu_0=\sqrt{\abar} \ell$.

\subsubsection{Squared running-coupling correction}\label{sec:nndl-squared-rc}
We first consider the case of \NNDL corrections proportional to
$\beta_0^2$. If we expand the running coupling according to
\begin{equation}\label{eq:1loop-alphas-2ndorder}
\as(k_t) =
  \dis\frac{\as}{1-2\as\beta_0\ln\frac{Q}{k_t}}\approx
  \as + 2\as^2\beta_0\ln{(Q/k_t)} + 4\beta_0^2\as^3 \ln{(Q/k_t)}^2
  + \order{\as^4},
\end{equation}
we see that we can get \NNDL corrections either by having a single
$\mathcal{O}\left([\as\beta_0\ln(Q/k_t)]^2\right)$ correction to
one of the \DL emissions, $h_{3,\order{\beta_0^2}}$, or by having
$\mathcal{O}\left(\as\beta_0\ln(Q/k_t)\right)$ corrections to
two emissions along the chain of nested emissions,
$h_{3,\beta_0 \times \beta_0}$, Fig.~\ref{fig:nndl-diagram-b0sqr}.
The two cases can be written as
\begin{align}
  \label{eq:nndl-beta0-squared} 
  \as h_{3,\order{\beta_0^2}}^{(i)}
  & = \frac{C_i}{C_A}
    \int_0^L \dd \l_1 \left[\delta(\ell_1)+n_g^{(\DL)}(\l_1)\right]\, 
    \abar \int_{\l_1}^L \dd \l_2 (2\as\beta_0 \l_2)^2\, (\l_2-\l_1)\,
    N^{(\DL)}_g(L;\l_2),\\
  \as h_{3,\beta_0 \times \beta_0}^{(i)}
  &= \frac{C_i}{C_A}
    \int_0^L \dd \l_1 \left[\delta(\ell_1)+n_g^{(\DL)}(\l_1)\right]\,
    \abar \int_{\l_1}^L \dd \l_2 (2\as\beta_0 \l_2)\, (\l_2-\l_1)\,
     N^{(\NDL)}_{g, \, \text{rc}}(L; \l_2).
\end{align}
For the first of the two contributions, a \DL emission, which was
created at scale $k_{t,1} = Qe^{-\ell_1}$, dresses a gluon with
a $\beta_0^2$ correction down to the scale $k_{t,2}= Qe^{-\ell_2}$.
For the second contribution, the $\beta_0$ correction
occurs at the scale $\ell_2$ and a second \NDL correction happens at a lower
$k_t$ scale as encoded by $N^{(\NDL)}_{g, \, \text{rc}}(L; \l_2)$, given by Eq.~\eqref{eq:n-ndl-rc}.
Evaluating the integrals using the known \DL and \NDL multiplicities, we
get 
\begin{equation}\label{eq:NNDL-beta0-squared-resummed}
  h_{3, \beta_0^2}^{(i)}(\nu)  = 
  \frac{C_i}{C_A}
  \frac{\pi \beta_0^2}{16C_A}
  \left[
    3\nu(2\nu^2-1) \sinhnu + (\nu^4+3\nu^2) \coshnu
  \right].
\end{equation}
%

\subsubsection{Running-coupling and hard-collinear 
correction}\label{sec:nndl-rc-times-hc}
We now consider the case where one emission in the chain of nested
emissions correctly receives running coupling correction
$\order{\beta_0}$, while also allowing one hard-and-collinear splitting.
As in Sec.~\ref{sec:hard-collinear-NDL}, we split the discussion in
terms of whether the hard-collinear splitting is flavour-conserving or
flavour-changing.  
Accordingly, we write the \NNDL correction as
\begin{align}
  \label{eq:nndl-beta0-HC}
  h_{3,\beta_0 \times \text{hc}} =
  h_{3,\beta_0 \times\text{hc,fd}} + h_{3,\beta_0 \times 
  \text{hc,fc}}.
\end{align}
\paragraph{Flavour-diagonal.}
In the case of a flavour-diagonal hard-collinear splitting, we can
write
\begin{align} 
  \as h_{3,\beta_0 \times \text{hc,fd}}^{(i)}
  =\frac{C_i}{C_A} { \bigg [ }
  &\int_0^L \dd \l_1\left[\delta(\ell_1)B_i+ n_g^{(\DL)}(\l_1)B_{gg}\right]\, 
  \abar
    \int_{\l_1}^L \dd \l_2 \,(2\as\beta_0\l_2)\,N^{(\DL)}_{g}(L;\l_2)
    \nonumber \\
  +&\int_0^L \dd \l_1 \left[\delta(\ell_1)B_i+n_g^{(\DL)}(\l_1)B_{gg}\right]\, 
  \abar
  \int_{\l_1}^L \dd \l_2 \, N^{(\NDL)}_{g, \, \text{rc}}(L; \l_2) \nonumber \\
  +&\int_0^L \dd \l_1 n^{(\NDL)}_{\text{RC}}(\l_1)\, \abar
     \int_{\l_1}^L \dd \l_2 B_{gg}\, N^{(\DL)}_{g}(L;\l_2) \bigg].
     \label{eq:nndl-beta0-HC-FD} 
\end{align}
where $n^{(\NDL)}_{\text{RC}}(\l_1)$ is defined as the derivative of
Eq.~\eqref{eq:n-ndl-rc} for gluons with respect to $L$.
This expression covers the three physical cases: if the hard-collinear
splitting occurs at a scale $\ell_2$, the running-coupling correction
can happen either for the same emission (the first line), or for a
later emission (the second line), or for an earlier emission (the last
line).  For the first two lines, the hard-collinear emission can
either be from the primary branch, proportional to $\delta(\ell_1)$
with $B_i=B_q$ for a quark and $B_i=B_{gg}$ for a gluon, or from a
subsequent gluon emission.
\paragraph{Flavour-changing.}
The case where the hard-collinear splitting is a $g\to q\bar q$
branching is computed the same way with the usual care that after a 
real $g\to q\bar q$ branching both branches behave like quarks while
the virtual corrections involve gluon-initiated emissions.
Modulo this detail, we still have to include three contributions for
the running-coupling NDL correction happening earlier, at the same
emission, or later than the $g\to q\bar q$ branching:
\begin{align} 
  \as h_{3,\beta_0 \times \text{hc,fc}}^{(i)} 
  & = \frac{C_i}{C_A} B_{gq}{ \bigg\{ }
  \int_0^L \dd \l_1 n^{(\NDL)}_{\text{rc}}(\l_1)\, \abar
    \int_{\l_1}^L \dd \l_2 \, \left[2N^{(\DL)}_{q}(L;\l_2) - 
    N^{(\DL)}_{g}(L;\l_2)\right]  \nonumber \\
  &+ \int_0^L \dd \l_1 \left[\delta_{ig}\delta(\ell_1)+n_g^{(\DL)}(\l_1)\right]\, 
  \abar
  \int_{\l_1}^L \dd \l_2 \,(2\as\beta_0 \l_2)\, \left[2N^{(\DL)}_{q}(L;\l_2) - 
  N^{(\DL)}_{g}(L;\l_2)\right]
  \nonumber \\
  &+\int_0^L \dd \l_1 \left[\delta_{ig}\delta(\ell_1)+n_g^{(\DL)}(\l_1)\right]\, 
  \abar
  \int_{\l_1}^L \dd \l_2 \, \frac{2C_F - C_A}{C_A}N^{(\NDL)}_{g,\text{rc}}(L; 
  \l_2)
   \bigg\}\,,\label{eq:nndl-beta0-HC-FC}
\end{align}
where we have used that
$N^{(\NDL)}_{q,\text{rc}}=C_F N^{(\NDL)}_{g,\text{rc}}/C_A$
in the last line.

\paragraph{Integrated contribution.}
Evaluating the integrals in the flavour-diagonal and flavour-changing
contributions, we get 
\begin{subequations}\label{eq:nndl-beta0-HC-FC-result}
  \begin{align}
    h_{3,\beta_0 \times \text{hc}}^{(q)}
    & = \frac{C_F}{C_A} \frac{\beta_0}{4} \big\{
      (B_{gg}+\cdiff B_{gq})\nu^3\sinhnu
      +\left[2B_q-2B_{gg}+(6-8\cdiff)B_{gq}\right]\nu\sinhnu
      \nonumber\\
    &\phantom{=\frac{C_F}{C_A} \frac{\beta_0}{4}}
      +2 (B_q+B_{gg}+B_{gq})\nu^2\coshnu
      -4(1-\cdiff)B_{gq}(2\coshnu-2+\nu^2)\big\}\,,\\
    h_{3,\beta_0 \times \text{hc}}^{(g)}
    & = \frac{\beta_0}{4} \big\{
      (B_{gg}+\cdiff B_{gq})\nu^3\sinhnu
      +6(1-\cdiff)B_{gq} \nu\sinhnu\nonumber\\
    &\phantom{=\frac{\beta_0}{4}}
      +2\left[2B_{gg}+(1+\cdiff)B_{gq}\right]\nu^2\coshnu
      -8B_{gq}(1-\cdiff)(\coshnu-1)\big\}\,.
  \end{align}
\end{subequations}

\subsection{Final result}\label{sec:final-result}

For completeness, we gather here the full expression for the \NNDL
multiplicity, as obtained from Eq.~(\ref{eq:nndl-master}) with the
individual contributions computed in the previous subsections (see
also Table~\ref{table:nndl-coefficients}).
We give results for both $\avnlpZ$ and $\avnlpH$ corresponding
respectively to rest-frame $e^+e^- \to Z \to q\qbar$ and
$e^+e^- \to H \to gg$ collisions with centre-of-mass energy $Q$.
We define $\as = \as(Q)$, $L = \ln{(Q/\ktcut)}$, $\xi=\as L^2$, 
$\nu = \sqrt{{2C_A\as L^2}/{\pi}}$ and $\cdiff = (2C_F-C_A)/C_A$.
The equations are written in terms of coefficients summarised in
Table~\ref{table:multiplicity-coefficients}.

The generic all-order structure of the Lund multiplicity is
\begin{equation}\label{eq:final-result-NLP}
  \langle N^{\text {(Lund)}}_{Z,H}(L;\as)\rangle
  = 2\big[h_1^{(q,g)}(\xi) 
  + \sqrt{\as} h_2^{(q,g)}(\xi) 
  + \as h_3^{(q,g)}(\xi) \big].
\end{equation}
Using $\nu=\sqrt{2C_A\xi/\pi}$, the \DL contribution is 
\begin{subequations}
  \label{eq:final-result-h1}
  \begin{align}
    h_1^{(q)} &= 1 + \frac{C_F}{C_A}(\coshnu-1)\\
    h_1^{(g)} &= \coshnu,
  \end{align}
\end{subequations}
the \NDL function $h_2$ is given by
\begin{subequations}
  \label{eq:final-result-h2}
  \begin{align}
    h_2^{(q)}  = \frac{C_F}{\sqrt{2\pi C_A}}
    & \Big\{\frac{\pi\beta_0}{2C_A}
      \big[(\nu^2-1)\sinhnu + \nu\coshnu \big] + (B_{gg}+ \cdiff B_{gq} ) \nu\coshnu\\
    & + \big[2B_q- B_{gg} + \left(2-3\cdiff\right)B_{gq}\big]\sinhnu
      + 2(\cdiff-1)B_{gq} \nu\Big\},\nonumber\\
    h_2^{(g)} = \sqrt{\frac{C_A}{2\pi}}
    &\Big\{ \frac{\pi\beta_0}{2C_A}
      \big[(\nu^2-1)\sinhnu + \nu\coshnu \big] \\
    & + (B_{gg}+ \cdiff B_{gq}) \nu\coshnu 
      + \big[B_{gg} + (2-\cdiff)B_{gq}\Big] \sinhnu\Big\}.\nonumber
  \end{align}
\end{subequations}
The \DL~\eqref{eq:final-result-h1} and \NDL~\eqref{eq:final-result-h2}
accurate expressions also apply to any $k_t$-based jet multiplicity,
for example Durham multiplicity~\cite{Catani:1991pm}.
The main new result of this paper is the \NNDL function $h_3$ which is
found to be
\begin{align}\label{eq:final-result-h3q}
  2\pi h_3^{(q)}
  & = \textcolor{colourend}{D_{\text{end}}^{q\to qg}+ 
    \left(D_{\text{end}}^{g\to gg}+D_{\text{end}}^{g\to q\bar q}\right)
    \frac{C_F}{C_A}(\coshnu-1)}
    + \textcolor{colourhme}{D^{qqg}_{\text{hme}} \coshnu}\\
  & 
    + \textcolor{colourpair}{\frac{C_F}{C_A}
    \Big[
    (1-\cdiff) D_{\text{pair}}^{q\qbar}
    (\coshnu-1)
    +\left(K + D_{\text{pair}}^{gg} + \cdiff D_{\text{pair}}^{q\qbar}\right)
    \frac{\nu}{2}\sinhnu
    \Big]}\nonumber\\
  & + \textcolor{colourclust}{C_F
    \Big[\Big(
    \coshnu-1 -\frac{1-\cdiff}{4}\nu^2\Big)
    D_\text{clust}^\text{(prim)} + (\coshnu-1)
    D_\text{clust}^\text{(sec)}\Big]}\nonumber\\
  & 
  + \textcolor{coloureloss}{\frac{C_F}{C_A}
    \Big[
    D_{\text{e-loss}}^g \, \frac{\nu}{2} \sinhnu
    + \left(D_{\text{e-loss}}^q  -
    D_{\text{e-loss}}^g\right)(\coshnu - 1) \Big]}
    \nonumber\\
  & + \textcolor{colourndlsqr}{\frac{C_F}{2}
    \Big\{
    (B_{gg}+\cdiff B_{gq})^2\nu^2 \coshnu
    +8\left[2\cdiff B_{gg}- 2\cdiff B_q
    -(1-3\cdiff^2)B_{gq}\right]B_{gq}\coshnu}\nonumber\\
  & \phantom{+C_F}\textcolor{colourndlsqr}{
    +\left[4 B_q(B_{gg}+(2 \cdiff+1)B_{gq})-(B_{gg}+\cdiff
    B_{gq})(B_{gg}+9\cdiff B_{gq})\right]\nu\sinhnu
    }\nonumber\\
  & \phantom{+C_F}\textcolor{colourndlsqr}{
    +4(1-\cdiff^2)B_{gq}^2\nu^2+8\left[2\cdiff B_q
    -2\cdiff B_{gg}+(1-3\cdiff^2)B_{gq}\right]B_{gq}
    \Big\}}\nonumber\\
  & +\textcolor{colourndlsqr}{
    \frac{C_F}{C_A} \frac{\pi\beta_0}{2} \Big\{
    (B_{gg}+\cdiff B_{gq})\nu^3\sinhnu
    +\left[2B_q-2B_{gg}+(6-8\cdiff)B_{gq}\right]\nu\sinhnu}
    \nonumber\\
  & \phantom{+ C_F\pi\beta_0}
    \textcolor{colourndlsqr}{
    +2 (B_q+B_{gg}+B_{gq})\nu^2\coshnu
    -4(1-\cdiff)B_{gq}(2\coshnu-2+\nu^2)\Big\}}
    \nonumber\\    
  & + \textcolor{colourndlsqr}{\frac{C_F}{C_A}
    \frac{\pi^2 \beta_0^2}{8C_A}
    \big[
    3\nu(2\nu^2-1) \sinhnu+(\nu^4+3\nu^2) \coshnu  \big]}
    \nonumber
\end{align}
\begin{table}
  \renewcommand{\arraystretch}{1.5}
  
  \begin{center}
    \begin{tabular}{l c}
      \toprule
      Contribution
      & Coefficient\\
      \midrule
      Running coupling
      & $\beta_0=\frac{11C_A - 4n_fT_R}{12\pi}$ \\
      Hard-collinear correction
      & $B_q=-\frac{3}{4}$, $B_{gg}=-\frac{11}{12}$, $B_{gq}=\frac{n_fT_R}{3C_A}$ \\
      Collinear endpoint
      & $D_{\text{end}}^{q \to qg}    = C_F(3+3\ln 2-\frac{\pi^2}{3})$ \\
      & $D_{\text{end}}^{g \to gg}    = C_A(\frac{137}{36}+\frac{11}{3}\ln 2-\frac{\pi^2}{3})$ \\
      & $D_{\text{end}}^{g \to q\qbar} = n_fT_R(-\frac{29}{18}-\frac{4}{3}\ln 2)$ \\
      Hard matrix-element
      & $D_{\text{hme}}^{qqg}=C_F(\frac{\pi^2}{6}-\frac{7}{4})$,
        $D_{\text{hme}}^{ggg}=C_A(\frac{\pi^2}{6}-\frac{49}{36})$,
        $D_{\text{hme}}^{gq\bar q}=n_fT_R\frac{2}{9}$\\
      Commensurate $k_t$ and angle
      & $D_{\text{pair}}^{q\qbar}=\frac{13}{9}n_fT_R$, $D_{\text{pair}}^{gg}=(\frac{\pi^2}{6}-\frac{67}{18})C_A$\\
      & $K = \left(\frac{67}{18}-\frac{\pi^2}{6} \right)C_A - \frac{10}{9} n_f T_R$\\
      Clustering
      & $D_\text{clust}^\text{(prim)}=-\frac{5\pi^2}{54}$,
        $D_\text{clust}^\text{(sec)}=\frac{\pi^2}{27}$\\
      Energy loss
      & $D_{\text{e-loss}}^q=\frac{7}{2}C_A+(\frac{5}{2}-\frac{2\pi^2}{3})C_F$\\
      & $D_{\text{e-loss}}^g=(\frac{67}{9}-\frac{2\pi^2}{3})C_A-\frac{26}{9}\frac{C_F}{C_A}n_fT_R$\\
      \bottomrule
    \end{tabular}
  \end{center}
  \caption{Coefficients entering the \NNDL $h_3$ function.}
  \label{table:multiplicity-coefficients}
\end{table}
for quarks, and
\begin{align}\label{eq:final-result-h3g}
  2\pi h_3^{(g)}
  & = \textcolor{colourend}{
    \left(
    D_{\text{end}}^{g\to gg} + D_{\text{end}}^{g\to q\qbar}
    \right) \coshnu}
    + \textcolor{colourhme}{
    \left[
    D^{ggg}_{\text{hme}} \coshnu+ D^{gq\qbar}_{\text{hme}}
    \left(\cdiff \coshnu + 1 - \cdiff  \right)
    \right]}\\
  & + \textcolor{colourpair}{
    \left[
    (1-\cdiff) D_{\text{pair}}^{q\qbar} (\coshnu-1)
    +\left(K + D_{\text{pair}}^{gg} + \cdiff D_{\text{pair}}^{q\qbar}\right)
    \frac{\nu}{2}\sinhnu
    \right]} \nonumber\\
  & + \textcolor{colourclust}{C_A
    \left(D_\text{clust}^\text{(prim)}+ D_\text{clust}^\text{(sec)}\right)
    (\coshnu-1)}
    + \textcolor{coloureloss}{
    D_{\text{e-loss}}^g \, \frac{\nu}{2} \sinhnu}\nonumber\\
  & + \textcolor{colourndlsqr}{
    \frac{C_A}{2}
    \big\{
    (B_{gg}+\cdiff B_{gq})^2\nu^2 \coshnu
    -8(1-\cdiff^2)B_{gq}^2(\coshnu-1)}\nonumber\\
  & \phantom{+C_A}\textcolor{colourndlsqr}{
    +\left[(B_{gg}+\cdiff B_{gq})(3B_{gg}-5\cdiff
    B_{gq})+4(1+\cdiff)B_{gq}B_q\right]\nu\sinhnu\big\}}\nonumber\\
  & + \textcolor{colourndlsqr}{
    \frac{\pi\beta_0}{2} \big\{
      (B_{gg}+\cdiff B_{gq})\nu^3\sinhnu +6(1-\cdiff)B_{gq} \nu\sinhnu
    +2\left[2B_{gg}+(1+\cdiff)B_{gq}\right]\nu^2\coshnu}
    \nonumber\\    
  & \phantom{+\pi\beta_0}\textcolor{colourndlsqr}{
    -8B_{gq}(1-\cdiff)(\coshnu-1)\big\}}
    + \textcolor{colourndlsqr}{\frac{\pi^2 \beta_0^2}{8C_A}
    \left[
    3\nu(2\nu^2-1) \sinhnu + (\nu^4+3\nu^2) \coshnu
    \right]}\nonumber
\end{align}
for gluons, with different contributions separated in different
colours, except for those involving ingredients already present
at \NDL accuracy which have been gathered (in orange).

\begin{figure}
  \begin{center}
    \includegraphics[scale=0.65,page=1]{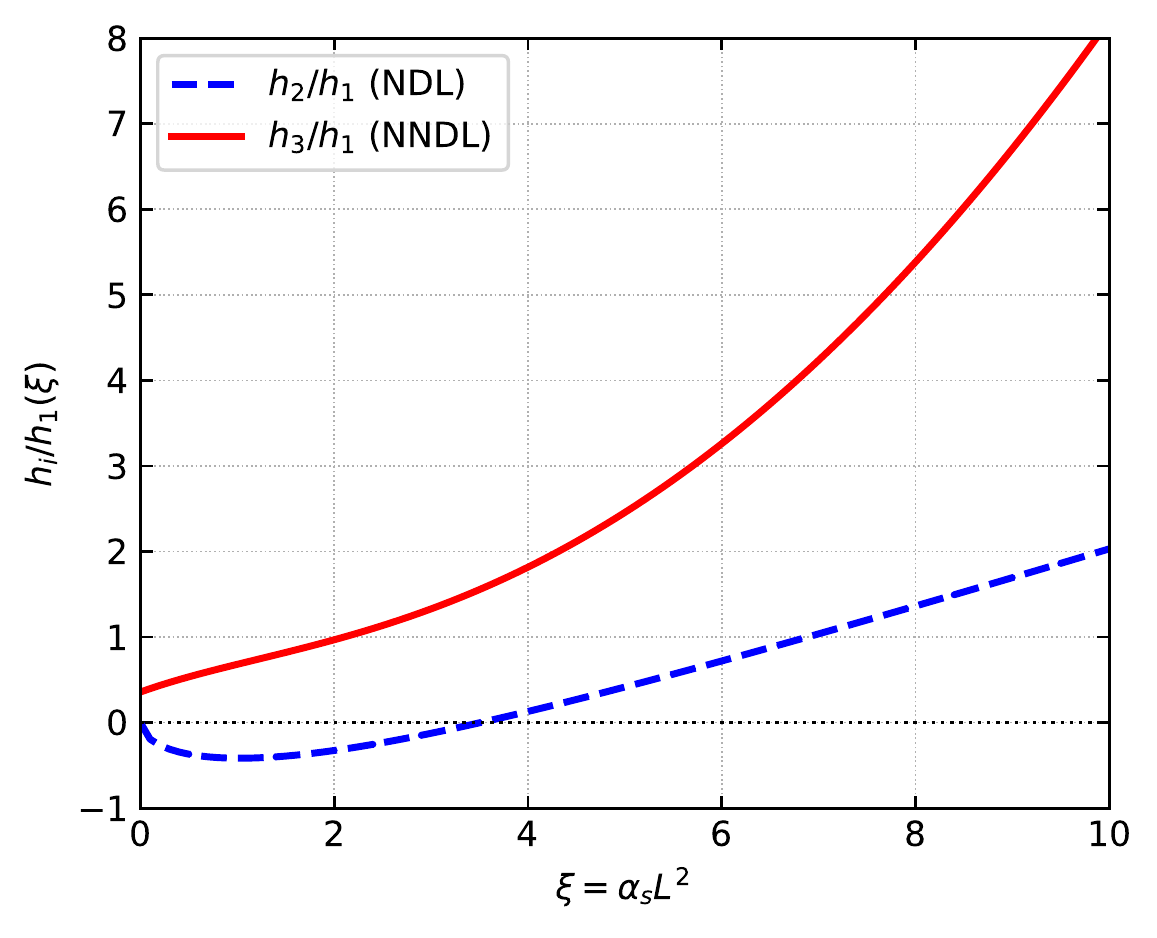}%
    \includegraphics[scale=0.65,page=2]{figs/plot-nndl-bits.pdf}%
    \\
    \includegraphics[scale=0.65,page=3]{figs/plot-nndl-bits.pdf}%
    \includegraphics[scale=0.65,page=4]{figs/plot-nndl-bits.pdf}%
    \caption{Top: 
      Ratio of \NDL (blue) and \NNDL (red) resummation functions to 
      the \DL result as a function of $\as L^2$ for quark
      (left) and gluon hemispheres (right).
      Bottom:
      Relative weight of the \NNDL contributions to $h_3$ for
      quark (left) and gluon hemispheres (right).}
    \label{fig:nndl-bits}
  \end{center}
\end{figure}
At \NNDL accuracy, the Cambridge multiplicity is simply
obtained from Eq.~\eqref{eq:final-result-NLP}:
\begin{equation}\label{eq:final-nndl-cam}
 \langle N^{\text{(Cam)}}(\as, L)\rangle = \langle N^{\text{(Lund)}}(\as, L)\rangle  
+ \frac{\as C_i}{2\pi} \frac{\pi^2}{6}.
\end{equation}
To gain insight into the relative sizes of the \NNDL contributions,
the top panels of Fig.~\ref{fig:nndl-bits} show the ratios
$h_{2,3}/h_1$ as a function of $\xi = \as L^2$ for quark (left) and
gluon (right) hemispheres.
We note that the \NDL corrections are negative at low values
of $\xi$, driven by hard-collinear corrections, while the \NNDL result
remains positive for all $\xi$. 
Unless we go to large values of $\xi$, the \NNDL function $h_3$ is
roughly of the same magnitude as the \NDL function $h_2$.
That said, for physical values of $\as\sim 0.1$, the size of the \NNDL
corrections, $\as h_3(\xi)$, would
be commensurate to the \NDL contribution $\sqrt{\as} h_2(\xi)$,
motivating the study of their phenomenological
impact, which we carry out in Sec.~\ref{sec:matching}.

In the bottom row of Fig.~\ref{fig:nndl-bits}, we investigate the
relative sizes of the \NNDL contributions to the total $h_3(\xi)$ for
quarks~(left) and gluons~(right), with line colours matching those
in Eqs.~(\ref{eq:final-result-h3q}) and~(\ref{eq:final-result-h3g}).
For conciseness, we have grouped the corrections stemming only from
\NDL hard-collinear or running coupling corrections into a unique
$(\NDL)^2$ contribution, i.e.\
$h_{3, \NDL^2} \equiv h_{3, \text{hc}^2} + h_{3, \beta_0 \times
  \text{hc}} + h_{3, \beta_0^2} $ (Sects.~\ref{sec:nndl-squared-hc}
and~\ref{sec:nndl-rc}).
We note that there are particularly large cancellations between the
 $h_{3,\beta_0 \times \text{hc}}$ and $h_{3, \beta_0^2}$ functions,
 further motivating their grouping. 

We observe that these $(\NDL)^2$ corrections become dominant  
when $\xi\gtrsim 3$ for both the quark and  gluon cases.
This can be traced to the running-coupling corrections which, relative
to $h_1$, scale like $\xi^2$ ($\xi$ for the \NDL contribution).
More generally, one should expect that pure running-coupling
corrections would bring contributions of order $\xi^{k-1}$ to
$h_k/h_1$, so that for $\sqrt{\xi}(\as L) =\sqrt{\as}\xi \sim 1$, the
N$^k$DL approach of expanding in series of $\sqrt{\as}$ at fixed
$\xi=\as L^2$, Eq.~(\ref{eq:log-counting}), should be revisited so as
to resum an arbitrary number of running-coupling
corrections.\footnote{In this case, mixed running-coupling and
  hard-collinear corrections are expected to be numerically
  significant.}
This is actually reminiscent of the LL, NLL, NNLL, ... logarithmic
counting where running-coupling corrections are included already in the
LL Sudakov factor~\cite{Banfi:2004yd}.
It is also interesting to recall that, while the original result from
Ref.~\cite{Catani:1991pm} is strictly speaking \NDL, the use of a
generating functional potentially includes an arbitrary number of
running-coupling corrections, partly resumming them in the above sense
to all orders.
In this context we have checked explicitly that parts of the
$h_{3, \NDL^2}$ function are already included in the expansion of the
result from~\cite{Catani:1991pm} to \NNDL order, in particular terms
which dominate at large values of $\xi$ (i.e.\ the terms proportional
to $\xi^2$, $\xi^{3/2}$ and $\xi$, multiplied by $\coshnu$
  or $\sinhnu$).
Further investigation on how to include this in a perturbatively
controlled way in our formalism is left for future work.

For phenomenologically-accessible values of $\xi$, $\xi\lesssim 5$, we
observe in Fig.~\ref{fig:nndl-bits} that
the contributions from the collinear endpoint
and energy loss are in the 20{-}40\% range\footnote{Or even more at
  small $\xi$, although this region would in practice be affected by
  fixed-order corrections.} for the quark case.
For gluons we observe a larger sensitivity, $\sim 10\%$, to other
contributions, like hard matrix-element, clustering and soft pairs.

The relative weight of each contribution to the \NNDL multiplicity is
an important piece of information to gauge the constraining power of
Lund multiplicity on parton shower validation, as it probes the full
multiple branching structure of QCD.
This was explicitly demonstrated in Ref.~\cite{Dasgupta:2020fwr} (and
Ref.~\cite{Hamilton:2020rcu} at full colour) where the NLL PanScales
family of showers were shown to successfully reproduce the \NDL
jet multiplicity analytic resummation~\cite{Catani:1991pm}, among
other observables.
Thus, the \NNDL analytic result presented in this paper would be a key
element towards testing a future generation of NNLL showers.
Since, as discussed above, some of the physical contributions to $h_3$
can be relatively small, an excellent control of the numerical
precision in the shower would be required.

\section{Validation and phenomenology}\label{sec:numerics}

In this section we first cross-check the fixed-order expansion of our
resummed result against the {\tt Event2}
generator~\cite{Catani:1996jh,Catani:1996vz}. We then discuss the
phenomenological impact of the \NNDL resummation at LEP energies,
including matching our resummed result to exact NLO from {\tt Event2}.
We finish this section with a brief comparison to existing OPAL data,
taking into account non-perturbative corrections obtained by means of
Monte Carlo simulations.

\subsection{Fixed-order validation}\label{sec:event2}

\begin{figure}
  \begin{subfigure}[t]{0.48\linewidth}
    \includegraphics[height=7cm,page=1]{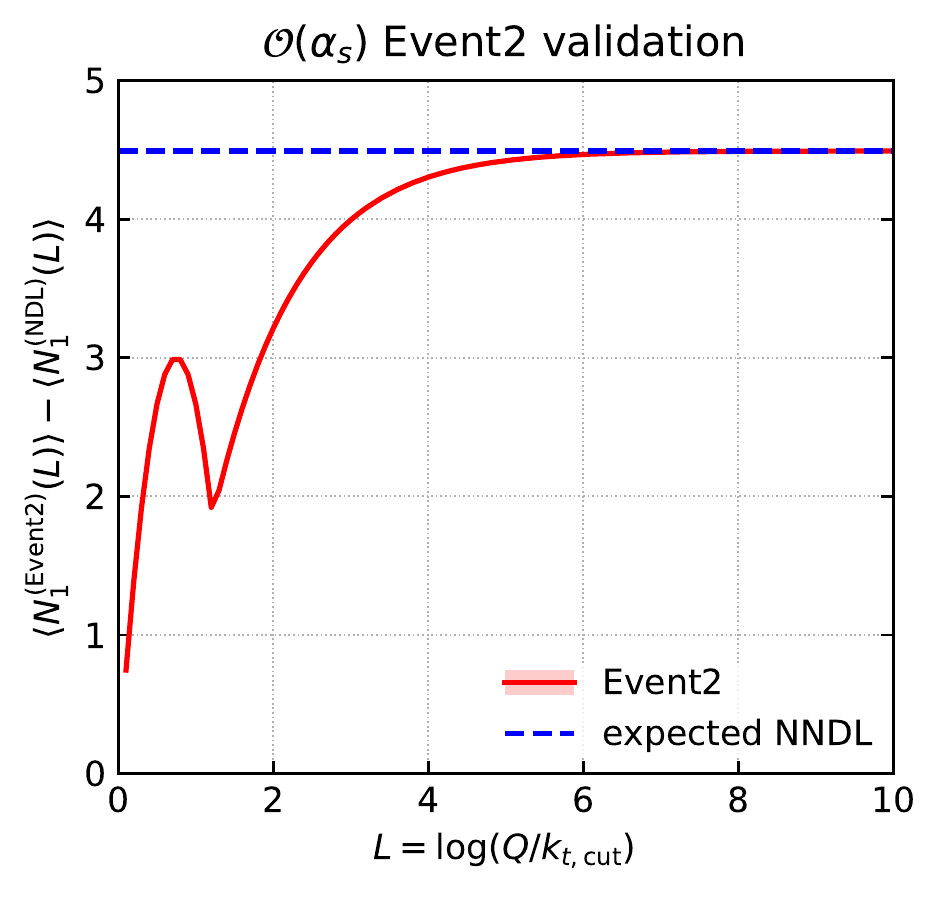}
    \caption{$\langle N\rangle$ test at $\order{\as}$}\label{fig:v-event2-order1}
  \end{subfigure}
  \begin{subfigure}[t]{0.48\linewidth}
    \includegraphics[height=7cm,page=2]{figs/NNDL-v-event2.pdf}
    \caption{$\langle n\rangle$ test at $\order{\as^2}$}\label{fig:v-event2-order2}
  \end{subfigure}
  \caption{Comparison between our \NNDL calculation expanded in series
    of $\as$ and exact fixed-order results obtained numerically
    with {\tt Event2}:
    (a) the multiplicity difference
    $\langle N^\text{\tt (Event2)}_1\rangle-\langle N^\text{(NDL)}_1\rangle$ at order 
    $\as$, expected to tend to the NNDL constant represented by the dashed
    (blue) line, cf.\ Eq.~\eqref{eq:nndl-fo-expansion-order1};
    (b) the differential multiplicity difference
    $\langle n^\text{\tt (Event2)}_2\rangle-\langle n^\text{(NNDL)}_2\rangle$
    at order $\as^2$ which is expected to go to a (N$^3$DL)
    constant if the results agree at NNDL, cf.\
    Eq.~\eqref{eq:nndl-fo-expansion-order2}.  }
  \label{fig:v-event2}
\end{figure}

The {\tt Event2} generator provides exact results for the multiplicity
in $e^+e^-\to Z\to q\bar q$ Born-level events up to $\order{\as^2}$:
\begin{equation}
  \langle\nlp(\as,L)\rangle
  = 2
  + \frac{\as}{2\pi}\langle\nlp_1(L)\rangle
  + \left(\frac{\as}{2\pi}\right)^2\langle\nlp_2(L)\rangle
  + \mathcal{O}(\as^3).
\end{equation}
The fixed-order coefficients have the following dependence on $L$:
\begin{subequations}\label{eq:fo-log-expansion}
  \begin{align}
    \frac{1}{2\pi}\langle\nlp_1(L)\rangle
    & = h_{11}L^2 + h_{21}L + h_{31} + \mathcal{O}(e^{-L}),
      \label{eq:nndl-fo-expansion-order1}\\
    \frac{1}{(2\pi)^2}\langle\nlp_2(L)\rangle
    & = h_{12}L^4 + h_{22}L^3 + h_{32}L^2 + h_{42}L + h_{52} + 
    \mathcal{O}(e^{-L}),
      \label{eq:nndl-fo-expansion-order2}
  \end{align}
\end{subequations}
where the  $h_{ij}$ coefficients correspond to the $j$-th order $\as$ expansion of the 
$h_i$ resummed expressions, e.g.\ $h_{11}$ is the first-order expansion of the 
\DL $h_1$ function, $h_{32}$ is the second-order expansion of the \NNDL $h_3$ 
function and so on. 
The extraction of the $\as$ and $\as^2L^2$ contributions
from {\tt Event2} therefore provides a validation of our \NNDL resummed
calculation.
More specifically, from the results in Sec.~\ref{sec:nndl}, we find
\begin{subequations}
  \begin{align}
    h_{31}
    & = \frac{1}{\pi}\left(D_{\text{hme}}^{qqg}+D_{\text{end}}^{q\to 
    qg}\right),
    \label{eq:nndl-coef-order1}\\
    h_{32}
    & = \frac{C_A}{\pi^2} D_{\text{hme}}^{qqg}
      + \frac{C_F}{\pi^2} \left(D_{\text{end}}^{g\to gg} + D_{\text{end}}^{g\to q\qbar}+ K + 
      D^{gg}_\text{pair} +
      D^{q\bar q}_\text{pair} + D_{\text{e-loss}}^{q} + C_F 
      D_\text{clust}^{\text{(prim)}} + C_A 
      D_\text{clust}^{\text{(sec)}}\right)
      \nonumber \\
     & +\frac{4 C_F C_A}{\pi^2}  B_q \left(B_{gg}+B_{gq}\right)
       + \frac{4 C_F}{\pi} B_q \beta_0 .
       \label{eq:nndl-coef-order2}
  \end{align}
\end{subequations}
In order to check that these coefficients are indeed correct, we have
performed two tests, shown in Fig.~\ref{fig:v-event2}.
First, focusing on the first order in $\as$, we subtract from the
{\tt Event2} multiplicity $\langle N^{\tt (Event 2)}_1\rangle$ the \NDL
expectation, $\langle N^{(\NDL)}_1\rangle$. This is plotted as
the solid (red) line in Fig.~\ref{fig:v-event2-order1}.
Our \NNDL expectation is that it should tend asymptotically to $2\pi
h_{31}$, plotted as the dashed (blue) line in
Fig.~\ref{fig:v-event2-order1}.
The solid curve indeed matches the dashed one at large $L$ as
expected.

Next, focusing on the second order, $\order{\as^2}$, we subtract from
the {\em differential} {\tt Event2} multiplicity distribution,
$\langle n^{\tt (Event 2)}_2\rangle$, the \NNDL expectation,
$\langle n^{(\NNDL)}_2\rangle$, obtained through a simple
derivative of our \NNDL result.
This is plotted in Fig.~\ref{fig:v-event2-order2} for both the full
multiplicity and for the contributions of each colour channel. 
The expectation is that the \NNDL-subtracted result tends to a constant
(associated with $h_{42}$ in Eq.~\eqref{eq:nndl-fo-expansion-order2})
at large $L$, which it does.
Similar tests, reported in Appendix~\ref{app:cambridge-vs-lund}, have
been performed for the Cambridge multiplicity.

These two tests show that our \NNDL calculation is correct, at least up
to order $\as^2$ for $e^+e^-\to Z\to q\bar q$ events.
If the tools become available, it would be interesting to extend these
tests to $e^+e^-\to H\to gg$ events, as well as one order higher in
$\as$.
This would for example allow us to probe the \NNDL contributions
starting at $\order{\as^2}$ beyond their first non-trivial
contribution, as well as probing the \NNDL coefficients which start
only at $\order{\as^3}$, such as the $\beta_0^2$ correction or the
vanishing of the contribution from three emissions at commensurate
angles.

Note that we have performed additional checks beyond the comparison
with {\tt Event2}. The fixed-order expansion up to $\order{\as^2}$ of
$h_{3,\text{end}}$ (Sec.~\ref{sec:nndl-end}), $h_{3, \text{e-loss}}$
(Sec.~\ref{sec:nndl-eloss}) and $h_{3, \text{hc}^2}$
(Sec.~\ref{sec:nndl-squared-hc}) has been checked against the {\tt
  MicroJet} results~\cite{Dasgupta:2014yra}. Finally, the large-angle
components of the calculation, see
Secs.~\ref{sec:nndl-large-angle-first} and \ref{sec:nndl-clustering},
have been cross-checked against the $\order{\as^2}$ expansion of a toy
dipole shower described in Ref.~\cite{Lifson:2020gua}.

\subsection{Matching with fixed-order}\label{sec:matching}

In order to produce reliable phenomenological predictions of the Lund
multiplicity in $e^+e^- \to Z$ events for all values of
$L=\ln(Q/\ktcut)$, the resummed distribution obtained in
Sec.~\ref{sec:nndl}, relevant at large $L$, needs to be matched with a
fixed-order calculation, relevant at small $L$.
For this observable, we use an additive matching scheme:
\begin{equation}
\label{eq:matching}
\avnlp_\text{match} = \avnlp_\text{fo} + \avnlp_\text{resum} - 
\avnlp_\text{resum,fo},
\end{equation} 
where $\avnlp_\text{fo}$ is the exact fixed-order result obtained with
{\tt Event2} as described in Sec.~\ref{sec:event2}, taken here at NLO
i.e.\ at order $\as^2$, $\avnlp_\text{resum}$ is the \NNDL-accurate
resummed result, Eq.~\eqref{eq:final-result-NLP}, and
$\avnlp_\text{resum,fo}$ avoids double counting by subtracting the
$\order{\as^2}$ expansion of the resummed result
$\avnlp_\text{resum}$.
At our accuracy, the fixed-order distribution can be normalised to the
NLO inclusive cross-section,
$\sigma_0+\sigma_1 = \sigma_0(1+\as/\pi)$.
The scale uncertainty is probed by varying the renormalisation scale
and the resummation scale.
For this, we introduce two dimensionless parameters, $x_R$ and $x_L$.
The renormalisation scale is taken as $\mu_R=x_R Q$, both in the
resummation and in the fixed-order results, and the resummation scale
is varied by redefining  $L$ as $\ln (x_LQ/k_t)$.
When varying $x_R$ and $x_L$, one must introduce counter-terms to
preserve the perturbative accuracy, both at fixed order and for the
resummed result.
For the resummed result, we get 
\begin{align}\label{eq:expressions-xR-xL}
  \avnlp_\text{resum}
  & = h_1(\xi)
  + \sqrt{\as} \big[h_2(\xi) - 2 \ln x_L \sqrt{\xi}\, h_1^\prime(\xi)\big]\\
  &+ \as \big[h_3(\xi)
        + (2 \beta_0 \ln x_R\, \xi + \ln^2 x_L) h_1^\prime(\xi)
        + 2 \ln^2 x_L\, \xi\, h_1^{\prime\prime}(\xi)
        - 2  \ln x_L \sqrt{\xi}\, h_2^\prime(\xi)\big],\nonumber
\end{align}
with $h_i$ defined in Sec.~\ref{sec:final-result}, $\as\equiv\as(\mu_R=x_RQ)$,
$\xi =\as L^2=\as \ln^2 (x_LQ/k_t)$, and $h_i^\prime$,
$h_i^{\prime\prime}$  the first and second derivatives of $h_i$
with respect to $\xi$.
In practice, we estimate the scale uncertainty by separately varying
either $x_L$ or $x_R$ between 1/2 and 2, taking the envelopes
of the variations, and summing in quadrature the renormalisation and resummation
uncertainties.

Finally, to account for the fact that the endpoints of the
resummed and fixed-order distributions do not coincide, we redefine
the $L$ in the argument of $\avnlp_\text{resum}$ according
to~\cite{Catani:1992ua}
\begin{equation}
  L = \ln \frac{Q}{\ktcut}
  \to
  L = \ln\left(\frac{x_L Q}{\ktcut} - \frac{x_L Q}{k_{t,\text{max}}}+1\right),
\end{equation}
with $k_{t,\text{max}}$ the smallest $k_t$ value at which the
fixed-order distribution is equal to its Born-level value, i.e.\
2.\footnote{At $\order{\as}$, we find
  $k_{t,\text{max}}^2=\tfrac{5\sqrt{5}-11}{2}Q^2\approx 0.0902\,Q^2$
  ($\ln Q/k_{t,\text{max}}\approx 1.203$). At $\order{\as^2}$ we determine
  $k_{t,\text{max}}$ numerically from the {\tt Event2}
  distribution, finding $k_{t,\text{max}}^2\approx 0.1062\,Q^2$, i.e.\
  $\ln Q/k_{t,\text{max}}\approx 1.121$.}
At large $L$, this replacement only generates power corrections in
$\ktcut$ and so does not spoil the logarithmic accuracy.

\begin{figure}
  \begin{subfigure}[t]{0.48\linewidth}
    \includegraphics[scale=0.73,page=1]{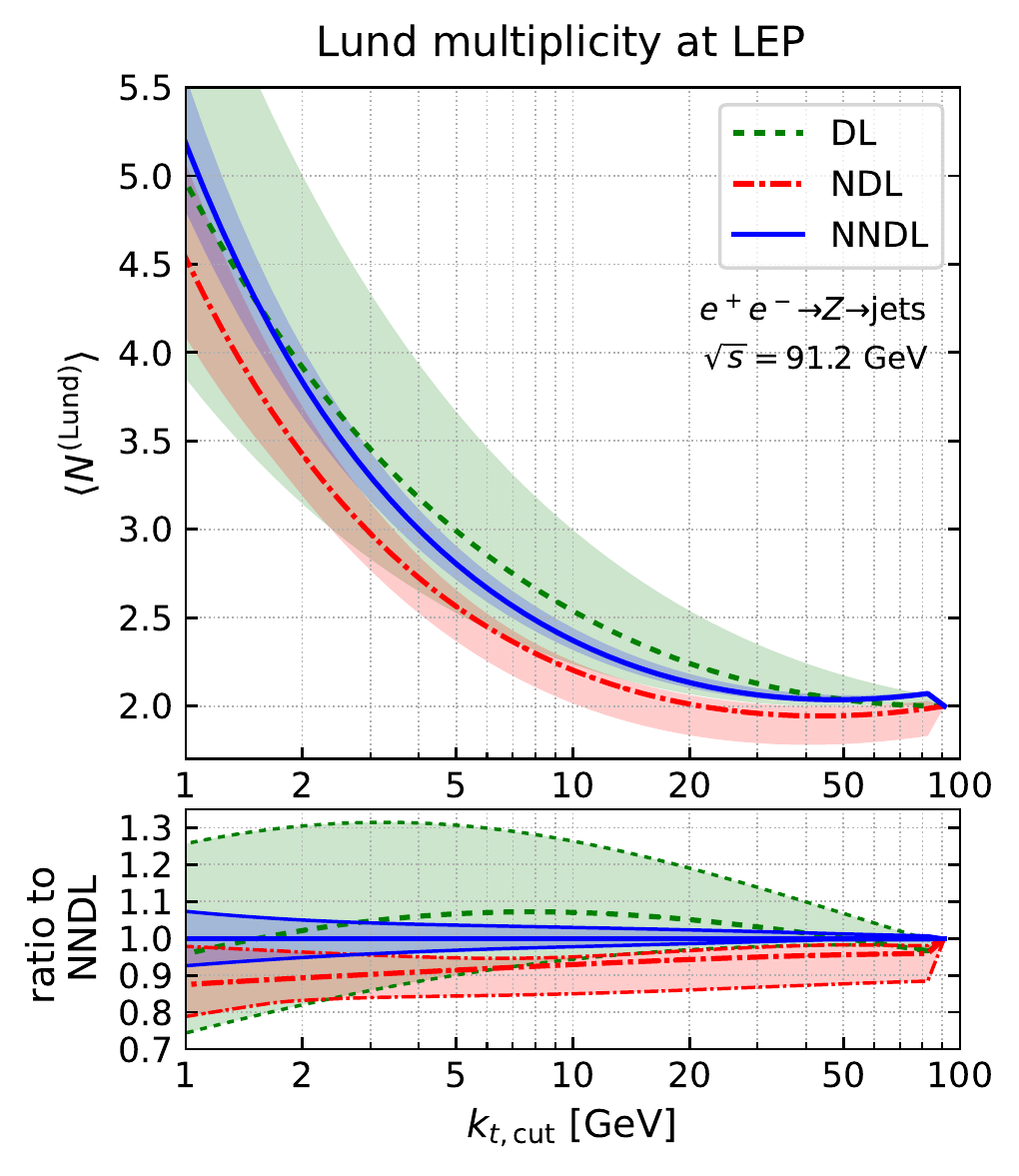}%
    \caption{}\label{fig:lep-plot-resum}
  \end{subfigure}
  \hfill%
  \begin{subfigure}[t]{0.48\linewidth}
    \includegraphics[scale=0.73,page=2]{figs/plot-match-paper.pdf}%
    \caption{}\label{fig:lep-plot-match}
  \end{subfigure}
  \caption{Left: Resummed Lund multiplicity at \DL (dashed, green),
    \NDL (dash-dotted, red) and \NNDL (solid, blue) as a function of
    $\ktcut$ for LEP kinematics.
    Right: Lund multiplicity after matching with NLO. The pure NLO
    result is shown in dotted (grey) for reference, together with
    matched NLO+\NDL (dash-dotted, red) and NLO+\NNDL (solid, blue).
    The bottom panels display the ratio to the most
    accurate result (\NNDL on the left and NLO+\NNDL on the right).}
  \label{fig:lep-plot}
\end{figure}

Fig.~\ref{fig:lep-plot} shows the Lund multiplicity, together with its
uncertainty,  as a function of $\ktcut$ for LEP $e^+e^-\to Z$ events at
$\sqrt{s}=91.2$~GeV.
A common feature to all curves is that the
Lund multiplicity increases significantly when $\ktcut$ decreases.
In Fig.~\ref{fig:lep-plot-resum}, we focus on the resummed results at
different degrees of logarithmic accuracy. We find that while \NDL
corrections reduce the multiplicity, \NNDL corrections cause a
non-negligible increase, as one would expect from
Fig.~\ref{fig:nndl-bits}.
Furthermore, increasing the logarithmic accuracy significantly reduces
the perturbative uncertainty. Working for example at $\ktcut=5$~GeV,
The \DL uncertainty of $\sim 20$\% is reduced to $5.6$\% at \NDL and
to $\sim 3.4$\% at \NNDL. For $\ktcut=1$~GeV, these numbers are around
27\%, 11\% and 7\%, showing a similar reduction from \NDL (the current
state-of-the-art) to \NNDL.
The \NNDL uncertainty band is also more symmetric than the \NDL one.
It is also interesting to notice that the \NDL and \NNDL uncertainty
bands only barely overlap.

Fig.~\ref{fig:lep-plot-match} shows the effect of matching with NLO
($\order{\as^2}$) fixed-order results.
We first see that the effect of the resummation is to increase the
average Lund multiplicity, even further so with \NNDL matching than
with \NDL matching, showing the necessity for the resummation at small
$\ktcut$.
Besides the sheer increase in multiplicity, NLO+\NNDL matched
results also bring a significant reduction in scale uncertainty
compared to what is observed at NLO+\NDL. This reduction is of about
50\% at almost all $\ktcut$ values, e.g.\ going from $\sim 9$\%
($2$\%) to $\sim 6$\% ($1.3$\%) at $\ktcut=5$~GeV (1~GeV).

\subsection{Comparison to LEP data}\label{sec:v-OPAL}

We conclude this section with a short comparison between our analytic
results and the existing measurement~\cite{JADE:1999zar} done by the
OPAL collaboration at LEP. Here, we focus solely on the results for
$\sqrt{s}=91.2$~GeV.

\begin{figure}
  \begin{subfigure}{0.48\textwidth}
    \centering
    \includegraphics[page=2,scale=0.75]{figs/plots-OPAL.pdf}
    \caption{Comparison of the data to standard Monte Carlo generators
      at parton (dashed) and hadron (solid) levels.}\label{fig:v-OPAL-mc}
  \end{subfigure}
  \hfill
  \begin{subfigure}{0.48\textwidth}
    \centering
    \includegraphics[page=1,scale=0.75]{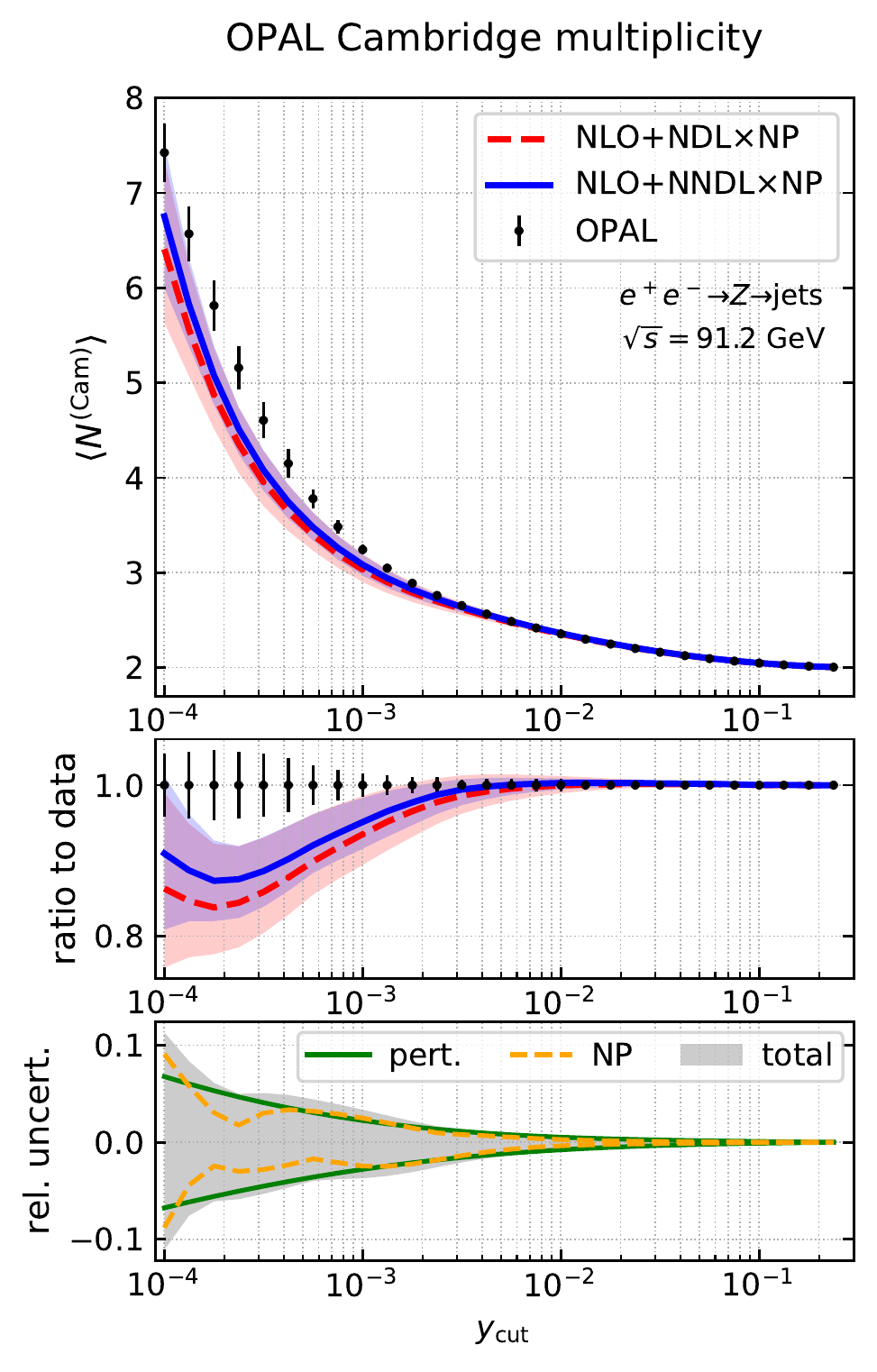}
    \caption{Comparison of the data to \NDL (red) and \NNDL (blue)
      resummed results matched to NLO and corrected for
      non-perturbative effects.}\label{fig:v-OPAL-nndl}
  \end{subfigure}
  \caption{Comparison of Monte Carlo (left) and analytic (right)
    results to the existing average Cambridge jet multiplicity
    measurement~\cite{JADE:1999zar} from the OPAL collaboration at
    LEP.
  }\label{fig:v-OPAL}
\end{figure}

From the viewpoint of perturbative QCD, the same approach we just
discussed for the Lund multiplicity can be applied to the Cambridge
multiplicity: our \NNDL resummed result,
Eq.~\eqref{eq:final-nndl-cam}, can be matched to {\tt Event2} exact
NLO multiplicities using additive matching, Eq.~(\ref{eq:matching}).
For a phenomenological comparison to the LEP data, we need to
supplement our perturbative result by non-perturbative corrections.
In this brief study, we have extracted these from standard
general-purpose Monte Carlo generators. Namely, we have run {\tt
  Pythia}(v8.306)~\cite{Bierlich:2022pfr}, {\tt
  Herwig}(v7.2.0)~\cite{Bahr:2008pv,Bellm:2015jjp}, and {\tt
  Sherpa}(v2.2.11)~\cite{Gerwick:2014gya,Sherpa:2019gpd}
each at parton level and hadron level separately.
This is plotted on the top panel of Fig.~\ref{fig:v-OPAL-mc} showing
that hadron-level predictions for the three generators do reproduce
the data, although their respective parton-level distributions are
quite different. 
Hadronisation corrections (and uncertainties) are then taken as the
average (and envelope) of the three ratios of hadron- to parton-level
distributions, cf.\ the lower panel of Fig.~\ref{fig:v-OPAL-mc}.

The non-perturbative corrections are applied as a multiplicative
factor to the matched perturbative results. The uncertainty is
obtained by adding the perturbative and non-perturbative uncertainties
in quadrature.
The resulting distributions are shown in Fig.~\ref{fig:v-OPAL-nndl}
for the \NDL and \NNDL resummations.
Globally speaking, we see a decent degree of agreement with the OPAL
data, although the predictions are systematically below the data for
$y_\text{cut}\lesssim 10^{-3}$, i.e.\ for $\ktcut\lesssim 3$~GeV.
Including the \NNDL corrections shows both an increase in
multiplicity, bringing the predictions closer to the data, and a
reduction in uncertainty.
This agrees with the observations made in the previous subsection.
We note as well that for the matched \NNDL results, the perturbative
and non-perturbative uncertainties are of the same order for
$y_\text{cut}\lesssim 0.002$ with the perturbative uncertainties
marginally dominating at larger values.

As a side comment, we observe that non-perturbative corrections become
larger than 10\% for $y_{\rm{cut}} \lesssim 10^{-4}$, corresponding to
$\ktcut\lesssim 1.5$~GeV, in quantitative agreement with Fig.~15 of
the original Cambridge paper~\cite{Dokshitzer:1997in}.
We also note that the uncertainty on the non-perturbative corrections
also grows significantly below $\ktcut\sim 1.5$~GeV.

From a phenomenological viewpoint, the significant reduction of the
perturbative uncertainties at \NNDL compared to \NDL, seen both in
this section and in the previous one, is the most important
phenomenological result of this paper.
It suggests that it would be interesting to perform a more extensive
comparison to the OPAL data.
Such a study could include matching to exact NNLO results, as well as
an improved resummation of the running-coupling corrections (see the
discussion at the end of Sec.~\ref{sec:final-result}).
These would hopefully further reduce the perturbative uncertainty and
produce a better agreement with the data at low values of $\ktcut$.
This could then potentially serve as a way to make an additional
extraction of the strong coupling constant $\as$.
It would also be of interest to measure the average Lund plane
multiplicity as a function of $\ktcut$ --- or the difference between
the Lund and Cambridge multiplicities ---, especially in the context
of the renewed interest in the community for the analysis of archived
ALEPH data to study new observables (see e.g.~\cite{Chen:2021uws}) or
phenomena (see e.g.~\cite{Badea:2019vey}).
From a theoretical viewpoint, one could expect that higher-order
running-coupling corrections would also increase the multiplicity at
small $y_\text{cut}$, reemphasising the interest already stated in
Sec.~\ref{sec:final-result}.

\section{A byproduct: multiplicity in hadron collisions at \NDL accuracy}\label{sec:ndl-pp}

In order to demonstrate the flexibility of the resummation strategy
that we introduced in Sec.~\ref{sec:recap-ndl}, we proceed to
extend the Lund multiplicity calculation to hadronic collisions or, 
equivalently, including initial-state radiation, up to \NDL accuracy
(see Sec.~\ref{sec:lund-mult-def} for how the definition of the
multiplicity is adapted to hadronic collisions).

We consider colour singlet production, i.e.\ Drell-Yan
($pp \to q\qbar \to Z$) and Higgs production via gluon fusion
($pp \to gg \to H$).
The colourless boson $X=Z,H$ is produced at a virtuality $Q^2=M^2_X$
and rapidity $y_{_X}=\tfrac{1}{2}\ln\tfrac{x_1}{x_2}$ from a pair of incoming partons $a,b$ with fixed
flavour and fixed longitudinal momentum fractions $x_1, x_2$. 
Since we are interested in the event-averaged Lund multiplicity,
for fixed $Q^2$ and $y_{_X}$, we normalise
the Lund multiplicity to the differential cross section for 
colour singlet production which schematically reads
\begin{align}
  \label{eq:colour-singlet-normalisation}
  \frac{\dd^2 \sigma_{pp \to X}}{\dd Q^2\dd y_{_X}}
  =
  \frac{\dd^2 \hat \sigma_{ab \to X}}{\dd Q^2 \dd y_{_X}}
  \times f_a(x_1, Q^2) f_{b}(x_2, Q^2)\, ,
\end{align}
where the differential cross section in
Eq.~\eqref{eq:colour-singlet-normalisation} is written in terms of a
hard partonic cross section $\dd^2 \hat \sigma_{ab \to X}$ and the
parton distribution functions (PDFs) $f(x_i, Q^2)$.

We note that as long as $\as(Q) \ln^2{(1-x_i)} \ll 1$, 
ensuring that we are sufficiently far from the quasi-elastic 
regime~\cite{Catani:1992rm}, both partons may be viewed, 
up to \NDL accuracy, as evolving independently from one 
another. We therefore analyse the multiplicity contribution of 
only one of the legs, akin to the procedure of Sec.~\ref{sec:recap-dl-ndl}.
We also note that the leading-order value of the multiplicity is
zero in this process, i.e.\ we do not count the colour singlet.

\paragraph{Differences with final-state multiplicity.}
In a colour singlet process we have to consider (primary) initial-state 
radiation from the incoming leg and their subsequent final-state evolution.
A first natural question is whether the jet radius, introduced in a
somewhat arbitrary way in the reconstruction of the Lund multiplicity
in the $pp$ case, affects the multiplicity calculation, preventing us
from recycling the $e^+e^-$ calculation.
In principle, radiation from the incoming parton with $\theta\sim R$
could be mistagged due to clustering as (secondary) final-state
radiation and vice-versa, similarly to the clustering correction in
Sec.~\ref{sec:nndl-clustering}.
For these emissions close to the jet boundary, the $k_t$ with respect
to the beam and the $k_t$ with respect to the jet axis differ by a
factor of $\order{R}$, so that for a finite jet radius, the difference
between the two is \NNDL~\cite{Lifson:2020gua,Caucal:2021bae}.
Therefore, up to \NDL accuracy we can neglect the effect of the jet
radius.

A direct consequence of the above discussion is that the only
differences between the Lund multiplicity measured in $pp$ and that
measured in $e^+e^-$ collisions come from radiation off the
leading partons, which, instead of evolving through final-state
radiation like in $e^+e^-$ collisions, now evolves
through initial-state radiation.
For concreteness, let us therefore write the Lund multiplicity for one
incoming leg of a $pp$ collision as a difference with respect to the
$e^+e^-$ case:
\begin{equation}
  N_{i,\text{IS}}(x,L) = N_{i,\text{FS}}(L) - 1 +
  \delta N_{i,\text{IS}}(x,L),
\end{equation}
where the `${-}1$' subtracts the incoming parton/beam --- which is
counted as part of the final state in $N_{i,\text{FS}}$ but should not
be counted in the initial-state case --- and we ought to compute
$\delta N_{i,\text{IS}}$ up to \NDL accuracy, including potential
differences already at \DL accuracy.
Let us therefore proceed as in Sec.~\ref{sec:recap-ndl} and list all
the possible contributions to $\delta N_{i,\text{IS}}$ up to \NDL
accuracy.
Two obvious candidates are the \NDL corrections already present in the
final-state case, namely one-loop running coupling corrections and
hard-collinear branchings.
Since the running of the coupling does not affect the kinematics of the incoming
parton, and, in particular, does not affect the longitudinal momentum
entering the PDFs, its \NDL correction is the same in the initial-
and final-state cases and its contribution to $\delta N_{i,\text{IS}}$
is zero.
Conversely, hard-collinear initial-state branchings differ from their
final-state equivalent in two ways: first, they lead to an increase of
the $x$-fraction of the incoming parton leading to a non-trivial
dependence on the PDF factor; second, the (flavour) structure of the
branching differs from that in the final state in a similar way
that initial-state backwards evolution in a parton shower differs from
forwards final-state evolution.

Besides these corrections, one should also account for the fact that
computing the multiplicity at a hadron collider involves a PDF factor.
When measuring the multiplicity of Lund declusterings with a relative
transverse momentum above $\ktcut$, this PDF factor has to be
evaluated at the scale $\ktcut$ which differs from the scale $Q^2$
used in Eq.~(\ref{eq:colour-singlet-normalisation}) for normalisation
of the Lund multiplicity.\footnote{ In practice, the average
  multiplicity can be written as the following cross-section-weighted
  average:
  \begin{equation}
    \avg{N_i(x,L)}= \left[f_i(x,Q^2) \frac{d\hat\sigma_i}{dQ^2}\right]^{-1}
    \left[
      \sum_j \int dx' f_j(x',\ktcut^2) \frac{d\hat\sigma_i}{dQ^2}
      \avg{N_{i|j}(x|x',L)}
    \right],
  \end{equation}
  where $\frac{d\hat\sigma_i}{dQ^2}$ is the Born-level hard partonic
  cross-section and we have normalised by the total cross-section at a
  fixed $x$ (i.e.\ the equivalent of
  Eq.~(\ref{eq:colour-singlet-normalisation}) for a single hemisphere
  or, equivalently, for DIS).
  In this expression, one integrates over all possible incoming
  partons of flavour $j$ and momentum fraction $x'$ and
  $\avg{N_{i|j}(x|x',L)}$ denotes the average Lund multiplicity with a
  given $\ktcut = Qe^{-L}$ for fixed $x'$ and $j$.
  In this case, the PDFs in the numerator have to be evaluated at the
  final scale $\ktcut^2$.}
Including both the primary hard-collinear splittings and the PDF
scale contributions, we can therefore write
\begin{equation}\label{eq:NDL-pp-multiplicity-setup}
  N_{i, \text{IS}}^{(\NDL)}(x, L)
  = N_{i, \text{FS}}^{(\NDL)}(L) - 1
  + \delta N_{i,\text{IS}}^{\text {prim}}(x,L)\, 
  + \delta N_{i, \text{IS}}^{\mu_F}(x, L).
\end{equation}

\paragraph{Factorisation scale contribution.}
Evaluating the PDF at the scale $\ktcut$ and taking into account the
normalisation in Eq.~(\ref{eq:colour-singlet-normalisation}), we can
write
\begin{equation}\label{eq:pp-ndl-muF-correction}
  \delta N_{i, \text{IS}}^{\mu_F}(x, L)
  \simeq \bigg[ \frac{f_i(x,\ktcut^2)}{f_i(x,Q^2)}-1\bigg]
  \big[N_{i, \text{FS}}^{(\DL)}(L)-1\big]
  \simeq -2L \frac{\partial \ln f_i(x,Q^2)}{\partial \ln Q^2}
  \big[N_{i, \text{FS}}^{(\DL)}(L)-1\big].
\end{equation}
When writing this equation we have made a few approximations valid at
\NDL accuracy.
For the second equality, we have used the DGLAP evolution
equation. Since $\partial\ln f_i(x,Q^2)/\partial {\ln Q^2}$ is of
order $\alpha_s$ with no logarithmic enhancement, it is sufficient to
expand this DGLAP evolution to first order at \NDL accuracy.
For the first equality, we have changed the scale of the PDFs, but kept
both the flavour and the longitudinal fraction $x$
constant.
For the first equality, since the change of PDF scale from $Q$ to
$\ktcut$ already induces an \NDL correction, we are allowed to keep
the flavour and $x$ fraction of the incoming parton unchanged, and to
evaluate the multiplicity in the second square bracket at \DL
accuracy.
Note that a direct consequence of Eq.~\eqref{eq:pp-ndl-muF-correction}
is that, at \DL accuracy we simply have
\begin{equation}\label{eq:pp-dl-result}
  N_{i, \text{IS}}^{(\DL)}(x, L) = N_{i, \text{FS}}^{(\DL)}(L) - 1.
\end{equation}

\paragraph{Hard-collinear initial-state branching.}
Let us now evaluate $\delta N_{i,\text{IS}}^{\text
  {prim}}(x,L)$. This takes into account differences between
hard-collinear branchings in the initial-state and in the final
state.
In particular, we should consider that such an initial-state splitting
affects the longitudinal momentum fraction entering the PDFs, taking
it from its initial value $x_\text{old}$ to a new value 
$x_\text{new} = x_\text{old}/\zeta$, where $\zeta$ 
is the transmitted collinear momentum fraction of the splitting.
Note that we now define the soft limit as corresponding to $\zeta\to 1$ 
and that the transmitted $\zeta$ and emitted $z$ momentum fractions are 
related by $\zeta = 1-z$.
Furthermore when hard-collinear branchings also change the flavour of
the incoming parton, the PDFs also have to be evaluated with the
new flavour.
As in the final-state case, hard-collinear branchings affect quarks
(i.e.\ Drell-Yan or Deep Inelastic Scattering (DIS)) and gluons (i.e.\
$gg\to H$) differently, so we discuss the two cases separately.

\subparagraph{\it Quark-initiated.} The contribution to the Lund multiplicity of a 
primary, hard-collinear splitting from an initial-state quark-leg at a scale 
$\ell$ is given by
\begin{align}
  \label{eq:DY-primary-HC-correction}
  \delta N_{q,\text{IS}}^{\text {prim}}(x,L)
  &= \frac{\as}{\pi}\int_0^L \!\!\dd \l
    \int_x^1\frac{\dd \zeta}{\zeta}
    P^{\text {IS}}_{qq}(\zeta)\,\frac{f_q(x/\zeta,Q^2)}{f_q\left(x,Q^2\right)}\,
    \left[N_{q,{\text {IS}}}^{(\DL)}(L)+ N_g^{(\DL)}(L-\l)\right]
  \nonumber \\
  & - \frac{\as}{\pi}\int_0^L \!\! \dd \l
    \int_0^1\dd \zeta\,
    P^{\text {IS}}_{qq}(\zeta)\,N_{q,{\text {IS}}}^{(\DL)}(L)\nonumber \\
  & - \frac{\as}{\pi} \int_0^L \!\! \dd \l \int_0^1 \!\! \dd \zeta\,
   P_{q \to qg}(1-\zeta) \, N_g^{(\DL)}(L-\l) \nonumber  \\
  & + \frac{\as}{\pi}
    \int_0^L \!\! \dd \l \int_x^1\frac{\dd \zeta}{\zeta}
    P^{\text{IS}}_{qg}(\zeta)\,\frac{f_g(x/\zeta,Q^2)}{f_q(x, Q^2)}\,
    \left[N_{q,{\text {IS}}}^{(\DL)}(L)+ N_g^{(\DL)}(L-\l)\right] \,,
\end{align}
where the standard, initial-state, DGLAP splitting functions read
\begin{align}
\label{eq:splitting-function-IS-DY}
  P^{\text{IS}}_{qq}(z) & = C_F\left(\frac{1+z^2}{1-z}\right),
  & P^{\text{IS}}_{qg}(z) &= T_R[z^2+(1-z)^2], 
\end{align}
In Eq.~\eqref{eq:DY-primary-HC-correction} the first line represents
the real emission of a gluon from a quark while backwards evolving
into a quark. All legs involved in the splitting are dressed by a
chain of soft-and-collinear emissions $N_i^{(\DL)}$,
Eq.~\eqref{eq:dl-resum} or Eq.~(\ref{eq:pp-dl-result}), evaluated at
$L-\ell$ for the final state gluon and at $L$ for the initial-state
quark.
In the second line, we consider the virtual correction for which only
the incoming quark leg is dressed with $N_{q,\text{IS}}^{(\DL)}(L)$.
The third line subtracts the hard-collinear splitting contribution in
the final-state case (after the real--virtual cancellation of the
radiation off the quark), remembering that we are computing the
difference between initial- and final-state multiplicities,
Eq.~\eqref{eq:NDL-pp-multiplicity-setup}.
There is no PDF ratio in this term.
Furthermore, $P_{q \to qg}(z) = P^{\text{IS}}_{qq}(1-z)$ and we have
used $z=1-\zeta$.
Finally, the fourth term in Eq.~\eqref{eq:DY-primary-HC-correction} accounts for 
the backwards evolution of a quark into a gluon by emitting a quark. Since this 
splitting does not exist in the final-state case we do not have to consider any 
subtraction term.

Grouping terms proportional to $N_q$ and $N_g$
and performing the $\l$ integrations we obtain 
\begin{align}
  \label{eq:DY-primary-HC-correction-two}
  \delta N_{q,\text{IS}}^{\text {prim}}(x,L)
  &=
  \frac{\as}{\pi} \int_x^1\frac{\dd \zeta}{\zeta}
    \left[P^{\text {IS}}_{qq,+}(\zeta)\,\frac{f_q(x/\zeta,Q^2)}{f_q(x,Q^2)} 
    +P^{\text {IS}}_{qg}(\zeta)\,\frac{f_g(x/\zeta,Q^2)}{f_q(x,Q^2)}\right]
   \, N_{q,{\text {IS}}}^{(\DL)} L  \nonumber \\
   &+ \frac{\as}{\pi} \int_x^1\frac{\dd \zeta}{\zeta}
    \left[P^{\text {IS}}_{qq,+}(\zeta)\, \frac{f_q(x/\zeta,Q^2)}{f_q(x,Q^2)} 
    +P^{\text {IS}}_{qg}(\zeta)\,\frac{f_g(x/\zeta,Q^2)}{f_q(x,Q^2)}\right]
    \, \frac{\sinhnu}{\sqrt{\abar}} \,
\end{align}
Where $P^{\text {IS}}_{qq,+}$ includes the `plus' prescription in the
$P^{\text {IS}}_{qq}$ splitting function of
Eq.~(\ref{eq:splitting-function-IS-DY}).
Using the DGLAP evolution equation
\begin{equation}
  \frac{\partial \ln f_q(x,Q^2)}{\partial \ln Q^2}
  = \frac{\as}{2\pi} \int_x^1\frac{\dd \zeta}{\zeta}
    \left[P^{\text {IS},+}_{qq}(\zeta)\frac{f_q(x/\zeta,Q^2)}{f_q(x,Q^2)} 
    +P^{\text {IS}}_{qg}(\zeta)\frac{f_g(x/\zeta,Q^2)}{f_q(x,Q^2)}\right]\, ,
\end{equation} 
we realise that the first line of
Eq.~\eqref{eq:DY-primary-HC-correction-two} is identical, but with
opposite sign, to the $\delta N_{q, \text{IS}}^{\mu_F}(x, L)$
contribution given by Eq.~(\ref{eq:pp-ndl-muF-correction}), cancelling
the $N_{q,\text{IS}}$ term in the first line
of~\eqref{eq:DY-primary-HC-correction-two}.
The final result for the quark-initiated Lund multiplicity in hadronic
collisions is then given by
\begin{align}
N_{q, \text{IS}}^{(\NDL)}(x, L) =  N_{q, \text{FS}}^{(\NDL)}(L) -1 + 
\frac{2\sinhnu}{\sqrt{\abar}} \frac{\partial \ln 
f_q(x,Q^2)}{\partial\ln{Q^2}} \, .
\label{eq:NDL-quark-pp-resummed}
\end{align}
This result agrees with the one obtained in 
Ref.~\cite{Catani:1993yx} using the initial-state generating functional 
formalism in the context of DIS. In fact, 
since the choice of clustering algorithm does not
impact the resummation at \NDL accuracy, the Lund multiplicity in
deep-inelastic scattering is identical to that of
Ref.~\cite{Catani:1993yx} which was instead defined using a
$k_t$-like algorithm.

Finally, we proceed to extend the calculation to Drell-Yan 
where the incoming parton entering into the hard 
process does not have a fixed flavour. In 
this case, the normalisation is given by 
\begin{align}
  \label{eq:colour-singlet-normalisation-flavours}
  \frac{\dd^2 \sigma_{pp \to X}}{\dd Q^2\dd y_{_X}}
  =
  \frac{\dd^2 \hat \sigma_{q\bar q \to X}}{\dd Q^2 \dd y_{_X}}
  \times \sum_q Q_q^2
  \left[
  f_q(x_1, Q^2) f_{\qbar}(x_2, Q^2) + (q \leftrightarrow \qbar)
  \right],
\end{align}
where an additional sum over all possible quark flavours weighted by their 
electric charge is included with respect to
Eq.~\eqref{eq:colour-singlet-normalisation}. This translates into our final 
formula for the \NDL accurate Drell-Yan multiplicity
\begin{align}\label{eq:NDL-resummed-Drell-Yan}
  \langle N^{\text{(Lund)}}(x_1,x_2, L) \rangle_{\text{DY},\NDL} 
  &=
    \langle N^{\text{(Lund)}}(L) \rangle_{e^+e^-,\NDL}-2
  \\
  &+
    \frac{2\sinhnu}{\sqrt{\abar}} 
    \,
    \frac{\partial}{\partial \ln Q^2} 
    \ln \sum_q Q_q^2
    \left[f_q(x_1, Q^2) f_{\qbar}(x_2, Q^2)
    + (q \leftrightarrow \qbar) \right].
    \nonumber
\end{align}

\subparagraph{\it Gluon-initiated} The contribution to the Lund multiplicity of a 
primary, hard-collinear splitting from an initial-state gluon-leg at a scale 
$\ell$ is given by 
\begin{align}\label{eq:ggH-primary-HC-correction}
  \delta N_{g,\text{IS}}^{\text {prim}}(x,L)
    & = \frac{\as}{\pi}
      \int_0^L \!\!\dd\l \int_x^1\frac{\dd \zeta}{\zeta}
      P^{\text {IS}}_{gg}(\zeta)\,\frac{f_g(x/\zeta,Q^2)}{f_g(x,Q^2)}
      \left[N_{g,{\text{IS}}}^{(\DL)}(L)+ N_g^{(\DL)}(L-\l)\right]   \nonumber \\
    & - \frac{\as}{\pi}
      \int_0^L \!\! \dd \l \int_0^1\dd \zeta
      \left[P^\text{IS}_{gg}(\zeta)+P_{qg}^\text{IS}(\zeta)\right] \,N_{g,\text{IS}}^{(\DL)}(L)\nonumber \\
    & - \frac{\as}{\pi}
      \int_0^L \!\! \dd \l \int_0^1 \!\! \dd \zeta\,
      P_{g\to gg}(1-\zeta)\, N_g^{(\DL)}(L-\l)
      \nonumber \\
    & + \frac{\as}{\pi}
      \int_0^L \!\!\dd \l \int_x^1\frac{\dd \zeta}{\zeta}
      P^{\text {IS}}_{gq}(\zeta)\,\frac{f_\Sigma(x/\zeta,Q^2)}{f_g(x,Q^2)}
      \left[N_{g,{\text {IS}}}^{(\DL)}(L)-N_g^{(\DL)}(L-\l)+2N_q^{(\DL)}(L-\l)\right] 
      \nonumber \\
    & - \frac{\as}{\pi}
      \int_0^L \!\! \dd \l \int_0^1 \!\! \dd \zeta\,
      P_{g \to q\qbar}(1-\zeta)\, \left[2N_q^{(\DL)}(L-\l) - N_{g}^{(\DL)}(L-\l)\right] \, ,
\end{align}
where the initial-state, DGLAP splitting functions read
\begin{align}
\label{eq:splitting-function-IS-two}
  P^{\text{IS}}_{gg}(z) &= 2C_A\left[\frac{z}{1-z} + \frac{1-z}{z}
                          +   z(1-z)\right],
  & P^{\text{IS}}_{gq}(z) & = C_F\left[\frac{1+(1-z)^2}{z}\right], 
\end{align}
and we have defined the singlet distribution $f_\Sigma(x, \mu_F^2) \equiv 
\sum_i^{n_f} [f_{q_i}(x, \mu_F^2) + f_{\qbar_i}(x, \mu_F^2)]$.
The ordering of the terms in Eq.~\eqref{eq:ggH-primary-HC-correction} follows 
the same logic as of the quark-initiated case.
The first three lines of Eq.~\eqref{eq:ggH-primary-HC-correction} are
related to the backwards evolution of a gluon into a gluon by emitting
another gluon: the first line describes real emissions, the
second line subtracts virtual corrections and the third one subtracts the
final-state contribution.
The last two lines describe the backwards evolution of a gluon into a
quark/anti-quark pair of any flavour.
Note that the final-state subtraction in this case has the usual
mismatch between real and virtual emissions, see e.g.\
Eq.~\eqref{eq:ndl-hc-fc-raw}, that leads to the
$[2N_q^{(\DL)}(L-\l) - N_g^{(\DL)}(L-\l)]$ term.

Recalling that DGLAP evolution for a gluon reads (using the
`plus' prescription for $P_{gg}$)
\begin{align}
\frac{\partial \ln f_g(x,Q^2)}{\partial \ln Q^2} = 
\frac{\as}{2\pi}\int_x^1\frac{\dd\zeta}{\zeta} 
\left[P^{\text{IS,+}}_{gg}\, \frac{f_g(x/\zeta,Q^2)}{f_g(x,Q^2)}+P^{\text{IS}}_{gq}
(\zeta)\frac{f_\Sigma(x/\zeta,Q^2)}{f_g(x,Q^2)})\right],
\end{align}
we can re-write Eq.~\eqref{eq:ggH-primary-HC-correction} as
\begin{align}
  \label{eq:ggH-primary-HC-correction-two}
  \delta N_{g,\text{IS}}^{\text {prim}}(x,L)
  &= \frac{\partial \ln f_g(x,Q^2)}{\partial \ln Q^2} \left[2L + \frac{2\sinhnu}{\sqrt{\abar}}\right] \nonumber \\
  & + \frac{2\as}{\pi}\int_0^L \!\dd \l\int \frac{\dd \zeta} {\zeta} P^{\text{IS}}_{gq}(\zeta)\frac{f_\Sigma(x/\zeta,Q^2)}{f_g(x,Q^2)}[N_q^{\rm DL}(L) - N_g^{(\DL)}(L-\l)] \nonumber \\
  & - \frac{2\as}{\pi}\int_0^L \!\dd \l\int \dd \zeta P_{g\to q\bar q}(1-\zeta) [N_q^{\rm DL}(L) - N_g^{(\DL)}(L-\l)]\, .
\end{align}
As in the Drell-Yan case, the first term of
Eq.~\eqref{eq:ggH-primary-HC-correction-two} exactly cancels
$\delta N_{g, \text{IS}}^{\mu_F}(x, L)$. After integrating over $\l$
we obtain the \NDL gluon-initiated Lund multiplicity in hadronic
collisions 
\begin{equation}\label{eq:NDL-gluon-pp-resummed}
  N_{g, \text{IS}}^{(\NDL)}(x, L)
  =  N_{g, \text{FS}}^{(\NDL)}(L) -1 + 
  \frac{2\sinhnu}{\sqrt{\abar}} \,
  \frac{\partial \ln f_g(x,Q^2)}{\partial\ln{Q^2}} 
  + \frac{4\as}{\pi} 
  (C_F-C_A)\left(\frac{\sinhnu}{\sqrt\abar}-L\right)
  (B_\Sigma-B_{gq}),
\end{equation}
with
\begin{align}
  B_\Sigma = \frac{1}{2C_A}\int_x^1\frac{\dd\zeta}{\zeta}
  P^{\text{IS}}_{gq}(\zeta)\,
  \frac{f_\Sigma(x/\zeta,Q^2)}{f_g(x,Q^2)},
\label{eq:Bsigma-def}
\end{align}
and $B_{gq}$ defined as in Eq.~\eqref{eq:B-coeffs}.  A clear
difference exists between Eq.~\eqref{eq:NDL-gluon-pp-resummed} and the
quark-initiated case, Eq.~(\ref{eq:NDL-quark-pp-resummed}):
besides the DGLAP evolution of the initiator's PDF, the
gluon-initiated case contains an additional term proportional to
$f_\Sigma$.
This second term can be traced back to the hard-collinear contribution
where a gluon backwards-evolve into a $q\bar q$ pair, creating a
mismatch between the $N_q$ and $N_g$ contributions.
As with Eq.~\eqref{eq:NDL-quark-pp-resummed},
Eq.~(\ref{eq:NDL-gluon-pp-resummed}) has been cross-checked against
the result from Ref.~\cite{Catani:1993yx} obtained with a
generating-functional approach.

Finally, the Lund multiplicity in Higgs production via gluon-gluon fusion can be 
constructed from Eq.~\eqref{eq:NDL-gluon-pp-resummed} by simply summing the 
contributions over both initial-state partons. We get
\begin{align}
  \label{eq:NDL-resummed-ggH}
  \langle N^{\text{(Lund)}}(x_1, x_2, L) \rangle_{gg \to H}
  & = \langle N^{\text{(Lund)}}(L) \rangle_{e^+e^-}-2  
    +
    \frac{2\sinhnu}{\sqrt{\abar}} 
    \,
    \frac{\partial}{\partial \ln{Q^2}} 
    \ln{\big[ 
    f_g(x_1, Q^2) f_{g}(x_2, Q^2)
    \big]}
    \nonumber\\
  &+
    \frac{4\as}{\pi}\left(C_F-C_A\right)
    \left(\frac{\sinhnu}{\sqrt{\abar}} - L\right)
    (B_{\Sigma,1}+B_{\Sigma,2}-2 B_{gq})\, ,
\end{align}
with $B_{\Sigma,i}$ given by Eq.~\eqref{eq:Bsigma-def} evaluated at
$x_i$ for each beam.

\section{Conclusions and outlook}\label{sec:conclusions}
This paper has presented the first theoretical determination of subjet
multiplicity for angular-ordered based clustering algorithms at \NNDL
accuracy in $e^+e^-$ collisions.
In particular, we have explored two definitions of subjet
multiplicity: (i) a novel procedure based on Lund declustering as
explained in Sec.~\ref{sec:lund-mult-def} and (ii) the more standard
average jet multiplicity obtained when running the Cambridge
algorithm~\cite{Dokshitzer:1997in} with $y_\text{cut}=\ktcut^2/Q^2$.
We have found that these two definitions are identical at \NNDL up to
a term related to the precise definition of $k_t$ for large-angle
emissions (see
Eqs.~(\ref{eq:final-result-h3q}),~(\ref{eq:final-result-h3g})
and~\eqref{eq:final-nndl-cam}).
Our calculation thus achieves an order higher in logarithmic accuracy
than the current state-of-the-art~\cite{Catani:1991pm}.

A key aspect allowing for this improvement is the use of an
angular-ordered clustering sequence instead of the typical Durham
clustering algorithm.
The latter receives additional contributions where one has a pair of
emissions with $k_t\simeq\ktcut$ and widely separated in angle
(Fig.~\ref{fig:nndl-diagram-Lsqr}).
At all orders, when adding an arbitrary number of soft-and-collinear
emissions, this would yield complex situations depending on fine details
of the clustering history (see App.~\ref{app:kperp-vs-cambridge}).
This configuration is absent when defining the multiplicity based on
the Cambridge clustering algorithm.
This is similar to the original argument motivating the Cambridge
algorithm~\cite{Dokshitzer:1997in}.

Another cornerstone of our work is the use of a novel resummation
approach which does not rely on generating functionals.
Instead, it is based on the observation that subleading contributions
can be obtained from configuration where the subleading part is
associated with only a handful of emissions --- a single one giving a
contribution proportional to $\as L$ at \NDL, one or two giving a
contribution proportional either to $\as$ or to $(\as L^2)$ at \NNDL
--- dressed with an arbitrary number of soft-and-collinear emissions.
Once these subleading contributions have been systematically
enumerated and computed (at fixed order), their all-order treatment
largely recycles results from lower orders and is therefore
straightforward (see e.g.\ Fig.~\ref{fig:ndl-sketch}).
This approach can be extended, at least in principle, to higher
resummation orders.

The final \NNDL formul\ae\ for quark ($e^+e^-\to Z\to q\bar q$) and
gluon ($e^+e^-\to H\to gg$) hemispheres, the main results of this
paper, can be found in Sec.~\ref{sec:final-result}. Besides its purely
theoretical interest, this compact formul\ae\ could be useful to test
the logarithmic accuracy of the next-generation of NNLL parton
showers.
As a proof of concept, in addition to the $e^+e^-$ result, we have
presented the extension of our formalism to initial-state radiation in
Sec.~\ref{sec:ndl-pp} and computed the Lund multiplicity at \NDL
accuracy for deep-inelastic scattering and colour singlet production in
hadronic collisions, i.e.\ Drell-Yan and Higgs production via gluon
fusion.

We have performed an exploratory study of the Lund multiplicity
at LEP energies.  To that end, the resummed predictions were
additively matched to the NLO result obtained via {\tt Event2}. We
have shown that the inclusion of \NNDL corrections leads to a strong
reduction of the theoretical uncertainty of about 50\% for almost all
$\ktcut$ values. Our work thus provides a precise pQCD benchmark for
eventual theory-to-data comparisons.
We have also shown that, after a simple inclusion of non-perturbative
corrections, we obtained a decent description of the measurements of
the average Cambridge multiplicity from the OPAL
collaboration~\cite{JADE:1999zar,OPAL:2005jle}.
In the future, the short analysis presented in this paper could
be extended into a deeper phenomenological study.
In this context, one could combine our increased perturbative accuracy
with recent efforts to reach NNLO fixed-order
accuracy~\cite{Gehrmann-DeRidder:2014hxk,DelDuca:2016ily,DelDuca:2016csb,Gehrmann:2017xfb}
and study if it could lead to a new determination of the strong coupling
constant.
Furthermore, it would be interesting to see if the Lund multiplicity
(or the difference between the Lund and Cambridge multiplicities which
may have a simple perturbative structure) could also be measured from
LEP data, following e.g.\ the recent interest for reanalyses of the
archived ALEPH data~\cite{Chen:2021uws}.

Besides these direct phenomenological application at LEP, there are
multiple possible extensions of this work.
On the $e^+e^-$ side, one first task would be to understand the
all-orders resummation of an arbitrary number of hard-collinear and
running-coupling corrections in a perturbatively-controlled manner,
since these two contributions are expected to dominate at large
$\as L^2$.
In this case, we expect that the expansion in
Eq.~\eqref{eq:log-counting} should be revisited.
Another option would be to generalise the proposed resummation scheme
so as to calculate higher-order moments of the Lund multiplicity
distribution. This would amount to accounting for correlations between
different splittings, an effect which may break the simplifications
arising from angular-ordering.
From a phenomenological viewpoint, a dedicated study of hadronisation
corrections is required for practical applications to precision
physics.
In this context, it would be interesting to see if recent approaches
either on the analytic front~\cite{Caola:2021kzt}, using Monte Carlo
techniques~\cite{Reichelt:2021svh}, or even methods based on
Deep-Learning~\cite{Ghosh:2022zdz}, could be helpful.

It would also be interesting to extend the calculation to other
processes like $e^+e^-\to W^+W^-$, relevant for FCC-ee studies, but
also to deep-inelastic scattering at the forthcoming Electron-Ion
Collider (EIC). 
Finally, a natural continuation of this work would be to compute the
Lund multiplicity at \NNDL for high-energy jets produced at hadron
colliders, in processes such as dijet or $Z+\text{jet}$ events at the
LHC (or for the average associated jet rate introduced in
Ref.~\cite{Bhattacherjee:2015psa}).
This calculation would be relevant for at least two reasons.
On the one hand, one could probe multiplicity distributions over an
energy range much larger than that is available at LEP.
This might ultimately provide a new handle on the measurement of the
strong coupling constant at the LHC.
In this context, one could fit the multiplicity together with
(an)other Lund-plane observable(s) (e.g. the primary Lund plane
density computed at NLL in~\cite{Lifson:2020gua}) so as to
simultaneously constrain the strong coupling constant and the
quark-gluon fraction.
On the other hand, the measurement of a quantity such as the
multiplicity, which is sensitive to the dynamics of the jet across all
scales from its transverse momentum down to a $\ktcut$ as low as a few
GeV, could help constraining Monte Carlo generators.
This could for example improve the determination of crucial
quantities such as the Jet Energy Scale for quark- and gluon-initiated
jets which is currently affected by large Monte Carlo uncertainties,
in particular from parton showers~\cite{CMS:2016lmd,ATLAS:2017bje}.
%

\section*{Acknowledgements}
We wish to thank Keith Hamilton and Gavin Salam for collaboration in the 
early stages of this work.
We are grateful to our PanScales collaborators (Melissa van Beekveld,
Mrinal Dasgupta, Fr\'ed\'eric Dreyer, Basem El-Menoufi, Silvia
Ferrario Ravasio, Keith Hamilton, Alexander Karlberg, Pier Monni,
Gavin Salam, Ludovic Scyboz and Rob Verheyen), as well as to Bryan
Webber for discussions and comments on this manuscript.
RM is grateful for the hospitality of Universit\'e Paris-Saclay, and
ASO and GS are grateful for the hospitality of the University of
Oxford  where part of this research was conducted. 
This work has been supported by the European Research Council (ERC) under
the European Union’s Horizon 2020 research and innovation programme
(grant agreement No.\ 788223, PanScales).
%

\appendix
%

\section{Subjet multiplicity definition and logarithmic accuracy}
\label{app:mult-def}
Here, we discuss why an angular-ordered clustering sequence
is essential for the viability of \NNDL resummation of the average
multiplicity (App.~\ref{app:kperp-vs-cambridge}).
We then show that the choice of recombination scheme in the initial
clustering does not affect the \NNDL accuracy of the resummed
average multiplicity (App.~\ref{app:recomb-scheme}).
Finally, we relate Lund and Cambridge multiplicity (App.~\ref{app:cambridge-vs-lund}). 

\subsection{Angular ordered vs $k_t$-based clustering algorithms}
\label{app:kperp-vs-cambridge}

As pointed out in Sec.~\ref{sec:nndl-list}, the main difference at
\NNDL between $k_t$-based~\cite{Catani:1991hj} and
angular-ordered~\cite{Dokshitzer:1997in} clustering algorithms comes
from configurations with multiple emissions with $k_t\sim\ktcut$,
Fig.~\ref{fig:nndl-diagram-Lsqr}.
We briefly discuss here how the presence of such a contribution
drastically complicates the calculation for $k_t$-based algorithms.

To illustrate this, consider the case of three primary emissions off a
leading quark, all with identical azimuthal angle relative to it.
Take their angles such that $\theta_{3q}\gg\theta_{1q}\gg\theta_{2q}$
and assume that $k_{t,1q}\gg \ktcut$ is in the double-logarithmic
region, while emissions 2 and 3 are just below $\ktcut$ such that
$\ktcut/2 < k_{t,2q}< k_{t,3q}<\ktcut$.
If we compute the multiplicity using a $k_t$-based algorithm, e.g.\
the Durham algorithm~\cite{Catani:1991hj}, emission 3 is first
clustered with 1 and does not contribute to the multiplicity as
$k_{t,3q}<\ktcut$; emission 2 is then clustered with the leading
parton and thus does not contribute to the multiplicity either.
If we instead consider the same situation in which emission 1 is
absent (e.g.\ virtual), emissions 2 and 3 would be clustered together
resulting in a $k_t$ of the pair above the $\ktcut$.
This mismatch creates a logarithmic contribution of order
$(\as L^2)_1 (\as L)_2 (\as L)_3 = \as^3L^4$, i.e.\
\NNDL.
Conversely, with Cambridge clustering, emissions 2 and 3 would be
clustered with the leading parton in both cases, leaving only an
effect beyond \NNDL accuracy when the emissions further satisfy 
$\theta_{23}\sim\theta_{2q}$.
More generally, computing the average multiplicity for $k_t$-based
algorithms would require knowledge of the full fragmentation process
for a pair of emissions with $k_t\sim \ktcut$ accompanied by an
arbitrary number of soft-collinear emissions. This is a very complex
task which depends in a non-trivial way on the geometry of all the
emissions and could most likely only be achieved (semi-)numerically.

\subsection{Recombination scheme independence}
\label{app:recomb-scheme}
We comment on why the choice of recombination scheme for the initial
Cambridge ($e^+e^-$) or C/A ($pp$) clustering affects the average Lund
multiplicity beyond \NNDL. 

As a case study, we contrast the usual $E$-scheme, where one simply
adds the 4-momenta of the two pseudo-jets, to the winner-takes-all (WTA)
recombination scheme~\cite{Bertolini:2013iqa,Larkoski:2014uqa}, where
the transverse momenta $p_t$ of the two pseudo-jets are added
linearly, but the direction of the recombined jet is determined by the
direction of the higher $p_t$ pseudo-jet.
As a case study, let us consider the case of two emissions.
When one emission is much softer than the other, their possible
recombination would yield the same result with $E$-scheme and WTA
recombination, up to power corrections in their relative energy
fractions. This would only generate effects beyond our accuracy,
including, for example, for the clustering correction of
Sec.~\ref{sec:nndl-clustering}.
When the emissions are strongly ordered in angles, the difference
between the recombination schemes would then be power-suppressed by
the ratio of the angles.
This only leaves a potential effect when the two emissions have both
commensurate angles and energies, i.e.\ the ``pair'' contribution from
Sec.~\ref{sec:nndl-double-soft}. In this case, the relative difference
between the two recombination schemes would be of order 1. This would
only be relevant for $k_t\sim\ktcut$ which would therefore contribute
only at the N$^3$DL.
To sum up, the $E$-scheme and WTA Lund multiplicities are the same at
\NNDL accuracy.
The above argument would hold for any infrared-and-collinear-safe
recombination scheme which would reproduce the $E$-scheme up to power
corrections in either the soft or collinear limit.

\subsection{Lund and Cambridge multiplicities}\label{app:cambridge-vs-lund}

In this Appendix, we discuss a specific aspect which allows us to
relate the Lund and Cambridge multiplicities.
Specifically, we argue that if we take the full clustering tree
obtained by running the Cambridge algorithm with $y_\text{cut}=1$,
the total number of clusterings for which $k_t$, defined in the Lund
sense, Eq.~(\ref{kt:def}), is above $\ktcut$ is equal to the Lund
multiplicity.

The only possible situation which could result in these two
multiplicities to differ is one where the Lund declustering
procedure hits a declustering $i\to j+k$ ($E_j>E_k$) with
$k_t<\ktcut$ such that one of the clusterings that led to the softer
branch, $k$, had a relative $k_t$ above $\ktcut$.
Indeed, in such a situation, this last clustering would be counted as
part of the Cambridge multiplicity, but missed in the Lund
declustering procedure as the latter would not recurse into subjet
$k$.
Since the energy can only grow throughout clustering, this situation
can only happen if the clustering angles are reduced.
We argue below that, even if the angles can decrease, it never creates
a situation where the Lund multiplicity would miss a Cambridge
clustering above $\ktcut$.

We start by considering a simple situation with 3 particles, $p_1$,
$p_2$, $p_3$, at small angles (with a fourth one recoiling against this
system for energy-momentum conservation).
For definiteness, let us assume that $p_1$ and $p_2$ cluster first,
followed by $p_3$ clustering with the $p_1+p_2$ combination.
We are interested in a configuration when $k_{t,12}>\ktcut$, but
$k_{t3,1+2} < \ktcut$.
Still without loss of generality, we can parameterise the energies as
$E_1=(1-z)z_pE$, $E_2=(1-z)(1-z_p)E$, $E_3=zE$ with $E=Q/2$,
and assume that the angle between particles 1 and 2 is
$\theta_{12}=\Delta$.
The Lund procedure would first obtain the $k_{t3,1+2}$ associated with
the last clustering. Next, if $z\leq 1/2$, $k_{t,12}$ would be
reconstructed by following the hard branch. However, since $k_{t3,1+2} <
\ktcut$, if $z>1/2$, the $p_1 + p_2$ clustering would be
ignored. 
Hence, we want to show that we can never have $k_{t,12}>k_{t3,1+2}$
if $z>1/2$, i.e.\ if the $p_1+p_2$ system is on the soft
branch of the final clustering.

\begin{figure}[t]
  \centering
  \begin{minipage}[c]{0.35\textwidth}
    \includegraphics[width=\textwidth]{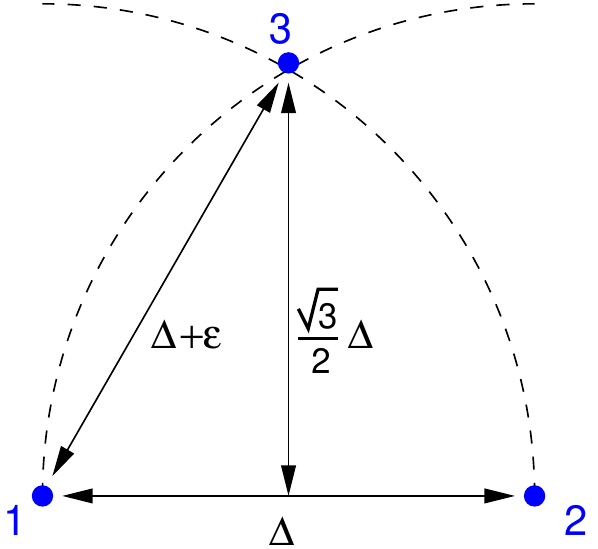}
  \end{minipage}
  \hfill
  \begin{minipage}[c]{0.55\textwidth}
    \includegraphics[width=\textwidth]{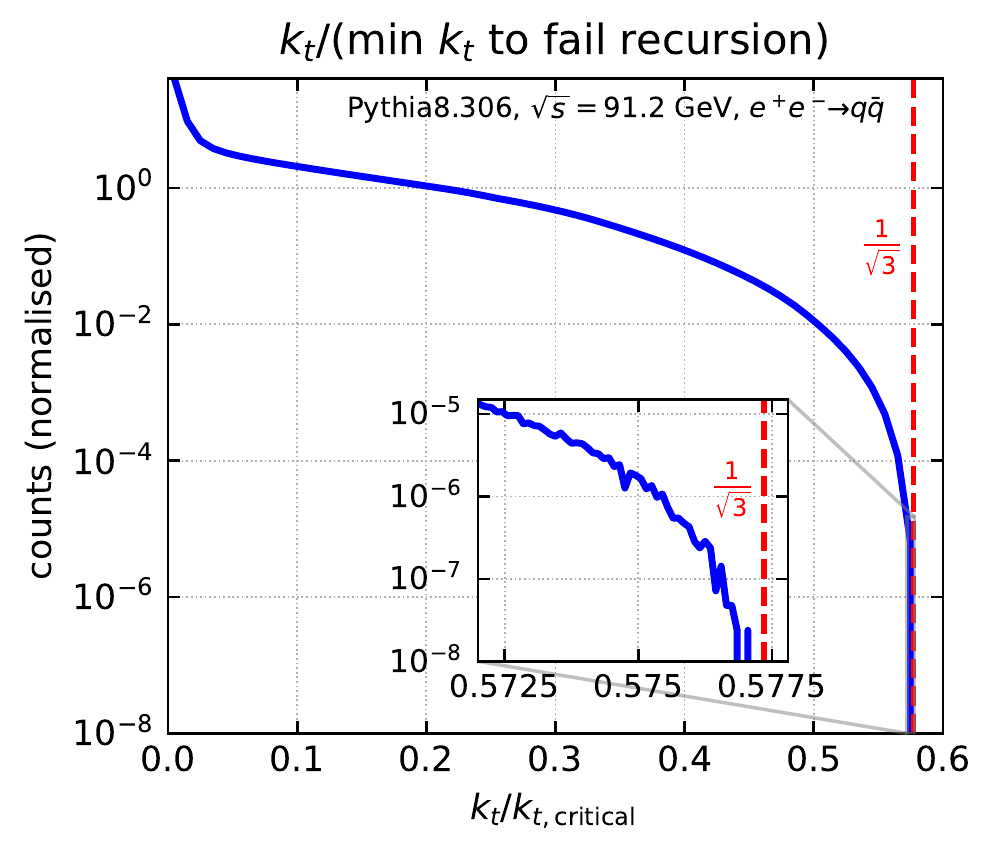}
  \end{minipage}
  \caption{
    Left: Sketch of a 3 particle system where angles between particles
    become smaller after clustering. Right: Number of events as a
    function of the ratio between the $k_t$ of each declustering and
    the critical $k_t$ as defined in the text. Since the number of
    events drops at $k_t/k_{t,\text{critical}} = 1/\sqrt{3} < 1$, this plot
    illustrates that the Lund multiplicity is equivalent to counting
    the Cambridge clusterings with a $k_t$ above a given $ \ktcut$.}
  \label{fig:clustering-triangle}
\end{figure}

Geometrically, the smallest possible $k_{t3,1+2}$ requires the
smallest possible angular distance $\Delta_{3,1+2}$, which is achieved
when $p_3$ is at a distance $\Delta+\varepsilon$ (with
$\varepsilon\ll\Delta$) from both $p_1$ and $p_2$, as depicted in
Fig.~\ref{fig:clustering-triangle}.
It is easy to show that, in such a configuration, 
\begin{equation}\label{eq:ktmax-recurse-soft}
  \frac{k_{t,21}}{k_{t3,1+2}} =
  \frac{z_p}{\sqrt{\frac{3}{4}+\big(\frac{1}{2}-z_p\big)^2}}\le \frac{1}{\sqrt{3}},
\end{equation}
where the maximum value is reached for $z_p=1/2$.
Since this maximal value is below 1, we never face a situation where a
soft branch can hide a clustering with a $k_t$ larger than the parent
one.
With a bit of an effort, the above argument can be shown to hold
beyond the small-angle limit, with the small-angle case still giving 
the extremal value.

We have not been able to prove that the upper bound derived in
Eq.~\eqref{eq:ktmax-recurse-soft} was still valid for an arbitrary
number of particles.
Instead, we approach the problem numerically.
We run a Pythia~\cite{Sjostrand:2014zea} simulation of $e^+e^-\to Z\to q\bar q$
events at $\sqrt{s}=91.2$ GeV and for each event reconstruct the
Cambridge clustering tree. 
For each branching, $\mathcal{B}$, associated with a relative transverse momentum
$k_t$, we look at all the earlier Lund declusterings for which
reaching $\mathcal{B}$ would, at some point, require following a soft
branch. Of these declusterings, we define $k_{t,\text{critical}}$ as
the one with the smallest $k_t$.
Seeing a value of $k_t/k_{t,\text{critical}}$ above 1 would mean that
for a value of $\ktcut$ between $k_t$ and $k_{t,\text{critical}}$, the
branching $\mathcal{B}$ would be counted as part of the Cambridge
multiplicity but not as part of the Lund multiplicity.
The distribution of $k_t/k_{t,\text{critical}}$ is plotted in
Fig.~\ref{fig:clustering-triangle} and clearly shows that the upper
bound in Eq.~(\ref{eq:ktmax-recurse-soft}), derived for 3 particles,
is still valid in general and that, in particular, the ratio never
exceeds 1.

The above discussion shows that, if an event is clustered with the
Cambridge algorithm with $y_\text{cut}=1$, counting Lund
declusterings with a $k_t$ (defined as in Eq.~(\ref{kt:def})) above
$\ktcut$, i.e.\ following the algorithm defined in
Sec.~\ref{sec:lund-mult-def}, is equivalent to counting declusterings
satisfying the same condition in the full clustering tree.
In practice, one could instead consider the Cambridge jet multiplicity,
obtained by running the Cambridge algorithm with
$y_\text{cut}=\ktcut^2/Q^2$ and counting the number of resulting jets.
Procedurally this is very similar to the Lund multiplicity, however it
differs from it in two ways, which we can account for within the Lund
multiplicity algorithm. 
First, we modify the definition of the relative transverse momentum to
use $k_t^\text{(Cam)}$, Eq.~(\ref{kt:def:cambridge}),
and, second, we run the initial Cambridge clustering sequence with the
recombination scheme 
\begin{align}
 \label{eq:cambridge-recombination-scheme}
 \text{recombine}\left(p_i, p_j \right) = 
 \left\{
 \begin{array}{lr}
   p_i + p_j, & \text{if } k_t < \ktcut\\
   \max{(p_i, p_j)}, & \text{if } k_t > \ktcut
 \end{array}
 \right.,
\end{align}
where $\max{(p_i, p_j)}$ is the most energetic of $p_i$ and  $p_j$. 
This describes the feature of the Cambridge algorithm that each newly
resolved jet is removed from the list of
particles~\cite{Dokshitzer:1997in}.
We have already discussed in the main text that trading the Lund
definition of $k_t$ for $k_t^\text{(Cam)}$ does impact the average
multiplicity starting from \NNDL accuracy, with the result in
Eq.~(\ref{eq:NNDL-cam-result}). 
Furthermore, the effect of the recombination
scheme in Eq.~\eqref{eq:cambridge-recombination-scheme} is beyond our
targeted \NNDL accuracy.\footnote{Note that the recombination scheme
  in Eq.~\eqref{eq:cambridge-recombination-scheme} does not satisfy the
  criteria discussed at the end of App.~\ref{app:recomb-scheme}.}
To see this, we need to consider a situation where the second line of
Eq.~\eqref{eq:cambridge-recombination-scheme} is used and see its
impact on subsequent clusterings.
For this to impact the counting (compared to an $E$-scheme approach),
one should be in a situation where $p_i$ and $p_j$ have commensurate
energies, i.e.\ in the case of a hard-collinear branching (with
$k_t\ge \ktcut$).
For the difference between $p_i$ (assuming $E_i>E_j$) and $p_i+p_j$ to
have an impact on the $\ktcut$ condition for a subsequent clustering,
this new clustering must have a $k_t$ commensurate to $\ktcut$ with
$p_i$ on its soft branch. This kinematic configuration is clearly
beyond \NNDL accuracy.

\begin{figure}
  \begin{subfigure}[t]{0.48\linewidth}
    \includegraphics[height=7cm,page=1]{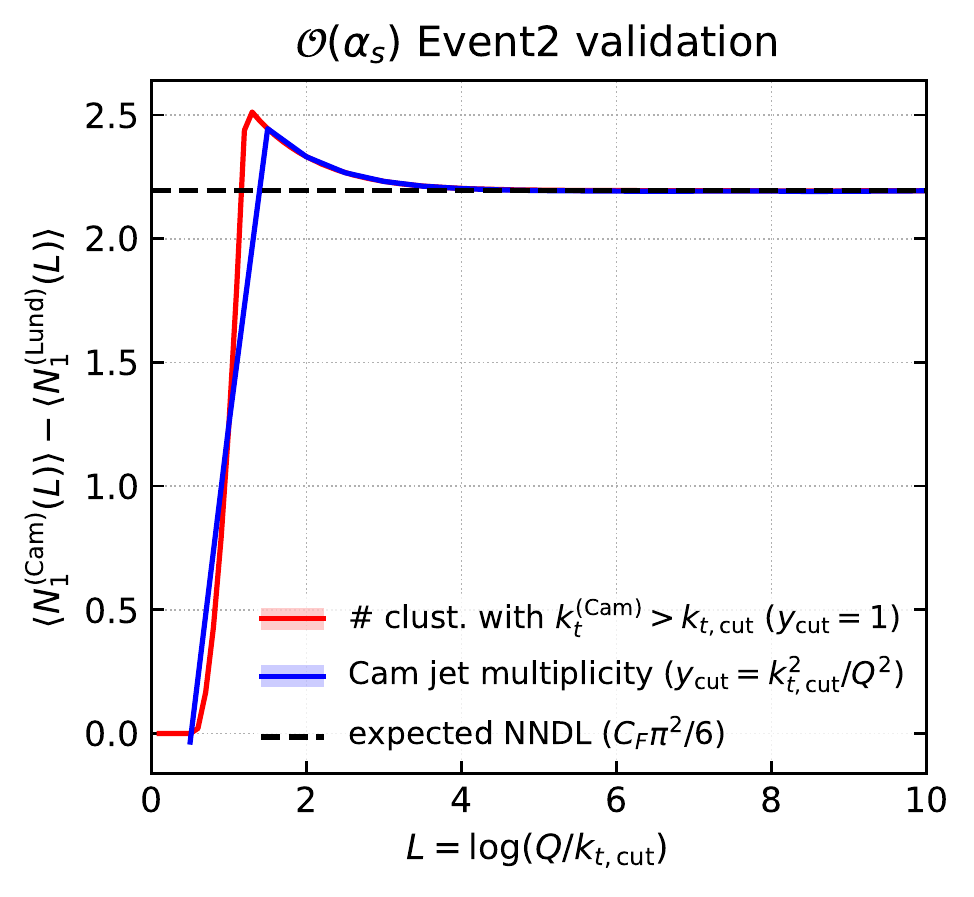}
    \caption{$\langle N\rangle$ test at $\order{\as}$}\label{fig:cam-v-lund-order1}
  \end{subfigure}
  \begin{subfigure}[t]{0.48\linewidth}
    \includegraphics[height=7cm,page=2]{figs/cam-v-lund.pdf}
    \caption{$\langle n\rangle$ test at $\order{\as^2}$}\label{fig:cam-v-lund-order2}
  \end{subfigure}
  \caption{Comparison between the Lund and Cambridge multiplicities at
    fixed-order with the {\tt Event2} program: %
    (a) the (cumulative) multiplicity difference
    $\langle N^\text{(Cam)}_1\rangle-\langle
    N^\text{(Lund)}_1\rangle$ at order $\as$,
    compared to the NNDL analytic expectation
    (b) the differential multiplicity difference
    $\langle n^\text{(Cam)}_2\rangle-\langle n^\text{(Lund)}_2\rangle$
    at order $\as^2$, which is expected to go to a (N$^3$DL)
    constant if the results agree at NNDL.}
  \label{fig:cam-v-lund}
\end{figure}

To further check the relation between the Lund and Cambridge
multiplicities, we run some fixed-order tests with
the {\tt Event2} program, similarly to what was done in
Sec.~\ref{sec:event2}.
We study the difference between the Lund multiplicity and one of the
following two Cambridge-based definitions: (i) running the Cambridge
algorithm with $y_\text{cut}=1$ and counting the
clusterings with $k_t^\text{(Cam)}>\ktcut$, i.e.\ using the Cambridge
definition of $k_t$ instead of the Lund definition, and (ii) using the
Cambridge jet multiplicity with $y_\text{cut}=\ktcut^2/Q^2$.
Our results are shown in Fig.~\ref{fig:cam-v-lund}.
In Fig.~\ref{fig:cam-v-lund-order1}, we plot differences between the
multiplicities at order $\as$.
We see that the Lund and Cambridge multiplicities indeed differ by a
constant which corresponds to the \NNDL expectation of $C_F\pi^2/6$
for the full event, cf.\ Sec.~\ref{sec:nndl-cam},
Eq.~(\ref{eq:NNDL-cam-result}).
At $\order{\as^2}$, Fig.~\ref{fig:cam-v-lund-order2}, we instead
consider the difference in the differential multiplicity (as done in
Fig.~\ref{fig:v-event2-order2}, Sec.~\ref{sec:event2}). The fact
that the differences tend to a constant at large $L$ indicates that
the two distributions agree at \NNDL accuracy, as expected from
our analytic calculations.

\section{$1\to 3$ splitting functions in the soft limit}\label{app:triple-collinear}

We provide explicit expressions for the key two ingredients in
Eq.~\eqref{eq:NNDL-double-soft-FO}: the (soft limit of the) $1\to 3$
splitting function $\hat P_{1\to 3}$ describing the emission of two
collinear particles at commensurate energies and angles from an
initial hard parton, and the corresponding \NDL subtraction term,
$\mathcal P$.
The $1\to 3$ splitting function were computed in
Ref.~\cite{Catani:1999ss}, where we can set $\varepsilon=0$ and take
the soft limit, $z_3\to 1$.
In this limit, the energy fractions and invariant masses entering the
splitting functions are related to the variables
introduced in Fig.~\ref{fig:double-soft-parametrisation}
through\footnote{In principle, the invariant masses are proportional
  to the square of the energy of the initial parton. Since we are
  always working with ratios of $s_{ij}$, these factors cancel and
  have been omitted for simplicity.}
\begin{subequations} \label{eq:web-catani}
  \begin{align}
    z_1 & = (1-z)z_p, \qquad \qquad \qquad \qquad z_2 = (1-z)(1-z_p), \\
    s_{12} & = (1-z)^2 z_p (1-z_p) \theta_a^2 \normtheta_{12}^2, \\
    s_{13} & = (1-z)\theta_a^2 z_p \left[1+2(1-z_p)\cos\phi_{12} 
    \normtheta_{12} +  (1-z_p)^2 \normtheta_{12}^2\right], \\ 
    s_{23} & = (1-z) \theta^2_a (1-z_p)\left[
      1-2z_p\cos\phi_{12}\normtheta_{12}+z_p^2\normtheta_{12}^2\right], \\
    s_{123} & = s_{12} + s_{13} + s_{23} \overset{z_3\to 1}{\approx} s_{13} + s_{23} = (1-z)\theta_a^2 
      \left[1+z_p(1-z_p)\normtheta_{12}^2\right], \\
    t_{12,3}& =2\frac{z_1s_{23}-z_2 s_{13}}{z_1+z_2}+
      \frac{z_1-z_2}{z_1+z_2}s_{12}
      \overset{z_3\to 1}{\approx} 2[z_ps_{23}-(1-z_p) s_{13}].
  \end{align}
\end{subequations}
Note that $s_{12}$ is proportional to $(1-z)^2$ while all other
invariants scale as $1-z$.
Also, only $s_{13}$ and $s_{23}$ (and, hence $t_{12,3}$) carry a
dependence on the azimuthal angle $\phi$.

The various $1\to 3$ splitting functions given below are taken with
the same normalisation as in the original
reference~\cite{Catani:1999ss}. When using them in practice in
Eq.~(\ref{eq:NNDL-double-soft-FO}), we need to insert the appropriate
symmetry factors.

\paragraph{Quark-initiated} The splitting of a quark can lead to two
different configurations: (i) a quark and two gluons,
$\hat P_{g_1g_2q_3}$ with an Abelian contribution proportional to
$C_F^2$ and a non-Abelian contribution proportional to $C_FC_A$, or
(ii) a quark plus a $q\bar q$ pair,
$\hat P_{\bar q_1 q_2 q_3} \propto C_F n_f T_R$.\footnote{The case in
  which all quarks have identical flavours includes a term
  proportional to $C_F(C_F-C_A/2)$. Since this term is finite
  in the $z\to 1$ limit (see e.g.~\cite{Dasgupta:2021hbh}) it is
  beyond our \NNDL accuracy.}
We begin by considering case (i).
If is straightforward to see that in the limit $z\to 1$ the $C_F^2$
contribution factorises in a product of independent soft-collinear
emissions $(2C_F)^2/(z_1z_2)$ and therefore does not
bring any \NNDL contribution.
For the non-Abelian contribution, the matrix element is given by
Eq.~(62) in Ref.~\cite{Catani:1999ss} and its soft limit reads:
\begin{align}\label{eq:P13-ggq}
  z_p(1-z_p)(1-z)^2\hat P_{g_1g_2q_3}=  C_F C_A
  &\bigg\lbrace
  2 \frac{[z_ps_{23}-(1-z_p)s_{13}]^2}{s_{12}\theta_a^2\normtheta_{12}^2}
    +\frac{s_{123}}{s_{13}} \left(z_pr-1\right) (2-z_p)  \nonumber \\
  &+\frac{s_{123}}{s_{23}}\left[(1-z_p)r-1\right](1+z_p)
    +r[1-8z_p(1-z_p)]\bigg\rbrace
\end{align}
with $r\equiv (1-z)s_{123}/s_{12}=[1+z_p(1-z_p)\normtheta_{12}^2]/[z_p(1-z_p)\normtheta_{12}^2]$. 
The \NDL contribution corresponds to the $P_{g\to gg}$ splitting function that, 
with our choice of variables, reads
\begin{align}
\dis\int \frac{\dd \phi_{12}}{2\pi} \mathcal{P}_{g_1g_2q_3} = 
4C_F C_A\frac{[1-z_p(1-z_p)]^2}{z_p(1-z_p)}\Theta\big(\normtheta_{12}< 1\big).
\label{eq:app-sub-ggg-corr}
\end{align}
where the factor $4$ appears due to the phase-space definition in 
Eq.~\eqref{eq:NNDL-double-soft-FO}. We have explicitly performed the 
$\phi$-integration to highlight that at \NDL the two emissions are angular 
ordered and as encompassed by the $\Theta(\normtheta_{12}< 1)$ 
constraint.

We now treat the case in which a quark emits a soft gluon that
subsequently decays into a quark/anti-quark pair with commensurate
energies and angle.
After taking the soft limit of the triple collinear splitting function
given in this case by Eq.~(57) in Ref.~\cite{Catani:1999ss}, we obtain
\begin{align}
\label{eq:P13-qqq}
z_p(1-z_p)(1-z)^2\hat P_{\bar q_1 q_2 q_3}&= 
2C _F n_f T_R 
\left[-\frac{[z_ps_{23}-(1-z_p)s_{13}]^2}{s_{12}\theta_a^2\normtheta_{12}^2}+z_p(1-z_p)r \right].
\end{align}
In this case, the \NDL subtraction term is given by 
\begin{align}
\int \frac{\dd \phi_{12}}{2\pi} \mathcal{P}_{\bar q_1 q_2 q_3} = 
4C_F n_f T_R[z^2_p+(1-z_p)^2]\Theta\big(\normtheta_{12}< 1\big)\, .
\label{app-sub-qqq}
\end{align}

\paragraph{Gluon-initiated} We have again two cases to consider for a gluon 
splitting since it can generate: (i) two more gluons, $\hat P_{g_1g_2g_3} 
\propto C^2_A$, or (ii) a $q\bar q$ pair, $\hat P_{q_1\bar q_2 g_3}$,
with a contribution proportional to $C_A n_f T_R$ which survives in
the soft limit $z_3\to 1$ and a contribution proportional to $C_FC_A$
which does not.
The $q\bar q$ channel is then given by the soft limit of Eq.~(69) in
Ref.~\cite{Catani:1999ss},\footnote{Compared to~\cite{Catani:1999ss},
  we have reordered the indices so that the gluon is the third parton
  and the soft limit still corresponds to $z_3\to 1$.} i.e.
\begin{equation}
\label{eq:P13-qqg}
  z_p(1-z_p)(1-z)^2\hat P_{q_1 \bar q_2 g_3}= 2 C_An_f T_R
  \left[
    -\frac{[z_ps_{23}-(1-z_p)s_{13}]^2}{s_{12}\theta_a^2\normtheta_{12}^2}
    +z_p(1-z_p)r \right]
\end{equation}
and the subtraction term is given by Eq.~\eqref{app-sub-qqq}
with the overall factor $C_F$ replaced by $C_A$.
Note that this result is equivalent to the production of a $q\bar q$
pair from an initial hard quark, up to the overall colour factor
($C_A$ or $C_F$).
This explains why the $D_\text{pair}^{g \to q\qbar g}$ and
$D_\text{pair}^{q \to q \qbar' q}$ coefficients in
Eq.~\eqref{eq:NNDL-double-soft-D-qq} are equal.

Finally, we tackle the case of a $g\to ggg$ splitting. Two
complications arise with respect to the previous cases due to the
indistinguishability of the gluons: (i) all six permutations of the
gluon momenta have to be taken into account in the splitting function
and (ii) the full $g\to g_1g_2g_3$ splitting function also includes
logarithmically-enhanced contributions where either $g_1$ or $g_2$ is
collinear to $g_3$. The corresponding \NDL contributions should
also be subtracted in order to obtain the \NNDL correction.
The triple collinear splitting function is provided in Eq.~(70) of
Ref.~\cite{Catani:1999ss}. After taking the $z_3 \to 1$ limit in all 6
permutations it reduces to
\begin{align}
\label{eq:P13-ggg}
  z_p(1-z_p)(1-z)^2 \hat P_{g_1 g_2 g_3} = C_A^2
  &\bigg\lbrace 
    2 \frac{[z_ps_{23}-(1-z_p)s_{13}]^2}{s_{12}\theta_a^2\normtheta_{12}^2}
    + \frac{s_{123}}{s_{13}}(z_pr-1)(2-z_p)  \\
  & + \frac{s_{123}}{s_{23}}[(1-z_p)r-1](1+z_p)
    +r\left[1-8z_p(1-z_p)\right]
    +\frac{4s_{123}^2}{s_{13}s_{23}}
    \bigg\rbrace \nonumber
\end{align}
We note that since we have made the explicit choice to take $z\equiv
z_3\to 1$, the symmetry factor associated with this contribution is
$1/2$ and not $1/3!$.

The \NDL subtraction term is a bit more involved than in the previous
cases. On top of the contribution where the first soft gluon undergoes
a hard-collinear splitting, we need to account for the case in which the two 
soft gluons are emitted off the hard one.
This contribution is more naturally first written in terms of the
$\theta_{13}$ and $\theta_{23}$ angles, followed by a
change in  variables to $\theta_a$ and $\normtheta_{12}$ by using
\begin{equation}
  \left(\frac{2\as C_A}{\pi}\right)^2
  \frac{\dd z_1}{z_1}\frac{\dd z_2}{z_2}
  \frac{\dd \theta^2_{13}}{\theta^2_{13}}
  \frac{\dd \theta^2_{23}}{\theta^2_{23}}
  \frac{\dd \phi}{2\pi}
  =
  \left(\frac{2C_A\as}{\pi}\right)^2
  \frac{\dd z}{1-z}\,\dd z_p
  \frac{\dd \theta^2_q}{\theta^2_g}
  \frac{\dd \normtheta^2_{12}}{\normtheta^2_{12}}
  \frac{\dd \phi}{2\pi}
  \frac{\normtheta_{12}^2}{[1+z_p(1-z_p)\normtheta_{12}^2]^2}\frac{s_{123}^2}{s_{13}s_{23}}.
\end{equation}
These steps yield the following \NDL subtraction term:
\begin{align}
  \label{eq:ggg-subtraction-term}
\mathcal P = \left(\frac{2\as C_A}{\pi}\right)^2 \left[P_{g\to gg} (1-z_p) + \frac{s^2_{123}}{s_{13}s_{23}}\frac{\normtheta_{12}^2}{[1+z_p(1-z_p)\normtheta_{12}^2]^2}\right].
\end{align}
It is interesting to note that the presence of the right-most term in
Eq.~\eqref{eq:ggg-subtraction-term} cancels exactly the last term in
Eq.~\eqref{eq:P13-ggg}, so 
that, up to the overall colour factor, the remaining part is the same
as for the emission of two soft gluons from a hard quark,
Eq.~\eqref{eq:P13-ggq}, explaining the equality between
$D_\text{pair}^{g \to ggg}$ and $ D_\text{pair}^{q\to ggq}$ in
\eqref{eq:NNDL-double-soft-D-gg}.

\bibliographystyle{JHEP}
\bibliography{multiplicity}

\end{document}